\documentclass[10pt,letterpaper]{article}\usepackage[]{graphicx}\usepackage[table]{xcolor}
\makeatletter
\def\maxwidth{ %
  \ifdim\Gin@nat@width>\linewidth
    \linewidth
  \else
    \Gin@nat@width
  \fi
}
\makeatother

\definecolor{fgcolor}{rgb}{0.345, 0.345, 0.345}
\newcommand{\hlkwb}[1]{\textcolor[rgb]{0.69,0.353,0.396}{#1}}%

\usepackage{framed}
\makeatletter
 {\par\unskip\endMakeFramed%
 \at@end@of@kframe}
\makeatother

\definecolor{shadecolor}{rgb}{.97, .97, .97}
\definecolor{messagecolor}{rgb}{0, 0, 0}
\definecolor{warningcolor}{rgb}{1, 0, 1}
\definecolor{errorcolor}{rgb}{1, 0, 0}
\newenvironment{knitrout}{}{} 

\usepackage{alltt}
\usepackage[top=0.85in,left=2.75in,footskip=0.75in]{geometry}

\usepackage{amsmath,amssymb}

\usepackage{changepage}

\usepackage{textcomp,marvosym}

\usepackage{cite}

\usepackage{nameref,hyperref}

\usepackage[nopatch=eqnum]{microtype}
\DisableLigatures[f]{encoding = *, family = * }

\usepackage{array}

\newcolumntype{+}{!{\vrule width 2pt}}

\newlength\savedwidth

\raggedright
\usepackage[aboveskip=1pt,labelfont=bf,labelsep=period,justification=raggedright,singlelinecheck=off]{caption}

\makeatletter
\renewcommand{\@biblabel}[1]{\quad#1.}
\makeatother

\usepackage{lastpage,fancyhdr,graphicx}
\usepackage{epstopdf}
\fancyheadoffset[L]{2.25in}
\fancyfootoffset[L]{2.25in}
\RequirePackage{amsthm,amsmath,amsfonts,amssymb,graphicx,enumerate,url,xr,lmodern}
\usepackage{url}
\usepackage{makecell}
\usepackage{multirow}
\usepackage{multicol}
\usepackage[ruled,noline,linesnumbered]{algorithm2e}
\usepackage{color}
\usepackage[normalem]{ulem}
\usepackage{bm}
\usepackage[mathscr]{euscript}

\newcommand\Wsat{W_{\mathrm{sat}}}
\newcommand\muIR{\mu_{IR}}
\newcommand\muEI{\mu_{EI}}
\newcommand\transmission{\beta}
\newcommand\seasAmplitude{a}
\newcommand\rainfallExponent{r}
\newcommand\muRS{\mu_{RS}}
\newcommand\vaccineEfficacy{\vartheta}
\newcommand\muBirth{\mu_S}
\newcommand\muDeath{\delta}
\newcommand\choleraDeath{\delta_{C}}
\newcommand\symptomFrac{f}
\newcommand\asymptomRelativeInfect{\epsilon}
\newcommand\asymptomRelativeShed{\epsilon_{W}}
\newcommand\Wbeta[1]{\beta_{W#1}}
\newcommand\Whur[1]{\beta_{W#1}^{hm}}
\newcommand\hHur[1]{h_{#1}^{hm}}
\newcommand\tHur{t_{hm}}
\newcommand\Iinit{I_{0,0}}

\newcommand\Wremoval{\delta_W}
\newcommand\Wshed{\mu_W}
\newcommand\mixExponent{\nu}
\newcommand\sigmaProc{\sigma_{\mathrm{proc}}}
\newcommand\reportRate{\rho}
\newcommand\obsOverdispersion{\psi}
\newcommand\phaseParm{\phi}
\newcommand\transmissionTrend{\zeta}

\newcommand\vaccClass{Z}
\newcommand\vaccCounter{z}

\newcommand\missing{}
\newcommand\fixed{^\dagger}
\newcommand\demography{\bullet}
\newcommand\code[1]{\texttt{#1}}
\newcommand\figTitle{\bf}
\newcommand\paramVec{\theta}
\newcommand\childReduce{q}
\newcommand\NBintercept{\alpha}
\newcommand\NBar{\beta}
\newcommand\NBsize{\varphi}

\DeclareSymbolFont{matha}{OML}{txmi}{m}{it}
\DeclareMathSymbol{\varv}{\mathord}{matha}{118}

\newcommand\myeqref[1]{(\ref{#1})}
\newcommand{\blind}{1}

\newcommand\seq[2]{{#1}\!:\!{#2}}

\newcommand\Var{\mathrm{Var}}

\newcolumntype{t}{>{\tiny}c}
\IfFileExists{upquote.sty}{\usepackage{upquote}}{}
\begin{document}
\vspace*{0.2in}

\begin{flushleft}
{\Large
\textbf\newline{Informing policy via dynamic models: Cholera in Haiti} 
}
\newline
\\
Jesse Wheeler\textsuperscript{1*},
AnnaElaine L. Rosengart\textsuperscript{2},
Zhuoxun Jiang\textsuperscript{1},
Kevin Tan\textsuperscript{3},
Noah Treutle\textsuperscript{1},
Edward L. Ionides\textsuperscript{1}
\\
\bigskip
\textbf{1} Statistics Department, University of Michigan, Ann Arbor, Michigan, USA
\\
\textbf{2} Statistics and Data Science, Carnegie Mellon University, Pittsburgh, Pennsylvania, USA
\\
\textbf{3} Wharton Statistics and Data Science, University of Pennsylvania, Philadelphia, Pennsylvania, USA
\\
\bigskip

%
%



* jeswheel@umich.edu

\end{flushleft}

\section*{Abstract}
Public health decisions must be made about when and how to implement interventions to control an infectious disease epidemic.
These decisions should be informed by data on the epidemic as well as current understanding about the transmission dynamics.
Such decisions can be posed as statistical questions about scientifically motivated dynamic models.
Thus, we encounter the methodological task of building credible, data-informed decisions based on stochastic, partially observed, nonlinear dynamic models.
This necessitates addressing the tradeoff between biological fidelity and model simplicity, and the reality of misspecification for models at all levels of complexity.
We assess current methodological approaches to these issues via a case study of the 2010-2019 cholera epidemic in Haiti.
We consider three dynamic models developed by expert teams to advise on vaccination policies.
We evaluate previous methods used for fitting these models, and we demonstrate modified data analysis strategies leading to improved statistical fit.
Specifically, we present approaches for diagnosing model misspecification and the consequent development of improved models.
Additionally, we demonstrate the utility of recent advances in likelihood maximization for high-dimensional nonlinear dynamic models, enabling likelihood-based inference for spatiotemporal incidence data using this class of models.
Our workflow is reproducible and extendable, facilitating future investigations of this disease system.

\section*{Author summary}
Quantitative understanding of infectious disease transmission dynamics relies upon mathematical models informed by scientific knowledge and relevant data.
The models aim to provide a statistical description of the trajectory of an epidemic and its uncertainty, together with a representation of the underlying biological mechanisms.
Evaluation of success at these goals is necessary in order for a model to provide a reliable tool for guiding evidence-based public policy interventions.
In this article, we conduct a re-analysis of the 2010-2019 cholera outbreak in Haiti. We use this case study to investigate current procedures for fitting mechanistic models to time series data, while identifying limitations of these methodologies and proposing remedies.
Our analysis presents methodology for diagnosing how well a model describes observed data. Using objective measures to assess model fit ensures that our evaluation is based on quantifiable criteria. Incorporating reproducibility into this assessment results in a framework that enables the validation or refinement of model based inferences when revisiting the data, facilitating scientific discovery.
Our data analysis workflow is supported by recent advances in algorithms, software and hardware, which facilitate statistical fitting of nonlinear stochastic dynamic models to observed incidence data.
However, inference for high-dimensional systems remains a methodological challenge.
One of the models under consideration involves spatially coupled stochastic meta-populations, and we demonstrate how a recently developed algorithm permits likelihood-based inference and model diagnostics in this setting.
We contend that raising the currently accepted standards of infectious disease modeling will result in a greater ability of scientists and policy makers to understand and respond to future infectious disease outbreaks.

\section*{Introduction}

Regulation of biological populations is a fundamental topic in epidemiology, ecology, fisheries and agriculture.
Population dynamics may be nonlinear and stochastic, with the resulting complexities compounded by incomplete understanding of the underlying biological mechanisms and by partial observability of the system variables.
Quantitative models for these dynamic systems offer potential for designing effective control measures \cite{vandermeer13,he10}.
Developing and testing models for dynamic systems, and assessing their fitness for guiding policy, is a challenging statistical task \cite{king16}.
Questions of interest include: What indications should we look for in the data to assess whether the model-based inferences are trustworthy?
What diagnostic tests and model variations can and should be considered in the course of the data analysis?
What are the possible trade-offs of increasing model complexity, such as the inclusion of interactions across spatial units?

\begin{figure}[!h]
\begin{knitrout}
\definecolor{shadecolor}{rgb}{0.969, 0.969, 0.969}\color{fgcolor}
\includegraphics[width=\maxwidth]{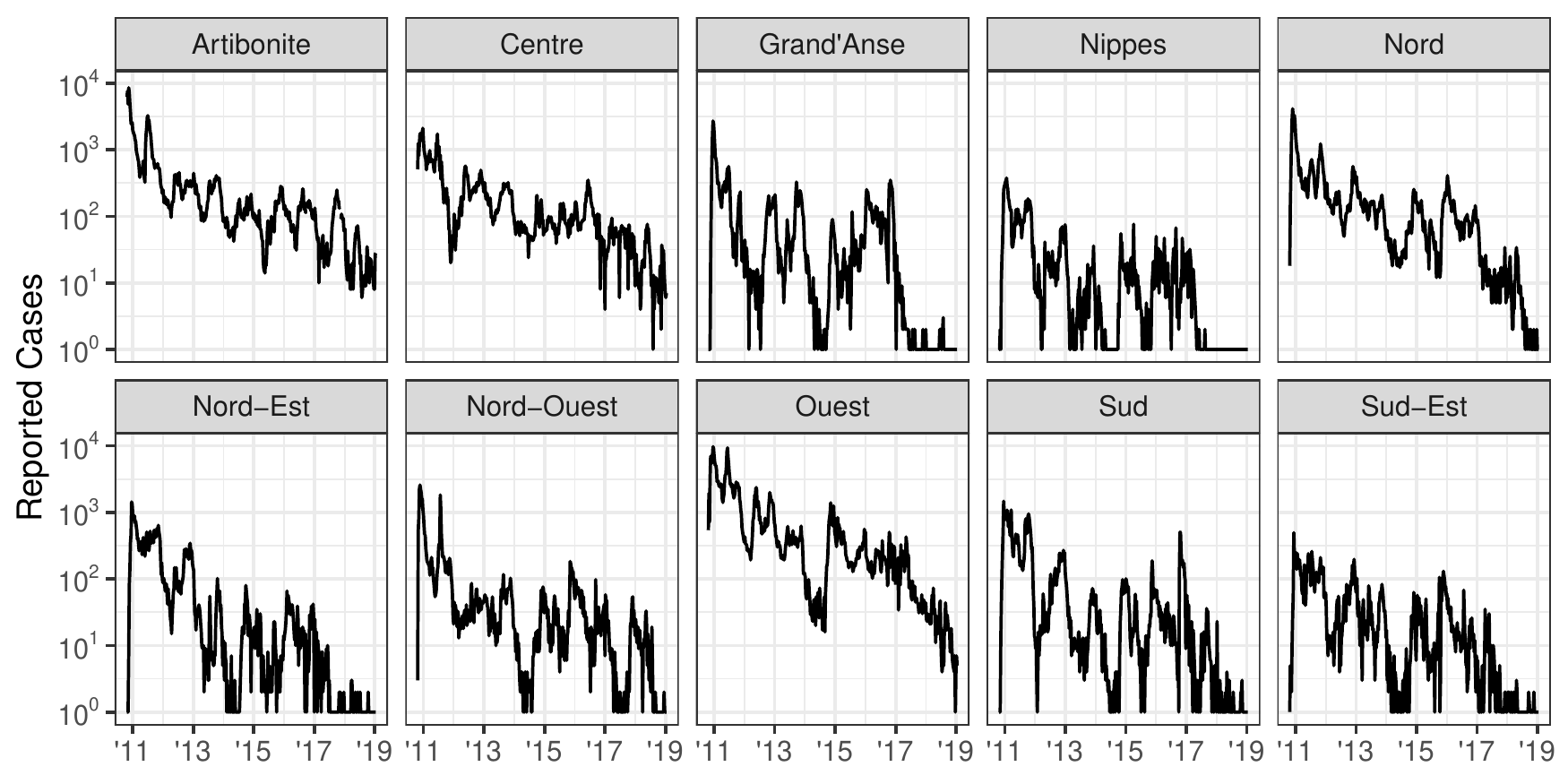} 
\end{knitrout}
\caption{\label{fig:CholeraData}
{\figTitle Weekly cholera cases.}
Weekly reported cholera cases in Haiti from October 2010 to January 2019 for each of the 10 administrative departments.
}
\end{figure}

This case study investigates the use of dynamic models and spatiotemporal data to inform public health policy in the context of the cholera outbreak in Haiti, which started in 2010.
Various dynamic models were developed to study this outbreak: searching PubMed with keywords ``Haiti, cholera, model'' we obtained 22 studies that utilized deterministic mechanistic dynamic models \cite{lee20,tuite11,andrews11,botelho21,fitzgibbon20,eisenberg13,rinaldo12,chao11,date11,lin19,abrams13,collins21,akman15,trevisin22,mavian20,collins14,kelly16,capone15,leung22,mari15,gatto12,kuhn14}
and 11 studies that used stochastic models \cite{kirpich16,lee20,pasetto18,mukandavire13,kirpich17,lewnard16,kunkel17,mukandavire15,sallah17,azman12,azman15}.
Incidence data on the outbreak are available at various spatial scales, motivating 17 studies in our literature review to consider spatially explicit dynamic models \cite{lee20,tuite11,pasetto18,fitzgibbon20,eisenberg13,rinaldo12,chao11,abrams13,trevisin22,sallah17,collins14,kelly16,azman12,leung22,kuhn14,mari15,gatto12}.
Here we focus on a multi-group modeling exercise by Lee et al.~\cite{lee20} in which four expert modeling teams developed models to the same dataset with the goal of comparing conclusions on the feasibility of eliminating cholera by a vaccination campaign.
Model~1 is stochastic and describes cholera at the national level;
Model~2 is deterministic with spatial structure, and includes transmission via contaminated water;
Model~3 is stochastic with spatial structure, and accounts for measured rainfall.
Model~4 has an agent-based construction, featuring considerable mechanistic detail but limited ability to calibrate these details to data.
These modeling strategies were selected by Lee et a.~\cite{lee20} to represent the range of approaches used in the research community.
We focus on Models~1--3, as the strengths and weaknesses of the agent-based modeling approach \cite{tracy18} are outside the scope of this article.
In addition, agent-based models were less widely used, the agent based model in \cite{lee20} being the only model of this class that was found in our literature review.
The data that were used in \cite{lee20}, and that we reanalyze, are displayed in Fig.~\ref{fig:CholeraData}.

The four independent teams were given the task of estimating the
potential effect of prospective oral cholera vaccine (OCV) programs.
While OCV is accepted as a safe and effective tool for controlling the spread of cholera, the global stockpile of OCV doses remains limited \cite{pezzoli20}.
Advances in OCV technology and vaccine availability, however, raised the possibility of planning a national vaccination program.
The possibility of controlling the Haiti cholera outbreak via OCV was considered by various research groups \cite{lee20,ivers17,andrews11,walton11,matias17,chao11,fung13,date11,leung22,azman15}.
In the Lee et al.~\cite{lee20} study, certain data were shared between the groups, including demography and vaccination history; vaccine efficacy was also fixed at a shared value between groups.
Beyond this, the groups made autonomous decisions on what to include and exclude from their models.
Despite their autonomy, the four independent teams obtained a consensus that an extensive nationwide vaccination campaign would be necessary to eliminate cholera from Haiti, estimating that a large number of cumulative cholera cases would be observed in the absence of additional vaccination efforts (Figure~3 and 4 of \cite{lee20}).
These forecasts are inconsistent with the prolonged period with no confirmed cholera cases between February, 2019 and September, 2022 \cite{trevisin22}.
Though cholera has recently reemerged in Haiti \cite{rubin22,PAHO23}, the inability to accurately forecast cholera incidence from 2019-2022 prompts us to consider retrospectively what may have been done differently in order to obtain more reliable conclusions, leading to recommendations for future studies.

The discrepancy between the model-based conclusions of Lee et al.~\cite{lee20} and the prolonged absence of cholera in Haiti has been debated \cite{francois20,rebaudetComment20,henrys20,leeReply20}.
Suggested origins of this discrepancy include the use of unrealistic models \cite{rebaudetComment20} and unrealistic criteria for cholera elimination \cite{henrys20}.
We find a more nuanced conclusion: attention to methodological details in model fitting, diagnosis and forecasting can improve each of the proposed model's ability to quantitatively describe observed data.
This improved ability may result in more accurate forecasts and facilitates the exploration of model assumptions.
Based on this retrospective analysis, we offer suggestions on fitting mechanistic models to dynamic systems for future studies.

Numerous guidelines have been proposed for using mechanistic models to inform policy, reviewed in \cite{behrend20}. Behrend et al.~\cite{behrend20} identify the importance of stakeholder engagement, transparency, reproducibility, uncertainty communication, and testable model outcomes. These and related principles are echoed by other influential articles \cite{saltelli20,donnelly18}. Additional literature emphasizes model calibration and evaluation techniques \cite{dahabreh17,egger17,penaloza15}. These guidelines often lack implementation specifics. As an example, \cite{lee20} largely adhere to the principles of \cite{behrend20}---though assessing the extent of stake-holder engagement is challenging---yet their projections are inconsistent with actual cholera incidence data from 2019 to 2022, demonstrating the limitations of current standards. We provide methodology for rigorous statistical calibration and evaluation of dynamic models (as advocated by \cite{saltelli19}), thereby expanding on the prevailing guidance. We specifically emphasize principles that prove essential in our case study. Complementary methodological suggestions arising from a spatio-temporal analysis of COVID-19 are detailed in \cite{li23}.

Our recommendations are presented in the context of a case study, with the goal of demonstrating how careful adherence to statistical principles may result in improved model fits.
We proceed by introducing the general modeling scheme employed by Models~1--3 and provide details of each individual model;
we then describe how each model is calibrated to data, and present a systematic approach to examining and refining these models.
Specifically, we focus on how to develop and test variations of the proposed models, as well as diagnosing the models once they have been assimilated to incidence reports.
This includes a comprehensive tutorial on performing inference with Model~3 (\nameref{S_mod3cal}), a highly non-linear, spatially explicit stochastic model, a challenging task that is possible due to recent methodological advancements.
We then use the improved model fits to project cholera incidence in Haiti under various vaccination scenarios considered by Lee et al.~\cite{lee20}.
Finally, we conclude with a discussion of the results, in which we relate our general recommendations for model based inference of biological systems to the case study of the Haiti cholera outbreak.

\section*{Materials and methods}
\subsection*{Mechanistic models for disease modeling}

Mechanistic models representing biological phenomena are valuable for epidemiology and consequently for public health policy \cite{lofgren14,mccabe21}. More broadly, they have useful roles throughout biology, especially when combined with statistical methods that properly account for stochasticity and nonlinearity \cite{may04}. In some situations, modern machine learning methods can outperform mechanistic models on epidemiological forecasting tasks \cite{lau22,baker18}. The predictive skill of non-mechanistic models can reveal limitations in mechanistic models, but cannot readily replace the scientific understanding obtained by describing the biological dynamics of the system in a mathematical model \cite{baker18,prosperi20}.

In this article, we refer to models that focus on learning relationships between variables in a dataset as {\it associative}, whereas models that incorporate a known scientific property of the system we call {\it causal} or {\it mechanistic}.
The danger in using forecasting techniques which rely on associative models to predict the consequence of interventions is called the Lucas critique in an econometric context.
Lucas et al.~\cite{lucas76} pointed out that it is naive to predict the effects of an intervention on a given system based entirely on historical associations.
To successfully predict the effect of an intervention, a model should therefore both provide a quantitative explanation of existing data and should have a causal interpretation: a manipulation of the system should correspond quantitatively with the corresponding change to the model.
This motivates the development of mechanistic models, which provides a statistical fit to the available data while also supporting a causal interpretation.
Despite their limited ability to project the effect of interventions on a system, associative models can be effectively used to make inference on an certain features of a system.
In our literature review, there were 14 studies that used associative models to describe various aspects of the cholera epidemic \cite{hulland19,moise20,piarroux11,matias17,eisenberg13,rebaudet19CATI,ivers15,raila17,cuneo17,charles14,richterman19,richterman19inverse,bengtsson15,li16adjusting}.

The four mechanistic models of Lee et al.~\cite{lee20} were deliberately developed with limited coordination.
This allows us to treat the models as fairly independently developed expert approaches to understanding cholera transmission.
However, it led to differences in notation, and in subsets of the data chosen for analysis, that hinder direct comparison.
Here, we have created a common notational framework that facilitates model comparison, and put all comparable model parameters---including parameters that were estimated or held constant---into Table~\ref{tab:params}.
Translations back to the original notation of Lee et al.~\cite{lee20} are given in the supplement (\nameref{S1_Table}).

\begin{table}
\begin{adjustwidth}{-2.25in}{0in}
\centering
\caption{\label{tab:params} {\bf Model parameters.}}
\begin{tabular}{|p{0.15\linewidth}|p{0.14\linewidth}t|p{0.15\linewidth}t|p{0.17\linewidth}t|}
\hline
Mechanism & Model 1 && Model 2 && Model 3 &
\\
\hline
\hline
Infection {\small (day)}
  & ${\muIR^{-1}=2.0}\fixed$
  & \eqref{model1:toR}
  & ${\muIR^{-1}=7.0}\fixed $
  & \eqref{model2:mu_IR}
  & ${\muIR^{-1}=5.0}\fixed $
  & \eqref{eq:model3:IR}
\\\hline
Latency {\small (day)}
  & ${\muEI^{-1}=1.4}\fixed$
  & \eqref{model1:EA}
  & ${\muEI^{-1}=1.3}\fixed$
  & \eqref{model2:mu_EI}
  & \missing
  &
\\
\hline
\multirow{3}{*}{Seasonality}
  & $\transmission_{1:6}=(
      1.4,\allowbreak
      1.2,\allowbreak
      1.1,\allowbreak
      1.1,\allowbreak
      1.4,\allowbreak
      1.0)$
  & \eqref{model1:beta}
  & \multirow{2}{*}{${\seasAmplitude=0.4}\fixed$}
  & \multirow{2}{*}{\eqref{model2:lambda}}
  & \multirow{2}{*}{$\seasAmplitude = 1.00$}
  & \multirow{2}{*}{\eqref{eq:model3:water}}
  \\
  & $\transmissionTrend = -0.04^*$
  & \eqref{model1:betat}
  & \multirow{-2}{*}{$\phaseParm = 0.97^*$}
  & \multirow{-2}{*}{\eqref{model2:lambda}}
  & \multirow{-2}{*}{$\rainfallExponent = 0.78$}
  & \multirow{-2}{*}{\eqref{eq:model3:water}}
  \\\hline
  \multirow{3}{*}{Immunity {\small (yr)}}
  & \multirow{3}{*}{${\muRS^{-1}=8.0}\fixed$}
  & \multirow{3}{*}{\eqref{model1:RS}}
  & $\muRS^{-1}= \ensuremath{1.4\times 10^{11}}$
  & \eqref{model2:RS}
  & \multirow{3}{*}{${\muRS^{-1}=8.0}\fixed$}
  & \multirow{3}{*}{\eqref{eq:model3:RRnext}}
\\
& & & ${\omega_1^{-1} = 1.0}\fixed$ & \eqref{model2:omega1} & &\\
& & & ${\omega_2^{-1} = 5.0}\fixed$ & \eqref{model2:omega2} & &\\
\hline
\multirow{2}{*}{Birth/death {\small (yr\textsuperscript{-1})}}
  & ${\muBirth =  10^{-2}\times 2.23}\fixed$
  & \eqref{model1:death}
  & \multirow{2}{*}{\missing}
  &
  & ${\muDeath = 10^{-2}\times1.59}\fixed$
  & \multirow{2}{*}{\eqref{eq:model3:IS}}
\\
& ${\muDeath = 10^{-3}\times7.5}\fixed$
& \eqref{model1:death}
&
&
& ${\choleraDeath = 1.46}\fixed$
&
\\\hline
Sympt. frac.
  & ${\symptomFrac_z(t)=c\vaccineEfficacy^*(t-\tau_d)}\fixed$
  & (\ref{model1:EI}-\ref{model1:EA})
  & ${\symptomFrac=0.2}\fixed$
  & \eqref{model2:mu_EI}
  & ${\symptomFrac=0.25}\fixed$
  & \eqref{eq:model3:SA}
\\\hline
Asympt.
  & \multirow{2}{*}{${\asymptomRelativeInfect=0.05}\fixed$}
  & \multirow{2}{*}{\eqref{model1:lambda}}
  & ${\asymptomRelativeInfect = 0.001}\fixed$
  & \eqref{model2:lambda}
  &  ${\asymptomRelativeInfect = 1}\fixed$
  & \eqref{eq:model3:foi}
\\
infectivity
&
&
& ${\asymptomRelativeShed = \ensuremath{10^{-7}}}\fixed$
& \eqref{model2:to_W}
& $\asymptomRelativeShed = 0.008$
& \eqref{eq:model3:water}
\\
\hline
\multirow{4}{*}{Human to human}
  & \multirow{4}{*}{$\transmission_{1:6}$ as above}
  & \multirow{4}{*}{\eqref{model1:lambda}}
  & \multirow{4}{*}{$\transmission=$\ensuremath{5.97\times 10^{-15}} {\small yr\textsuperscript{-1}}}
  & \multirow{4}{*}{\eqref{model2:lambda}}
  & $\transmission_{1:10}=(
  0.82, \allowbreak
	0.02, \allowbreak
	0.38, \allowbreak
  0.21, \allowbreak
	0.51, \allowbreak
	0.51, \allowbreak
	0.35, \allowbreak
	0.12, \allowbreak
	0.26, \allowbreak
	0.10
	) \allowbreak \times 10^{-6}$ {\small yr\textsuperscript{-1}}
  & \multirow{4}{*}{\eqref{eq:model3:foi}}
\\
\hline
\multirow{5}{*}{Water to human}
  & \multirow{5}{*}{\missing}
  &
  & \multirow{3}{*}{${\Wsat = \ensuremath{10^{5}}}\fixed$}
  & \multirow{3}{*}{\eqref{model2:lambda}}
  & $\Wbeta{_{1:10}}= (
        4.70, \allowbreak
        21.00, \allowbreak
        24.97, \allowbreak
        27.14, \allowbreak
        5.28, \allowbreak
        30.70, \allowbreak
        10.17, \allowbreak
        0.99, \allowbreak
        11.89, \allowbreak
        12.82)$ {\small yr\textsuperscript{-1}}
  & \multirow{4}{*}{\eqref{eq:model3:foi}}
\\
& & & \multirow{-3}{*}{$\beta_W= 1.1$ {\small yr\textsuperscript{-1}}} & \multirow{-3}{*}{\eqref{model2:lambda}} & &
\\
\hline
Human to water
  & \missing
  &
  & $\Wshed = $ 179 {\small wk\textsuperscript{-1}}
  & \eqref{model2:to_W}
  & $\Wshed= \ensuremath{9.77\times 10^{-7}}$ {\tiny	$\frac{\mathrm{km^2}}{\mathrm{wk}}$}
  & \eqref{eq:model3:water}
\\
Water survival {\footnotesize (wk)}
  & \missing
  &
  & ${\Wremoval^{-1} = 3}\fixed$
  & \eqref{model2:from_W}
  & $\Wremoval^{-1}=0.11$
  & \eqref{eq:model3:Decay}
\\\hline
Mixing exponent
  & $\mixExponent=0.98$
  & \eqref{model1:lambda}
  & \missing
  &
  & \missing
  &
\\
\hline
\raggedright Process noise {\small (wk\textsuperscript{1/2})}
  & \raggedright $\sigmaProc = (0.09,
  0.12)^*$
  & \multirow{2}{*}{\eqref{model1:lambda}}
  & \multirow{2}{*}{\missing}
  &
  & \multirow{2}{*}{$\sigmaProc= 0.218$}
  & \multirow{2}{*}{\eqref{eq:model3:SA}}
\\\hline
Reporting rate
  & $\reportRate=0.679$
  & (\nameref{S_meas})
  & ${\reportRate=0.20}\fixed$
  & (\nameref{S_meas})
  & $\reportRate=0.98$
  & (\nameref{S_meas})
\\\hline
\raggedright Observation variance
  & \raggedright $\obsOverdispersion = (279.15, 78.33)$
  & \multirow{2}{*}{(\nameref{S_meas})}
  & \multirow{2}{*}{$\obsOverdispersion = 1.319$}
  & \multirow{2}{*}{(\nameref{S_meas})}
  & \multirow{2}{*}{$\obsOverdispersion = 88.58$}
  & \multirow{2}{*}{(\nameref{S_meas})}
\\
\hline
\multirow{2}{*}{Initial Values} & $\Iinit = 7298$
  &
  & \multirow{2}{*}{\missing}
  &
  & \multirow{2}{*}{$\Iinit^{3,4} = (
        21, 6)^*$}
  &  \multirow{2}{*}{(\nameref{S_init})}
  \\
  & $E_{0,0} = 350$ & & \multicolumn{1}{c}{}
  &
  &
  &
\\\hline
\multirow{3}{*}{Hurricane}
  & \multirow{4}{*}{\missing}
  &
  & \multirow{4}{*}{\missing}
  &
  & \raggedright $\Whur{3, 9} = (36.88, 31.64)^*$
  & \multirow{2}{*}{\eqref{eq:model3:foi}}
\\
 \multirow{-1}{*}{Parameters} &
  &
  &
  &
  & \raggedright $\hHur{3, 9} = (98.98, 58.43)^*$
  & \multirow{2}{*}{\eqref{eq:model3:foi}}
\\\hline
\end{tabular}
\begin{flushleft} References to the relevant equation are given in parentheses.
Parameters that were fixed and not calibrated using the data are indicated with $\fixed$; all fixed parameters values were chosen to match the fixed parameter values of \cite{lee20}.
Parameters that were added during our re-analysis and were not considered by Lee et al. are indicated with $^*$.
Confidence intervals for model parameters are given in the supplement (\nameref{S_CI}).
Translations back into the notation of \cite{lee20} are given in \nameref{S1_Table}.
\end{flushleft}
\end{adjustwidth}
\end{table}

Each model describes the cholera dynamics as a partially observed Markov process (POMP) with a latent state vector $\bm{X}(t)$ for each continuous time point $t$.
$N$ observations on the system are collected at time points $t_1,\dots,t_N$, written as $t_{1:N}$.
The observation at time $t_n$ is modeled by the random vector $\bm{Y}_n$.
While the latent process exists between observation times, the value of the latent state at observations times is of particular interest.
We therefore write $\bm{X}_n = \bm{X}(t_n)$ to denote the value of the latent process at the $n$th observation time, and $\bm{X}_{1:N}$ is the collection of latent state values for all observed time points.
The observable random variables $\bm{Y}_{1:N}$ are assumed to be conditionally independent given $\bm{X}_{0:N}$.
Together, with the density for the initial value of the latent state $\bm{X}_{0} = \bm{X}(t_0)$, each model defines a joint density $f_{\bm{X}_{0:N}, \bm{Y}_{1:N}}\big(\bm{x}_{0:N}, \bm{y}_{1:N};\paramVec\big)$, where $\paramVec$ is a parameter vector that indexes the model.
The observed data $\bm{y}_{1:N}^*$, along with the unobserved true value of the latent state, are modeled as a realization of this joint distribution.

Because of the probabilistic nature of both the unobserved latent state and the observable random variables, it is possible to consider various marginal and conditional densities of these two jointly random vectors.
An important example is the marginal density of the observed random vector $\bm{Y}_{1:N}$, evaluated at the observed data $\bm{y}_{1:N}^*$, as shown in Eq.~\myeqref{eq:likedef}:

\begin{eqnarray}
\label{eq:likedef}
f_{\bm{Y}_{1:N}}\big(\bm{y}_{1:N}^*; \paramVec\big) = \int f_{\bm{X}_{0:N}, \bm{Y}_{1:N}}\big(\bm{x}_{0:N}, \bm{y}_{1:N}^*;\paramVec\big) d\bm{x}_{0:N}.
\end{eqnarray}

\noindent When treated as a function of the parameter vector $\paramVec$, this marginal density is called the \emph{likelihood function}, which is the basis of likelihood based statistical inference.

Using the conditional independence of $\bm{Y}_{1:N}$ given $\bm{X}_{0:N}$ and the Markov property of $\bm{X}_{0:N}$, the joint density can be re-factored into the useful form given in Eq.~\myeqref{eq:jointLik}:

\begin{eqnarray}
\label{eq:jointLik}
f_{\bm{X}_{0:N}, \bm{Y}_{1:N}}\big(\bm{x}_{0:N}, \bm{y}_{1:N};\paramVec\big) = f_{\bm{X}_0}\big(\bm{x}_0;\paramVec\big)\prod_{n = 1}^N f_{\bm{X}_n|\bm{X}_{n-1}}\big(\bm{x}_{n}|\bm{x}_{n-1}; \paramVec\big)f_{\bm{Y}_n|\bm{X}_{n}}\big(\bm{y}_n|\bm{x}_{n}\big).
\end{eqnarray}

\noindent This factorization is useful because it demonstrates that POMP models may be completely described using three parts: the \emph{initialization model} for the latent states $f_{\bm{X}_0}\big(\bm{x}_0;\paramVec\big)$; the \emph{one-step transition density}, or \emph{the process model} $f_{\bm{X}_n|\bm{X}_{n-1}}\big(\bm{x}_{n}|\bm{x}_{n-1}; \paramVec\big)$; and the \emph{measurement model} $f_{\bm{Y}_n|\bm{X}_{n}}\big(\bm{y}_n|\bm{x}_{n}\big)$.
In the following subsections, we describe Models 1--3 in terms of these three components.
The latent state vector $\bm{X}(t)$ for each model consists of individuals labeled as susceptible (S), infected (I), asymptomatically infected (A), vaccinated (V), and recovered (R), with various sub-divisions sometimes considered.
The observable random vector $\bm{Y}_{1:N}$ represents the random vector of cholera incidence data for each model;
Models~2 and~3 have metapopulation structure, meaning that each individual is a member of a spatial unit, denoted by a subscript $u\in \seq{1}{U}$, in which case we denote the observed data for each unit using $\bm{y}_{1:N}^* = y_{1:N,1:U}^*$.
Here, the spatial units are the $U=10$ Haitian administrative d\'{e}partements (henceforth anglicized as departments).

While the complete model description is scientifically critical, as well as necessary for transparency and reproducibility, the model details are not essential to our methodological discussions of how to diagnose and address model misspecification with the purpose of informing policy.
A first-time reader may choose to skim through the rest of this section, and return later.
Additional details about the numeric implementation of these models are provided in a supplemental text (\nameref{S1_Text}).
While each of the dynamic models considered in this manuscript can be fully described using the mathematical equations provided in the following section, diagrams of dynamic systems can be helpful to understand the equations.
For this reason, we provide flow chart diagrams for Models~1--3 in supplement figures (\nameref{S_Mod1}, \nameref{S_Mod2} and \nameref{S_Mod3}).

\subsubsection*{Model~1}
\label{sec:model1}
The latent state vector $\bm{X}(t) = \big(S_{\vaccCounter}(t),  E_{\vaccCounter}(t), I_{\vaccCounter}(t), A_{\vaccCounter}(t), R_{\vaccCounter}(t), \vaccCounter \in 0:\vaccClass\big)$ describes susceptible, latent (exposed), infected (and symptomatic), asymptomatic, and recovered individuals in vaccine cohort $\vaccCounter$ at time $t$.
Here, $\vaccCounter=0$ corresponds to unvaccinated individuals, and $\vaccCounter \in \seq{1}{\vaccClass}$ describes hypothetical vaccination programs. Each program $\vaccCounter$ indexes differences in both the number of doses administered (one versus two doses per individual) and the round of vaccine administration, separating individuals into compartments with distinct dynamics based on vaccination status.
The force of infection is
\begin{equation}
\label{model1:lambda}
\lambda(t) = \Big(\sum_{\vaccCounter=0}^{\vaccClass}  I_{\vaccCounter}(t) + \asymptomRelativeInfect \sum_{\vaccCounter=0}^{\vaccClass} A_{\vaccCounter}(t) \Big)^\nu \frac{d\Gamma(t)}{dt} \transmission(t)/N,
\end{equation}
where $\transmission(t)$ is a periodic cubic spline representation of seasonality, given in terms of a B-spline basis $\{ s_j(t), j\in \seq{1}{6}\}$ and parameters $\transmission_{1:6}$ as
\begin{equation}
\label{model1:beta}
\transmission(t) = \bar{\transmission} \exp\Big(\sum_{j=1}^6 \transmission_j s_j(t)\Big),
\end{equation}
where $\bar{\transmission} = 1$ (wk)\textsuperscript{-1} is a dimensionality constant.
The process noise $d\Gamma(t)/dt$ is multiplicative Gamma-distributed white noise, with infinitesimal variance parameter $\sigmaProc^2$.
Lee et al.~\cite{lee20} included process noise in Model~3 but not in Model~1, i.e., they fixed $\sigmaProc^2 = 0$.
Gamma white noise in the transmission rate gives rise to an over-dispersed latent Markov process \cite{breto11} which has been found to improve the statistical fit of disease transmission models \cite{stocks20,he10}.

For any time point in $t_{1:N}$, the process model $f_{\bm{X}_n|\bm{X}_{n-1}}\big(\bm{x}_{n}|\bm{x}_{n-1}; \paramVec\big)$ is defined by describing how individuals move from one latent state compartment to another.
Per-capita transition rates are given in Eqs.~\myeqref{model1:SE}-\myeqref{model1:birth}:
\begin{eqnarray}
\label{model1:SE}
\mu_{S_{\vaccCounter}E_{\vaccCounter}} &=& \lambda(t),
\\
\label{model1:EI}
\mu_{E_{\vaccCounter}I_{\vaccCounter}} &=& \muEI\big(1-\symptomFrac_\vaccCounter(t)\big),
\\
\label{model1:EA}
\mu_{E_{\vaccCounter}A_{\vaccCounter}} &=& \muEI\, \symptomFrac_\vaccCounter(t),
\\
\label{model1:toR}
\mu_{I_{\vaccCounter}R_{\vaccCounter}} &=& \mu_{A_{\vaccCounter}R_{\vaccCounter}} = \muIR,
\\
\label{model1:RS}
\mu_{R_{\vaccCounter}S_{\vaccCounter}} &=& \muRS,
\\
\label{model1:vacc}
\mu_{S_0S_{\vaccCounter}} &=& \mu_{E_0E_{\vaccCounter}} = \mu_{I_0I_{\vaccCounter}} = \mu_{A_0A_{\vaccCounter}} = \mu_{R_0R_{\vaccCounter}} = \eta_{\vaccCounter}(t),
\\
\label{model1:death}
\mu_{S_{\vaccCounter}\demography} &=& \mu_{E_{\vaccCounter}\demography} = \mu_{I_{\vaccCounter}\demography} = \mu_{A_{\vaccCounter}\demography}=\mu_{R_{\vaccCounter}\demography} = \delta,
\\
\label{model1:birth}
\mu_{\demography S_0} &=& \muBirth,
\end{eqnarray}
where $\vaccCounter\in \seq{0}{\vaccClass}$.
Here, $\mu_{AB}$ is a transition rate from compartment $A$ to $B$.
We have an additional demographic source and sink compartment $\demography$ modeling entry into the study population due to birth or immigration, and exit from the study population due to death or immigration.
Thus, $\mu_{A\demography}$ is a rate of exiting the study population from compartment $A$ and $\mu_{\demography B}$ is a rate of entering the study population into compartment $B$.

In Model~1, the advantage afforded to vaccinated individuals is an increased probability that an infection is asymptomatic.
Conditional on infection status, vaccinated individuals are also less infectious than their non-vaccinated counterparts by a rate of $\asymptomRelativeInfect = 0.05$ in Eq.~\eqref{model1:lambda}.
In Eqs.~\myeqref{model1:EA} and~\myeqref{model1:EI} the asymptomatic ratio for non-vaccinated individuals is set $\symptomFrac_0(t)=0$, so that the asymptomatic route is reserved for vaccinated individuals.
For $\vaccCounter\in\seq{1}{\vaccClass}$, the vaccination cohort $\vaccCounter$ is assigned a time $\tau_{\vaccCounter}$, and we take $\symptomFrac_\vaccCounter(t) = c \, \vaccineEfficacy^*(t-\tau_{\vaccCounter})$
where $\vaccineEfficacy^*(t)$ is efficacy at time $t$ since vaccination for adults, a step-function represented in Table~S4 of \cite{lee20}, and $c=\big(1-(1-0.4688)\times 0.11\big)$ is a correction to allow for reduced efficacy in the 11\% of the population aged under 5 years.
Single and double vaccine doses were modeled by changing the waning of protection; protection was modeled as equal between single and double dose until 52 weeks after vaccination, at which point the single dose becomes ineffective.

The latent state vector $\bm{X}(t)$ is initialized by setting the counts for each compartment and vaccination scenario $\vaccCounter \neq 0$ as zero, and introducing initial-value parameters $I_{0,0}$ and $E_{0, 0}$ such that $R_{0}(0) = 0$, $I_{0}(0) = \mathrm{Pop} \times I_{0, 0}$, $E_{0}(0) = \mathrm{Pop} \times E_{0, 0}$ and $S_{0}(0) = \mathrm{Pop} \times (1 - I_{0, 0} - E_{0, 0})$, where $\mathrm{Pop}$ is the total population of Haiti.
The measurement model describes reported cholera cases at time point $n$ come from a negative binomial distribution, where only a fraction ($\reportRate$) of new weekly cases are reported.
More details about the initialization model $f_{\bm{X}_0}\big(\bm{x}_0;\paramVec\big)$ and the measurement model $f_{\bm{Y}_n|\bm{X}_{n}}\big(\bm{y}_n|\bm{x}_{n}\big)$ for Models~1--3 are provided a supplement text (\nameref{S_init} and \nameref{S_meas}).

\subsubsection*{Model~2}
\label{sec:model2}
Susceptible individuals are in compartments $S_{u\vaccCounter}(t)$, where $u\in\seq{1}{U}$ corresponds to the $U=10$ departments, and $\vaccCounter\in\seq{0}{4}$ describes vaccination status:
\begin{itemize}
  \item[$\vaccCounter=0$:] Unvaccinated or waned vaccination protection.
  \item[$\vaccCounter=1$:] One dose at age under five years.
  \item[$\vaccCounter=2$:] Two doses at age under five years.
  \item[$\vaccCounter=3$:] One dose at age over five years.
  \item[$\vaccCounter=4$:] Two doses at age over five years.
\end{itemize}

Like Model~1, the process model $f_{\bm{X}_n|\bm{X}_{n-1}}\big(\bm{x}_{n}|\bm{x}_{n-1}; \paramVec\big)$ is primarily defined via the description of movement of individuals between compartments, however Model~2 also includes a dynamic description of a latent bacterial compartment as well.
Individuals can progress to a latent infection $E_{u\vaccCounter}$ followed by symptomatic infection $I_{u\vaccCounter}$ with recovery to $R_{u\vaccCounter}$ or asymptomatic infection $A_{u\vaccCounter}$ with recovery to $R^A_{u\vaccCounter}$.
The force of infection depends on both direct transmission and an aquatic reservoir, $W_u(t)$, and is given by
\begin{equation}
\label{model2:lambda}
\lambda_{u}(t) = 0.5\big(1+\seasAmplitude \cos(2\pi t + \phaseParm)\big)
\frac{\beta_W\, W_u(t)}{ \Wsat  + W_u(t)} +
\transmission \left\{\sum_{\vaccCounter=0}^4 I_{u\vaccCounter}(t) + \asymptomRelativeInfect \sum_{\vaccCounter=0}^4 A_{u\vaccCounter}(t) \right\}.
\end{equation}
The latent state is therefore described by the vector $\bm{X}(t) = \big(S_{u\vaccCounter}(t),\allowbreak E_{u\vaccCounter}(t),\allowbreak I_{u\vaccCounter}(t),\allowbreak A_{u\vaccCounter}(t),\allowbreak R_{u\vaccCounter}(t),\allowbreak R_{u\vaccCounter}^A(t),\allowbreak W_u,\allowbreak u \in \seq{1}{U},\allowbreak \vaccCounter \in \seq{0}{4}\big)$.
The cosine term in Eq.~\myeqref{model2:lambda} accounts for annual seasonality, with a phase parameter $\phaseParm$.
The Lee et al.~\cite{lee20} implementation of Model~2 fixes $\phaseParm = 0$.

Individuals move from department $u$ to $v$ at rate $T_{uv}$, and aquatic cholera moves at rate $T^W_{uv}$.
The nonzero transition rates are
\begin{eqnarray}
\label{model2:mu_SE}
\mu_{S_{u\vaccCounter}E_{u\vaccCounter}} &=& (1 - \vaccineEfficacy_\vaccCounter) \, \lambda_u(t),
\\
\label{model2:mu_EI}
\mu_{E_{u\vaccCounter}I_{u\vaccCounter}} &=& \symptomFrac\muEI, \quad \mu_{E_{u\vaccCounter}A_{u\vaccCounter}} = (1-\symptomFrac)\muEI,
\\
\label{model2:mu_IR}
\mu_{I_{u\vaccCounter}R_{u\vaccCounter}} &=& \mu_{A_{u\vaccCounter}R^A_{u\vaccCounter}} = \muIR,
\\
\label{model2:RS}
\mu_{R_{u\vaccCounter}S_{u\vaccCounter}} &=& \mu_{R^A_{u\vaccCounter}S_{u\vaccCounter}} = \muRS,
\\
\label{model2:transport}
\mu_{S_{u\vaccCounter}S_{{\varv}\vaccCounter}} &=& \mu_{E_{u\vaccCounter}E_{{\varv}\vaccCounter}} = \mu_{I_{u\vaccCounter} I_{{\varv}\vaccCounter}} = \mu_{A_{u\vaccCounter}A_{{\varv}\vaccCounter}} = \mu_{R_{u\vaccCounter}R_{{\varv}\vaccCounter}} = \mu_{R^A_{u\vaccCounter} R^A_{{\varv}\vaccCounter}} = T_{u\varv},
\\
\label{model2:omega1}
\mu_{S_{u1}S_{u0}} &=& \mu_{S_{u3}S_{u0}} = \omega_1,
\\
\label{model2:omega2}
\mu_{S_{u2}S_{u0}} &=& \mu_{S_{u4}S_{u0}} = \omega_2,
\\
\label{model2:to_W}
\mu_{\demography W_u} &=& \Wshed \left\{ \sum_{\vaccCounter=0}^4 I_{u\vaccCounter}(t) + \asymptomRelativeShed \sum_{\vaccCounter=0}^4 A_{u\vaccCounter}(t) \right\},
\\
\label{model2:from_W}
\mu_{W_u\demography} &=& \Wremoval,
\\
\label{model2:water_transport}
\mu_{W_uW_{\varv}} &=& w_r T^W_{u\varv}.
\end{eqnarray}
In Eq.~\myeqref{model2:transport} the spatial coupling is specified by a gravity model,
\begin{equation}
\label{model2:gravity}
T_{u\varv} = v_{\mathrm{rate}} \times \frac{\mathrm{Pop}_u \mathrm{Pop}_{\varv}}{D_{u\varv}^2},
\end{equation}
where $\mathrm{Pop}_u$ is the mean population for department $u$,
$D_{u\varv}$ is a distance measure estimating average road distance between randomly chosen members of each population, and $v_{\mathrm{rate}}= 10^{-12} \,\mbox{km$^2$yr$^{-1}$}$ was fixed at the value used in \cite{lee20}.
In Eq.~\myeqref{model2:water_transport}, $T^W_{u\varv}$ is a measure of river flow between departments.
The unit of $W_u(t)$ is cells per ml, with dose response modeled via a saturation constant of $\Wsat$ in Eq.~\myeqref{model2:lambda}.
In Eq.~\myeqref{model2:mu_SE}, $\vaccineEfficacy_\vaccCounter$ denotes the vaccine efficacy for each vaccination campaign $\vaccCounter \in \vaccClass$, with $\vaccineEfficacy_0 = 0$, $\vaccineEfficacy_1 = 0.429 \childReduce$, $\vaccineEfficacy_2 = 0.519 \childReduce$, $\vaccineEfficacy_3 = 0.429$, and $\vaccineEfficacy_4 = 0.519$
Here, $\childReduce = 0.4688$ represents the reduced efficacy of the vaccination for children under the age of five years, and the values $0.429$ and $0.519$ are the median effectiveness of one and two doses over their effective period respectively, according to Table S4 in the supplement material of Lee et al.~\cite{lee20}.
Because vaccine efficacy remains constant, individuals in this model transition from a vaccinated compartment to the susceptible compartment at the end of the vaccine coverage period.

The starting value for each element of the latent state vector $\bm{X}(0)$ are set to zero except for $I_{u0}(0) = y_u^*(0)/\reportRate$ and $R_{u0}(0) = \mathrm{Pop}_u - I_{u0}(0)$, where $y_u^*(0)$ is the reported number of cholera cases in department $u$ at time $t = 0$.
Reported cases are described using a log-normal distribution, with the log-scale mean equal to the reporting rate $\reportRate$ times the number of newly infected individuals.
See the supplement material on model initializations for more details (\nameref{S_init}).

\subsubsection*{Model~3}
\label{sec:model3}

The latent state is described as $\bm{X}(t) = \big(S_{u\vaccCounter}(t),\allowbreak I_{u}(t),\allowbreak A_{u}(t),\allowbreak R_{u\vaccCounter k}(t),\allowbreak W_u(t),\allowbreak u \in \seq{0}{U},\allowbreak \vaccCounter \in \seq{0}{4},\allowbreak k \in \seq{1}{3}\big)$.
Here, $\vaccCounter=0$ corresponds to unvaccinated, $\vaccCounter=2j-1$ corresponds to a single dose on the $j$th vaccination campaign in unit $u$ and $\vaccCounter=2j$ corresponds to receiving two doses on the $j$th vaccination campaign.
$k\in\seq{1}{3}$ models non-exponential duration in the recovered class before waning of immunity.
The processes model $f_{\bm{X}_n|\bm{X}_{n-1}}\big(\bm{x}_n|\bm{x}_{n-1};\paramVec\big)$ describes the movement of individuals between latent compartments, as well as the birth and death process of local, unobserved bacterial compartments $W_{u}(t)$.
The force of infection is
\begin{equation}
\label{eq:model3:foi}
\lambda_u(t) = \left(\Wbeta{_u} + 1_{(t \geq \tHur)}\Whur{_u} e^{-\hHur{u}\left(t - \tHur\right)}\right) \frac{W_u(t)}{1+W_u(t)} + \transmission_u \sum_{{\varv}\neq u}\big(I_{\varv}(t)+ \asymptomRelativeInfect A_{\varv}(t)\big),
\end{equation}
where $\tHur$ is the time Hurricane Matthew struck Haiti \cite{ferreirai16}, and $1_{(A)}$ is the indicator function for event $A$.
In \cite{lee20}, $\Whur{_u}$ and $\hHur{u}$ were set to zero for all $u$;
the need to account for the effect Hurricane Matthew had on cholera transmission for this model is explored in Sec.~S5 of the supplement.

Per-capita transition rates are used for both compartments representing human counts and the aquatic reservoir of bacteria; these rates  are given in Eqs.~\myeqref{eq:model3:SI}--\myeqref{eq:model3:Decay}.
\begin{eqnarray}
\label{eq:model3:SI}
\mu_{S_{u\vaccCounter}I_{u}} &=& \symptomFrac \,  \lambda_u \big(1-\vaccineEfficacy_{u\vaccCounter}(t)\big) \, d\Gamma/dt,
\\
\label{eq:model3:SA}
\mu_{S_{u\vaccCounter}A_{u}} &=& (1-\symptomFrac) \,  \lambda_u \big(1-\vaccineEfficacy_{u\vaccCounter}(t)\big) \,  d\Gamma/dt,
\\
\label{eq:model3:IR}
\mu_{I_{u}R_{u\vaccCounter 1}} &=& \mu_{A_{u}R_{u\vaccCounter 1}} = \muIR,
\\
\label{eq:model3:IS}
\mu_{I_{u}S_{u0}} &=& \muDeath + \choleraDeath, \hspace{2mm} \mu_{A_{u}S_{u0}} = \muDeath
\\
\label{eq:model3:RRnext}
\mu_{R_{u\vaccCounter 1}R_{u\vaccCounter 2}} &=& \mu_{R_{u\vaccCounter 2}R_{u\vaccCounter 3}} = 3\muRS,
\\
\label{eq:model3:RS}
\mu_{R_{u\vaccCounter k}S_{u0}} &=& \muDeath + 3\muRS \, \mathbf{1}_{\{k=3\}},
\\
\label{eq:model3:water}
\mu_{{\demography}W_u} &=& \big[1 + \seasAmplitude \big(J_u(t)\big)^r \big] \mathrm{Den}_u \, \Wshed \big[ I_{u}(t)+ \asymptomRelativeShed A_{u}(t) \big],
\\
\label{eq:model3:Decay}
\mu_{W_u\demography} &=& \Wremoval.
\end{eqnarray}
As with Model~1, $d\Gamma_u(t)/dt$ is multiplicative Gamma-distributed white noise in Eqs.~\myeqref{eq:model3:SI} and \myeqref{eq:model3:SA}.
In Eq.~\myeqref{eq:model3:water}, $J_u(t)$ is a dimensionless measurement of precipitation that has been standardized by dividing the observed rainfall at time $t$ by the maximum recorded rainfall in department $u$ during the epidemic, and $\mathrm{Den}_u$ is the population density.
Demographic stochasticity is accounted for by modeling non-cholera related death rate $\muDeath$ in each compartment, along with an additional death rate $\choleraDeath$ in Eq.~\myeqref{eq:model3:IS} to account for cholera induced deaths among infected individuals.
All deaths are balanced by births into the susceptible compartment in Eqs.~\myeqref{eq:model3:IS} and \myeqref{eq:model3:RS}, thereby maintaining constant population in each department.

Similar to Model~1, there are no distinct compartments for individuals under five years of age, and the vaccination efficacy is taken as a age adjusted weighted average of the efficacy for individuals both over and under five years of age:  $\vaccineEfficacy_{u \vaccCounter}(t) = c\vaccineEfficacy^*(t - \tau_{uz})$, where $\tau_{uz}$ is the vaccination time for unit $u$ and vaccination campaign $z$.
The value $c$ and the function $\vaccineEfficacy^*$ are equivalent to those described in the Model~1 description.

Latent states are initialized using an approximation of the instantaneous number of infected, asymptomatic, and recovered individuals at time $t_0$ by using the first week of cholera incidence data.
Specifically, we set $I_{u0}(0) = \frac{y^*_{1u}}{\reportRate(\muDeath + \choleraDeath + \muIR)}$, $A_{u0}(0) = \frac{1 - \symptomFrac}{\symptomFrac}I_{u0}(0)$, $R_{u0k} = y^*_{1u} - I_{u0}(0) - A_{u0}(0)$, and we initialize $W_{u}(0)$ by enforcing the rainfall dynamics supposed by the one step transition model;
all other compartments that represent population counts are set to zero at time $t_0$.
For each unit $u$ with zero case counts at time $t_1$, this initialization scheme results in having zero individuals in the Infected and Asymptomatic compartments, as well as having no bacteria in the aquatic reservoir.
In reality, it is plausible that some bacteria or infected individuals were present in unit $u$ but went unreported.
Therefore, for departments with zero case counts in week 1, we estimate the number of infected individuals rather then treating this value as a constant (\nameref{S_init}).
Finally, reported cholera cases are modeled using a negative binomial distribution with mean equal to a fraction ($\reportRate$) of individuals in each unit who develop symptoms and seek healthcare, and with over-dispersion parameter $\obsOverdispersion$ (\nameref{S_meas}).

\subsection*{Model Fitting}\label{sec:model_fitting}

Each of the three models considered in this study describes cholera dynamics as a partially observed Markov process (POMP) \cite{king16}, with the understanding that the deterministic Model~2 is a special case of a Markov processes solving a stochastic differential equation in the limit as the noise parameter goes to zero.
Each model is indexed by a parameter vector, $\paramVec$, and different values of $\paramVec$ can result in qualitative differences in the predicted behavior of the system.
Therefore, the choice of $\paramVec$ used to make inference about the system can greatly affect model-based conclusions \cite{saltelli20}.
Elements of $\paramVec$ can be fixed at a constant value based on scientific understanding of the system, but parameters can also be calibrated to data by maximizing a measure of congruency between the observed data and the model's mechanistic structure.
Calibrating model parameters to observed data does not guarantee that the resulting model successfully approximates real-world mechanisms, since the model description of the dynamic system may be incorrect and does not change as the model is calibrated to data.
However, the congruency between the model and observed data serves as a proxy for the congruency between the model and the true underlying dynamic system.
As such, it is desirable to obtain the best possible fit of the proposed mechanistic structure to the observed data.

In this article we follow \cite{lee20} by calibrating the parameters of each of our models using maximum likelihood, as described in Eq.~\myeqref{eq:likedef}.
The likelihood for each of the fitted models---and the corresponding AIC values for model comparisons that include an adjustment for the number of calibrated parameters---is provided in Table~\ref{tab:likes}.
In the following subsections we describe in detail our approach to calibrating the three proposed mechanistic models to observed cholera incidence data.
The main alternative to maximum likelihood estimation is Bayesian inference via Markov chain Monte Carlo, used to analyze the Haiti cholera epidemic by \cite{andrews11,pasetto18,rinaldo12,lewnard16,trevisin22,sallah17,azman12,kuhn14,mari15,gatto12}.

\begin{table}[!h]
\centering
\caption{\label{tab:likes}{\bf AIC values for Models~1--3 and their benchmarks.}}
\begin{tabular}{|c|c|c|c|}
\hline
 & \thead{Model~1} & \thead{Model~2} & \thead{Model~3}
\\
\hline
\hline
\multirow{2}{*}{Log-likelihood} &
  $-2728.1$ &
  $-21957.3$ &
  $-17332.9$ \\
    & ($-3030.9$)\footnotemark[1] &
  ($-29367.4$) &
  ($-33832.6$)\footnotemark[2]
\\
\hline
Number of &
  $15$ &
  $6$ &
  $34$ \\
 Fit Parameters &
 (20) &
 ($6$) &
 ($29$)
\\
\hline
\multirow{2}{*}{AIC} &
  $5486.3$ &
  $43926.5$ &
  $34733.9$ \\
  & ($6101.8$)\footnotemark[1] &
  ($58746.9$) &
  ($67723.2$)\footnotemark[2]
\\
\hline
\thead{Benchmark \\ AIC} &
  $5585.3$ &
  $36961.0$ &
  $35945.2$
\\
\hline
\end{tabular}
\begin{flushleft}
Values in parentheses are corresponding values obtained using the models of \cite{lee20}.  \textsuperscript{1}The reported likelihood is an upper bound of the likelihood of the model in \cite{lee20}. \textsuperscript{2}In \cite{lee20}, Model~3 was fit to a subset of the data (March 2014 onward, excluding data from Ouest in 2015-2016).
On this subset, their model has a likelihood of $-9721.2$.
On this same subset, our model has a likelihood of $-7219.5$.
Details of estimating the likelihood of the models used in \cite{lee20} are provided in the supplement (\nameref{S_lee20}).
\end{flushleft}
\end{table}

\subsubsection*{Calibrating Model~1 Parameters}\label{sec:fit1}

Model~1 is a highly nonlinear over-dispersed stochastic dynamic model, favoring a scientifically plausible description of cholera dynamics rather than one that is statistically convenient \cite{he10}.
This results in the inability to obtain a closed form expression of the joint model density---described in Eq.~\myeqref{eq:jointLik}.
Therefore in order to perform likelihood based inference on this model, we are restricted to use parameter estimation techniques that have the \emph{plug-and-play} property, which is that the fitting procedure only requires the ability to simulate the latent process rather than evaluating transition densities \cite{breto09,he10};
in the context of the notation and definitions employed in this article, this means that we only require the ability to simulate from $f_{\bm{X}_0}\big(\bm{x}_0;\paramVec\big)$ and $f_{\bm{X}_n|\bm{X}_{n-1}}\big(\bm{x}_{n}|\bm{x}_{n-1}; \paramVec\big)$ rather than needing to evaluate these densities.
Plug-and-play algorithms include Bayesian approaches like ABC and PMCMC \cite{toni09,andrieu10}, but here we use algorithms that enable maximum likelihood estimation.
To our knowledge, the only plug-and-play methods that have been effectively used to maximize the likelihood for arbitrary nonlinear POMP models are iterated filtering algorithms \cite{ionides15}, which modify the well-known \emph{particle filter} \cite{arulampalam02}.

The particle filter, also referred to as sequential Monte Carlo, is a simulation based method that is frequently used in Bayesian inference to approximate the posterior distribution of latent states.
This algorithm can also be used to accurately approximate the log-likelihood of a POMP model, defined as the integral in Eq.~\myeqref{eq:likedef}.
Iterated filtering algorithms, such as IF2 \cite{ionides15}, extend the particle filter by performing a random walk for each parameter and particle;
these perturbations are carried out iteratively over multiple filtering operations, using the collection of parameters from the previous filtering pass as the parameter initialization for the next iteration, and decreasing the random walk variance at each step.
With a sufficient number of iterations, the resulting parameter values converge to a region of the parameter space that maximizes the model likelihood.

The ability to maximize the likelihood allows for likelihood-based inference, such as performing statistical tests for potential model improvements.
We demonstrate this capability by proposing a log-linear trend $\transmissionTrend$ in transmission in Eq.~\myeqref{model1:beta}:
\begin{equation}
\label{model1:betat}
\transmission(t) = \bar{\transmission} \exp\Big(\sum_{j=1}^6 \transmission_s s_j(t) + \transmissionTrend\bar{t}\Big),
\end{equation}
where $\bar{t} = \frac{t - (t_N + t_0) / 2}{t_N - (t_N + t_0) / 2}$, so that $\bar{t} \in [-1, 1]$.
The proposal of a trend in transmission is a result of observing an apparent decrease in reported cholera infections from 2012-2019 in Fig.~\ref{fig:CholeraData}.
While several factors may contribute to this decrease, one explanation is that case-area targeted interventions (CATIs), which included education sessions, increased monitoring, household decontamination, soap distribution, and water chlorination in infected areas \cite{rebaudet19CATI}, may have substantially reduced cholera transmission over time \cite{rebaudet21}.

\begin{figure}[!h]
\centering
\begin{knitrout}
\definecolor{shadecolor}{rgb}{0.969, 0.969, 0.969}\color{fgcolor}

{\centering \includegraphics[width=\maxwidth]{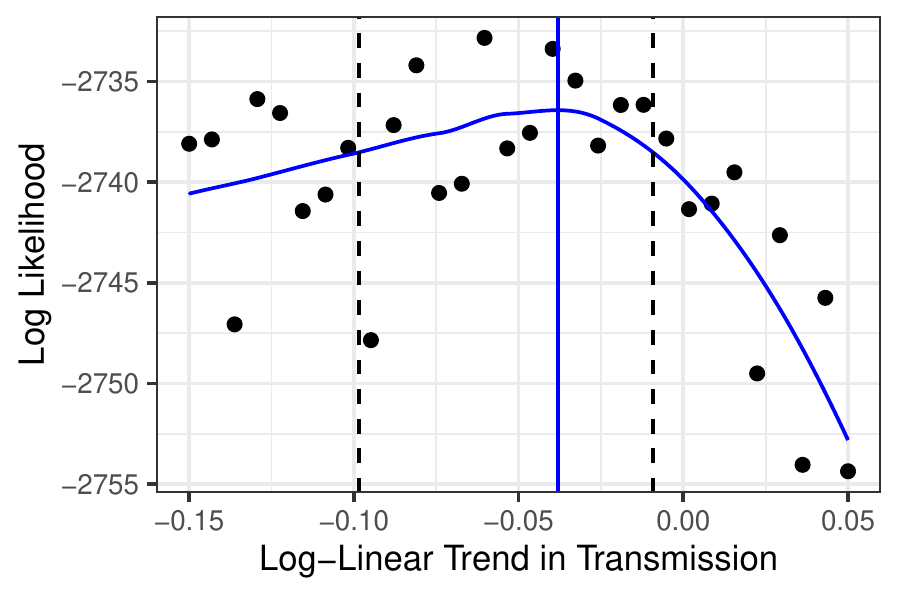} 

}

\end{knitrout}
\caption{\label{fig:betat}
{\figTitle Confidence interval for the log-linear trend in transmission.}
Monte Carlo adjusted profile (MCAP) of $\transmissionTrend$ for Model~1.
The blue curve is the MCAP, the vertical blue line indicates the MLE, and the vertical dashed lines indicate the $95\%$ confidence interval.
}
\end{figure}

We perform a statistical test to determine whether or not the data indicate the presence of a trend in transmissibility.
To do this, we perform a profile-likelihood search on the parameter $\transmissionTrend$ and obtain a $95\%$ confidence interval via a Monte Carlo Adjusted Profile (MCAP) \cite{ionides17}.
Lee et al.~\cite{lee20} implemented Model~1 by fitting two distinct phases: an epidemic phase from October 2010 through March 2015, and an endemic phase from March 2015 onward.
We similarly allow the re-estimation of process and measurement overdispersion parameters ($\sigmaProc^2$ and $\obsOverdispersion$), and require that the latent Markov process $X(t)$ carry over from one phase into the next.
The resulting 95\% confidence interval for $\transmissionTrend$ is $(-0.098, -0.009)$, with the full results displayed in Fig.~\ref{fig:betat}.
These results are suggestive that the inclusion of a trend in the transmission rate improves the quantitative ability of Model~1 to describe the observed data.
The maximum likelihood estimate for $\transmissionTrend$ corresponds to a 7.3\% reduction to the transmission rate over the course of the outbreak, with a $95\%$ confidence interval of (1.8\%, 17.9\%) for the overall reduction in transmission.
The reported results for Model~1 in the remainder of this article were obtained with the inclusion of the parameter $\transmissionTrend$.
The inclusion of a trend in transmission rate demonstrates a class of model variation that can be highly beneficial to consider: the model variation has a plausible scientific justification, and is easily testable using likelihood based methods.

If a mechanistic model including a feature (such as a representation of a mechanism, or the inclusion of a covariate) fits better than mechanistic models without that feature, and also has competitive fit compared to associative benchmarks, this may be taken as evidence supporting the scientific relevance of the feature.
As for any analysis of observational data, we must be alert to the possibility of confounding.
For a covariate, this shows up in a similar way to regression analysis: the covariate under investigation could be a proxy for some other unmodeled phenomenon or unmeasured covariate.

The statistical evidence of a trend in transmission rate in this model could be explained by any trending variable (such as hygiene improvements, or changes in population behavior), resulting in confounding from collinear covariates. Alternatively, it is possible that the negative trend observed in the incidence data could be attributed to a decreasing reporting rate rather than decreasing transmission rate. This could be formally tested by comparing models with either trend specification. We did not do this because evidence suggests that reporting rate was maintained or increased (Figure~1 of \cite{rebaudet21}). We instead argue that a decreasing transmission rate is a plausible way to explain the decrease in cases over time, as there is alternative evidence that supports this model \cite{rebaudet19CATI,rebaudet21,michel19}. It is not practical to test all remotely plausible model variations, yet a strongly supported conclusion should avoid ruling out untested hypotheses. The robust statistical conclusion for our analysis is that a model which allows for change fits better than one which does not, and a trend in transmission is a plausible way to do this.

\begin{figure}[!h]
\centering
\begin{knitrout}
\definecolor{shadecolor}{rgb}{0.969, 0.969, 0.969}\color{fgcolor}

{\centering \includegraphics[width=\maxwidth]{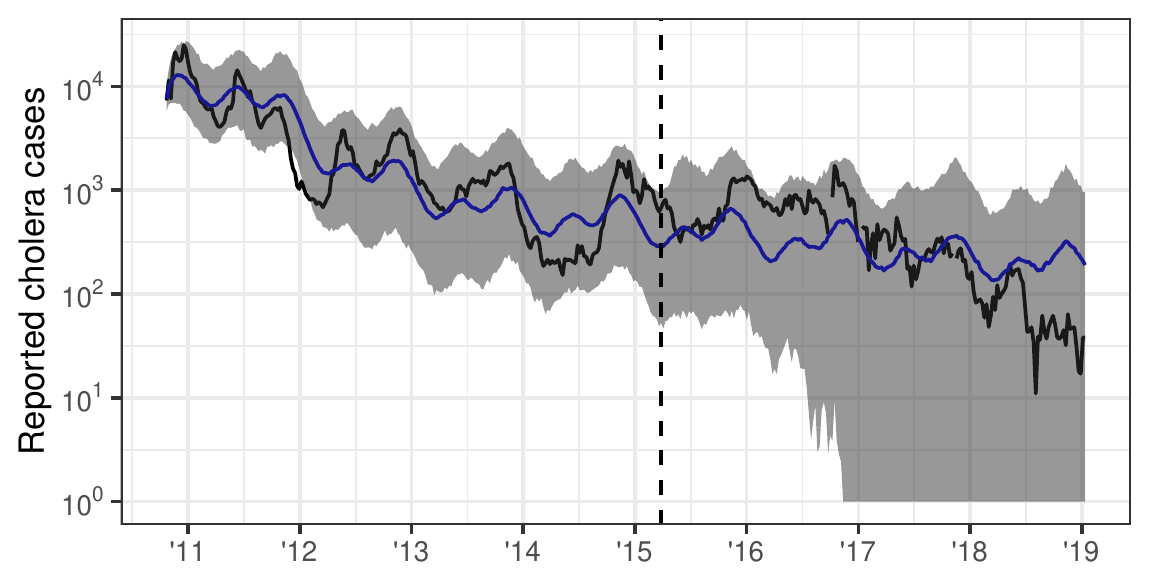} 

}

\end{knitrout}
\caption{\label{fig:mod1fit}
{\figTitle Simulations from Model~1 compared to reported cholera cases.}
The black curve is observed data, the blue curve is median of 500 simulations from initial conditions using estimated parameters, and the vertical dashed line represents break-point when parameters are refit.
}
\end{figure}

We implemented Model~1 using the \code{pomp} package \cite{king16}, relying heavily on the source code provided by Lee et al.~\cite{lee20}.
Both analyses used the \code{mif2} implementation of the IF2 algorithm to estimate $\paramVec$ by maximum likelihood.
One change we made in the statistical analysis that led to larger model likelihoods was increasing the computational effort in the numerical maximization.
While IF2 enables parameter estimation for a large class of models, the theoretic ability to maximize the likelihood depends on asymptotics in both the number of particles and the number of filtering iterations.
Many Monte Carlo replications are then required to quantify and further reduce the error.
The large increase in the log-likelihood for Model~1 (Table~\ref{tab:likes}) can primarily be attributed to increasing the computational effort used to calibrate the model.
This result highlights the importance of carefully determining the necessary computational effort needed to maximize model likelihoods and acting accordingly.
In this case study, this was done by performing standard diagnostics for the IF2 and particle filter algorithms\cite{king16,li23,pons-salort18,laneri10}. Given the considerable computational costs of simulation-based algorithms, we find it useful to perform an initial assessment using hyperparameter values---such as the number of particles, filtering iterations, and replicates based on different parameter initializations---that enable relatively quick calculations. The insights obtained from this preliminary analysis help in accurately determining the amount of computation that is required to achieve reliable outcomes.
Simulations from the initial conditions of our fitted model are plotted against the observed incidence data in Fig.~\ref{fig:mod1fit}.

\subsubsection*{Calibrating Model~2 Parameters}\label{sec:fit2}

Model~2 is a deterministic compartmental model defined by a set of coupled differential equations.
The use of deterministic compartment models have a long history in the field of infectious disease epidemiology \cite{kermack1927,brauer2017,giordano20},
and can be justified by asymptotic considerations in a large-population limit \cite{dadlani2020,ndii17}.
Because the process model of Model~2 is deterministic, maximum likelihood estimation reduces to a least squares calculation when combined with a Gaussian measurement model (\nameref{S_meas}).
Lee et al.~\cite{lee20} fit two versions of Model~2 based on a presupposed change in cholera transmission from a epidemic phase to endemic phase that occurred in March, 2014.
The inclusion of a change-point in model states and parameters increased the flexibility of the model and hence the ability to fit the observed data.
The increase in model flexibility, however, resulted in hidden states that were inconsistent between model phases.
The inclusion of a model break-point by Lee et al.~\cite{lee20} is perhaps due to a challenging feature of fitting a deterministic model via least squares: discrepancies between model trajectories and observed case counts in highly infectious periods of a disease outbreak will result in greater penalty than the discrepancies between model trajectories and observed case counts in times of relatively low infectiousness. This results in a bias towards accurately describing periods of high infectiousness.
This bias is particularly troublesome for modeling cholera dynamics in Haiti: the inability to accurately fit times of low infectiousness may result in poor model forecasts, as few cases of cholera were observed in the last few years of the epidemic.

To combat this issue, we fit the model to log-transformed case counts, since the log scale stabilizes the variation during periods of high and low incidence.
An alternative solution is to change the measurement model to include overdispersion, as was done in Models~1 and~3.
This permits the consideration of demographic stochasticity, which is dominant for small infected populations, together with log scale stochasticity (also called multiplicative, or environmental, or extra-demographic) which is dominant at high population counts.
Here we chose to fit the model to transformed case counts rather than adding overdispersion to the measurement model with the goal of minimizing the changes to the model proposed by Lee et al.~\cite{lee20}.

\begin{figure}[!h]
\begin{knitrout}
\definecolor{shadecolor}{rgb}{0.969, 0.969, 0.969}\color{fgcolor}

{\centering \includegraphics[width=\maxwidth]{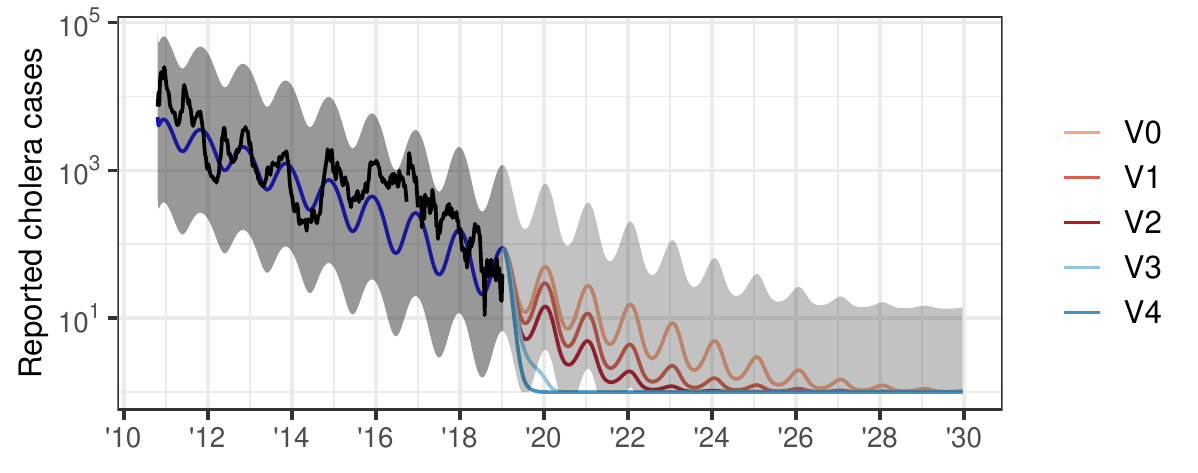} 

}

\end{knitrout}
\caption{\label{fig:mod2Traj}
{\figTitle Simulated trajectory of Model~2.}
The black line shows the nationally aggregated weekly cholera incidence data.
The blue curve from 2012-2019 is the trajectory of the calibrated version of Model~2.
Projections under the various vaccination scenarios, which are discussed in detail in the  {\bf Forecasts} subsection are also included. The gray ribbons represent a $95\%$ interval obtained from the log-normal measurement model. To avoid over-plotting, measurement variance is only plotted for the V0 vaccination scenario.}
\end{figure}

We implemented this model using the \code{spatPomp} \code{R} package \cite{asfaw23arxiv}.
The model was then fit using the subplex algorithm \cite{king2020Subplex}.
A comparison of the trajectory of the fitted model to the data is given in Fig.~\ref{fig:mod2Traj}.

\subsubsection*{Calibrating Model~3 Parameters}\label{sec:fit3}
Model~3 describes cholera dynamics in Haiti using a metapopulation model, where the hidden states in each administrative department has an effect on the dynamics in other departments.
The decision to address metapopulation dynamics using a spatially explicit model, rather than to aggregate over space, is double-edged.
Evidence for the former approach has been provided in previous studies \cite{king15}, including the specific case of heterogeneity between Haitian departments in cholera transmission \cite{collins14}.
However, a legitimate preference for simplicity can support a decision to consider nationally aggregated models \cite{saltelli20,green15}.

In our literature review, 17 articles considered dynamic models that incorporate spatial heterogeneity \cite{lee20,tuite11,pasetto18,fitzgibbon20,eisenberg13,rinaldo12,chao11,abrams13,trevisin22,sallah17,collins14,kelly16,azman12,leung22,kuhn14,mari15,gatto12}. All but four \cite{lee20,pasetto18,sallah17,azman12} of these studies used deterministic dynamic models: this greatly simplifies the process of calibrating model parameters to incidence data, though deterministic models can struggle to describe complex stochastic dynamics. The model in \cite{pasetto18} was fit using an Ensemble Kalman Filter (EnKF) \cite{evensen09}; though EnKF scales favorably with the number of spatial units, it relies on linearization of latent states which can be problematic for highly nonlinear systems \cite{evensen22,ionides21}. Alternative approaches used to fit stochastic models included making additional simplifying assumptions to aid in the fitting process \cite{lee20}, and using MCMC algorithms \cite{sallah17,azman12} which require specific structures in the latent dynamics, making these algorithms non plug-and-play. In this subsection, we present how the recently developed iterated block particle filter (IBPF) algorithm \cite{ning23,ionides22} can be used to fit a spatially explicit stochastic dynamic model to incidence data.

One issue that arises when fitting spatially explicit models is that parameter estimation techniques based on the particle filter become computationally intractable as the number of spatial units increases.
This is a result of the approximation error of particle filters growing exponentially in the dimension of the model \cite{rebeschini15,park20}.
To avoid the approximation error present in high-dimensional models, Lee et al.~\cite{lee20} simplified the problem of estimating the parameters of Model~3 by creating an approximate version of the model where the units are independent given the observed data.
Reducing a spatially coupled model to individual units in this fashion requires special treatment of any interactive mechanisms between spatial units, such as found in Eq.~\myeqref{eq:model3:foi}.
Because the simplified, spatially-decoupled version of Model~3 implemented in \cite{lee20} relies on the observed cholera cases, the calibrated model cannot readily be used to obtain forecasts.
Therefore, in order to obtain model forecasts, Lee et al.~\cite{lee20} used the parameters estimates from the spatially-decoupled approximation of Model~3 to obtain forecasts using the fully coupled version of the model.
This approach of model calibration and forecasting avoids the issue of particle depletion, but may also be problematic.
One concern is that cholera dynamics in department $u$ are highly related to the dynamics in the remaining departments;
calibrating model parameters while conditioning on the observed cases in other departments may therefore lead to an over-dependence on observed cholera cases.
Another concern is that the two versions of the model are not the same, resulting in sub-optimal parameter estimates for the spatially coupled model,
as parameters that maximize the likelihood of the decoupled model almost certainly do not maximize the likelihood of the fully coupled model.
These two concerns may explain the unrealistic forecasts and low likelihood of Model~3 in \cite{lee20} (Table~\ref{tab:likes}).

\begin{figure}[!h]
\begin{knitrout}
\definecolor{shadecolor}{rgb}{0.969, 0.969, 0.969}\color{fgcolor}
\includegraphics[width=\maxwidth]{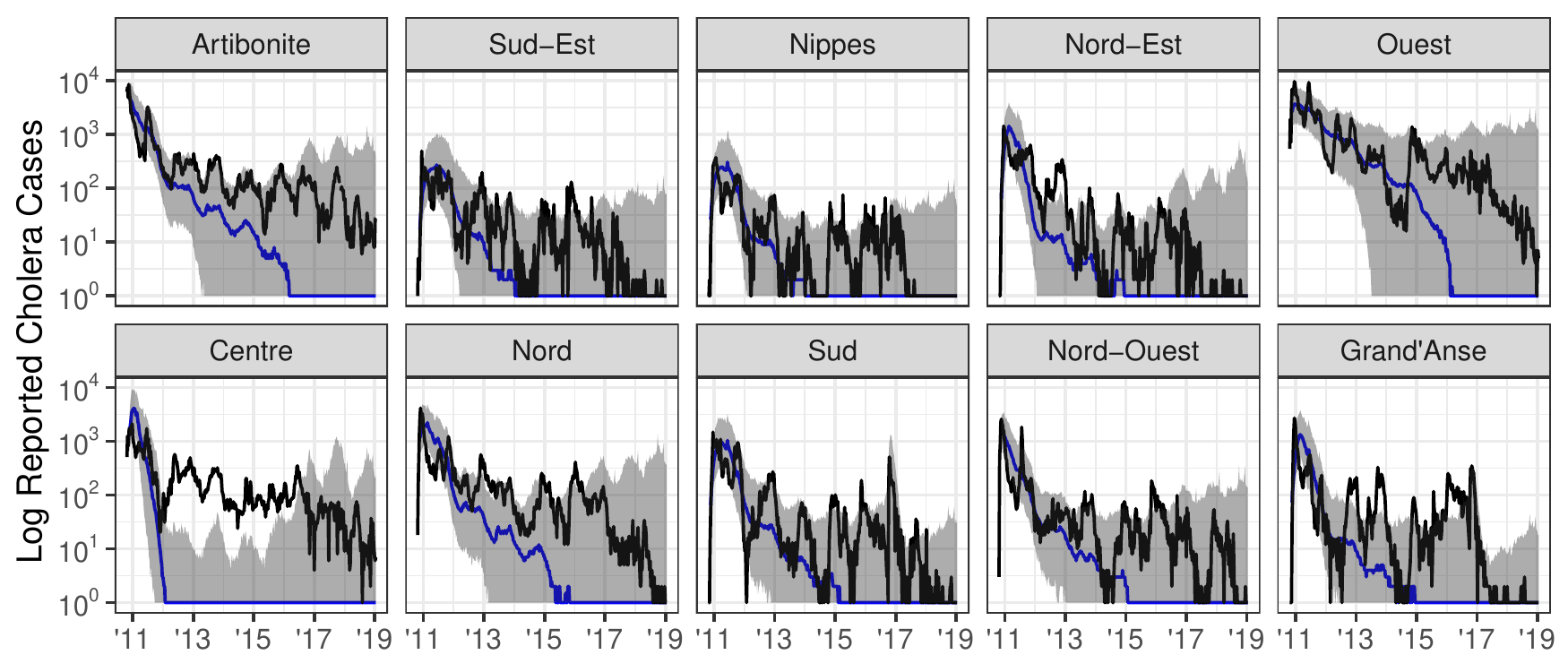} 
\end{knitrout}
\caption{\label{fig:h3spatsims}
{\figTitle Simulations from Model~3 compared to reported cholera cases.}
Simulations from initial conditions using the spatially coupled version of Model~3.
The black curve represents true case count, the blue line the median of 500 simulations from the model, and the gray ribbons representing $95\%$ confidence interval.
}
\end{figure}

At the time Lee et al.~\cite{lee20} conducted their study, there was no known algorithm that could readily be used to maximize the likelihood of an arbitrary meta-population POMP model with coupled spatial dynamics, which justifies the spatial decoupling approximation that was used to calibrate model parameters.
For our analysis, we calibrate the parameters of the spatially coupled version of Model~3 using the IBPF algorithm \cite{ionides22}.
This algorithm extends the work of Ning and Ionides \cite{ning23}, who provided theoretic justification for the version of the algorithm that only estimates unit-specific parameters.
The IBPF algorithm enables us to directly estimate the parameters of models describing high-dimensional partially-observed nonlinear dynamic systems via likelihood maximization.
The ability to directly estimate parameters of Model~3 is responsible for the large increase in model likelihoods reported in Table~\ref{tab:likes}.
Simulations from the fitted model are displayed in Fig.~\ref{fig:h3spatsims}.

\subsection*{Model Diagnostics}\label{sec:model_diagnostics}

The goal of parameter calibration---whether done using Bayesian or frequentist methods---is to find the best description of the observed data in the context of the model.
Obtaining the best fitting set of parameters for a given model does not, however, guarantee that the model provides an accurate representation of the system under investigation.
Model misspecification, which may be thought of as the omission of a mechanism in the model that is an important feature of the dynamic system, is inevitable at all levels of model complexity.
To make progress, while accepting proper limitations, one must bear in mind the much-quoted observation of George Box \cite{box79} that ``all models are wrong but some are useful.''
Beyond being good practical advice for applied statistics, this assertion is relevant for the philosophical justification of statistical inference as severe testing \cite{mayo18}. 
In this section, we discuss some tools for diagnosing mechanistic models with the goal of making the subjective assessment of model ``usefulness'' more objective.
To do this, we will rely on the quantitative ability of the model to match the observed data, which we call the model's {\it goodness-of-fit}, with the guiding principle that a model which cannot adequately describe observed data may not be reliable for useful purposes.
Goodness-of-fit may provide evidence supporting the causal interpretation of one model versus another, but cannot by itself rule out the possibility of alternative explanations.

One common approach to assess a mechanistic model's goodness-of-fit is to compare simulations from the fitted model to the observed data.
Visual inspection may indicate defects in the model, or may  suggest that the observed data are a plausible realization of the fitted model.
While visual comparisons can be informative, they provide only a weak and informal measure of the goodness-of-fit of a model.
The study by Lee et al.~\cite{lee20} provides an example of this: their models and parameter estimates resulted in simulations that visually resembled the observed data, yet resulted in model likelihoods that were considerably smaller than likelihoods that can be achieved (see Table~\ref{tab:likes}).
Alternative forms of model validation should therefore be used in conjunction with visual comparisons of simulations to observed data.

Another approach is to compare a quantitative measure of the model fit (such as MSE, predictive accuracy, or model likelihood) among all proposed models.
These comparisons, which provide insight into how each model performs relative to the others, are quite common \cite{rinaldo12,sallah17}.
To calibrate relative measures of fit, it is useful to compare against a model that has well-understood statistical ability to fit data, and we call this model a {\it benchmark}.
Standard statistical models, interpreted as associative models without requiring any mechanistic interpretation of their parameters, provide suitable benchmarks.
Examples include linear regression, auto regressive moving average (ARMA) time series models, or even independent and identically distributed measurements.
Benchmarks enable us to evaluate the goodness of fit that can be expected of a suitable mechanistic model.

Associative models are not constrained to have a causal interpretation, and typically are designed with the sole goal of providing a statistical fit to data.
Therefore, we should not require a candidate mechanistic model to beat all benchmarks.
However, a mechanistic model which falls far short against benchmarks is evidently failing to explain some substantial aspect of the data.
A convenient measure of fit should have interpretable differences that help to operationalize the meaning of far short.
Ideally, the measure should also have favorable theoretical properties.
Consequently, we focus on log-likelihood as a measure of goodness of fit, and we adjust for the degrees of freedom of the models to be compared by using the Akaike information criterion (AIC) \cite{aic74}.

In some cases, a possible benchmark model could be a generally accepted mechanistic model, but often no such model is available.
Because of this, we use a simple negative binomial model with an auto regressive mean as our associative benchmark; this model is described in \myeqref{eq:AR_NB}.
\begin{eqnarray}
\label{eq:AR_NB}
Y_n|Y_{n - 1} \sim \text{NB}\big(\NBintercept + \NBar Y_{n-1}, \NBsize \big),
\end{eqnarray}
where $\mathrm{E}\big(Y_n|Y_{n - 1}\big) = \NBintercept + \NBar Y_{n-1}$, and $\Var\big(Y_n|Y_{n - 1}\big) = \mathrm{E}\big(Y_n|Y_{n - 1}\big) + \mathrm{E}\big(Y_n|Y_{n - 1}\big)^2 / \NBsize$.
To obtain a benchmark for models with a meta-population structure, we fit independent auto-regressive negative binomial models to each spatial unit. Under the assumption of independence, the log-likelihood of the benchmark on the entire collection of data can be obtained by summing up the log-likelihood for each independent model. In general, a spatially explicit model may not have well-defined individual log-likelihoods, and, in this case, comparisons to benchmarks must be made at the level of the joint model.

In the case where the case counts are large, an alternative benchmark recommended by He et al.~\cite{he10} is a log-linear Gaussian ARMA model;
the theory and practice of ARMA models is well developed, and these linear models are appropriate on a log scale due to the exponential growth and decay characteristic of biological dynamics.
We use the auto regressive negative binomial model, however, because the large number of weeks with zero recorded cholera cases in department level data makes a benchmark based on a continuous distribution problematic.
Log-likelihoods and AIC values of Models~1--3 and of their respective benchmark models are provided in Table~\ref{tab:likes}.
Models that are fit to the same datasets can be directly compared using AIC values, making it a useful tool to compare to benchmark models. Though Models~2 and 3 are both fit to department level incidence reports, their AIC values are not directly comparable due to the way Model~3 initializes latent states (\nameref{S_init}).

It should be universal practice to present measures of goodness of fit for published models, and mechanistic models should be compared against benchmarks.
In our literature review of the Haiti cholera epidemic, no non-mechanistic benchmark models were considered in any of the 32 papers that used dynamic models to describe cholera in order to obtain scientific conclusions.
Including benchmarks would help authors and readers to detect and confront any major statistical limitations of the proposed mechanistic models.
In addition, the published goodness of fit provides a concrete point of comparison for subsequent scientific investigations.
When combined with online availability of data and code, objective measures of fit provide a powerful tool to accelerate scientific progress, following the paradigm of the {\it common task framework} \cite{donoho17}. 

The use of benchmarks may also be beneficial when developing models at differing spatial scales, where a direct comparison between model likelihoods is meaningless.
In such a case, a benchmark model can be fit to each spatial resolution being considered, and each model compared to their respective benchmark.
Large advantages (or shortcomings) in model likelihood relative to the benchmark for a given spatial scale that are not present in other spatial scales may provide weak evidence for (or against) the statistical fit of models across a range of spatial resolutions.

Comparing model log-likelihoods to a suitable benchmark may not be sufficient to identify all the strengths and weaknesses of a given model.
Additional techniques include the inspection of conditional log-likelihoods of each observation given the previous observations in order to understand how well the model describes each data point (\nameref{S_mod3cal}).
Other tools include plotting the effective sample size of each observation \cite{liu01}; plotting the values of the hidden states from simulations (\nameref{S_mod3cal});
and comparing summary statistics of the observed data to simulations from the model \cite{wood10,king15}.

\subsection*{Corroborating Fitted Models with Scientific Knowledge}

The resulting mechanisms in a fitted model can be compared to current scientific knowledge about a system.
Agreement between model-based inference and our current understanding of a system may be taken as a confirmation of both model-based conclusions and our scientific understanding.
On the other hand, comparisons may generate unexpected results that have the potential to spark new scientific knowledge \cite{ganusov16}.

\begin{figure}[!h]
\begin{knitrout}
\definecolor{shadecolor}{rgb}{0.969, 0.969, 0.969}\color{fgcolor}

{\centering \includegraphics[width=\maxwidth]{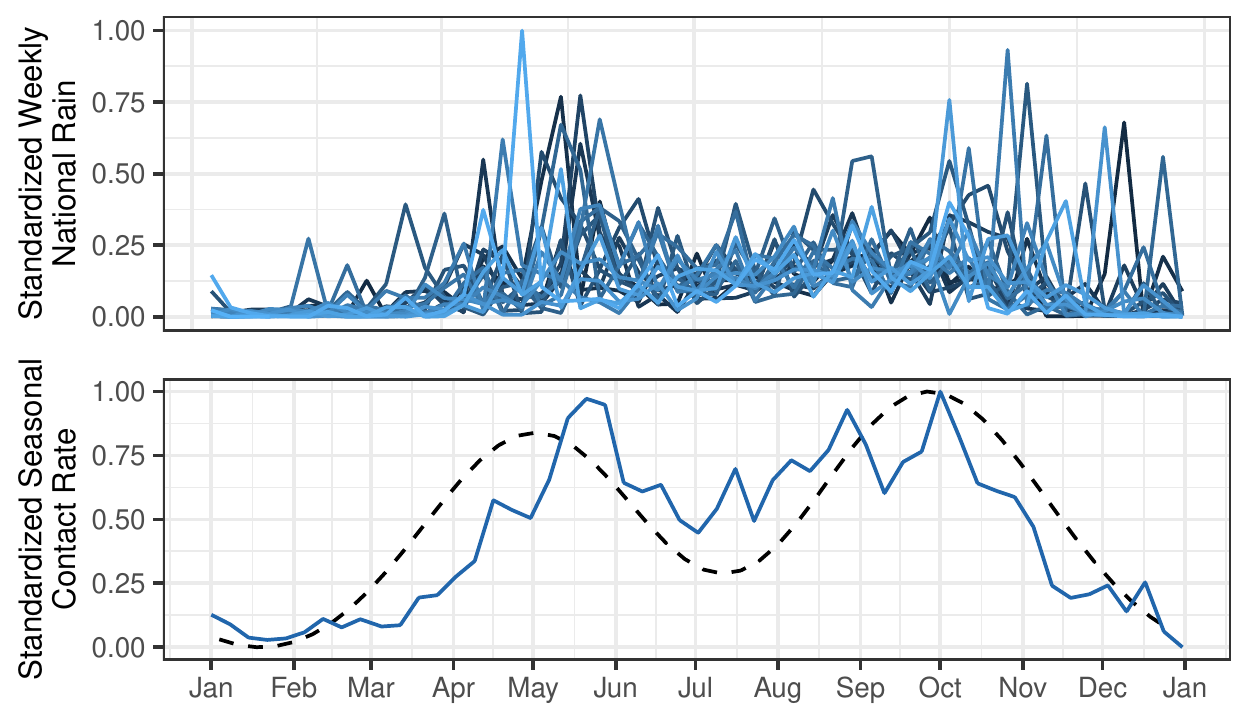} 

}

\end{knitrout}
\caption{\label{fig:h1SeasRain}
{\figTitle Seasonality of Model~1 transmission compared to rainfall data.}
(Top) weekly rainfall in Haiti, lighter colors representing more recent years.
(Bottom) estimated seasonality in the transmission rate (dashed line) plotted alongside mean rainfall (solid line).
The outsized effect of rainfall in the fall may be due to Hurricane Matthew, which struck Haiti in October of 2016 and resulted in an increase of cholera cases in the nation.
}
\end{figure}

In the context of our case study, we demonstrate how the fit of Model~1 corroborates other evidence concerning the role of rainfall in cholera epidemics.
Specifically, we examine the results of fitting the flexible cubic spline term in Model~1 (Eqs.~\myeqref{model1:lambda}--\myeqref{model1:beta}).
The cubic splines permit flexible estimation of seasonality in the force of infection, $\transmission(t)$.
Fig.~\ref{fig:h1SeasRain} shows that the estimated seasonal transmission rate $\transmission$ mimics the rainfall dynamics in Haiti, despite Model~1 not having access to rainfall data.
This is consistent with previous studies that incorporated rainfall as an important part of their mechanistic model or otherwise argue that rainfall is an important driver of cholera dynamics in Haiti \cite{hulland19,kirpich17,lee20,moise20,pasetto18,kirpich16,eisenberg13,rinaldo12,mavian20}.
The estimated seasonality also features an increased transmission rate during the fall, which was noticed at an earlier stage of the epidemic \cite{rinaldo12}. The high transmission rate in the fall may be a result of the increase transmission that occurred in the fall of 2016, when hurricane Matthew struck Haiti \cite{ferreirai16}.

For any model-based inference, it is important to recognize and assess the modeling simplifications and assumptions that were used in order to arrive at the conclusions.
In epidemiological studies, for example, quantitative understanding of individual-level processes may not perfectly match model parameters that were fit to population-level case counts, even when the model provides a strong statistical fit \cite{he10}.
This makes direct interpretation of estimated parameters delicate.

Our case study provides an example of this in the parameter estimate for the duration of natural immunity due to cholera infection, $\muRS^{-1}$.
Under the framework of Model~2, the best estimate for this parameter is $\ensuremath{1.4\times 10^{11}}$ yr, suggesting that individuals have permanent immunity to cholera once infected.
Rather than interpreting this as scientific evidence that individuals have permanent immunity from cholera, this result suggests that Model~2 favors a regime where reinfection events are a negligible part of the dynamics. 
The depletion of susceptible individuals may be attributed to confounding mechanisms---such as localized vaccination programs and non-pharmaceutical interventions that reduce cholera transmission \cite{trevisin22, rebaudet21}---that were not accounted for in the model.
Perhaps the best interpretation of the estimated parameter, then, is that under the modeling framework that was used, the model most adequately describes the observed data by having a steady decrease in the number of susceptible individuals.
The weak statistical fit of Model~2 compared to a log-linear benchmark (see Table~\ref{tab:likes}) cautions us against drawing quantitative conclusions from this model.
A model that has a poor statistical fit may nevertheless provide a useful conceptual framework for thinking about the system under investigation.
However, a claim that the model has been validated against data should be reserved for situations where the model provides a statistical fit that is competitive against alternative explanations.

A model which aspires to provide quantitative guidance for assessing interventions should provide a quantitative statistical fit for available data.
However, strong statistical fit does not guarantee a correct causal structure: it does not even necessarily require the model to assert a causal explanation.
A causal interpretation is strengthened by corroborative evidence.
For example, reconstructed latent variables (such as numbers of susceptible and recovered individuals) should make sense in the context of alternative measurements of these variables \cite{grad12}.
Similarly, parameters that have been calibrated to data should make sense in the context of alternative lines of evidence about the phenomena being modeled, while making allowance for the possibility that the interpretations of parameters may vary when modeling across differing spatial scales.

In the supplement material (\nameref{S_mod3cal}), we explore in more detail the process of model fitting and diagnostics for Model~3. Here we demonstrate that the model outperforms its benchmark model on the aggregate scale.
However, when focusing on the spatial units with the highest incidence of cholera, Model~3 performs roughly the same as a simple benchmark. By comparing simulations from the fitted model to the filtering distribution, we see that the reconstructed latent states of the model favor higher levels of cholera transmission than what is typically observed in the incidence data.
These results hint at the possibility of model mispecification, and warrant a degree of caution in interpreting the model's outputs.

\section*{Results}

\subsection*{Forecasts}\label{sec:filter}

Forecasts are an attempt to provide an accurate estimate of the future state of a system based on currently available data, together with an assessment of uncertainty.
Forecasts from mechanistic models that are compatible with current scientific understanding may also provide estimates of the future effects of potential interventions.
Further, they may enable real-time testing of new scientific hypotheses \cite{lewis22}.

Forecasts of a dynamic system should should be consistent with the available data. It is particularly important that forecasts are consistent with the most recent information available, as recent data is likely to be more relevant than older data. While this assertion may seem self-evident, it is not the case for deterministic models, for which the initial conditions together with the parameters are sufficient for forecasting, and so recent data may not be consistent with model trajectories. Epidemiological forecasts based on deterministic models are not uncommon in practice, despite their limitations \cite{king15}. Lee et al.~\cite{lee20} chose to obtain forecasts from all of their models by simulating forward from initial conditions, rather than conditioning forecasts based on the available data.
This decision is possibly as a result of using a deterministic model, as forecasts from different models may only be considered comparable if they are obtained in the same way, which is most easily done be by simulating from initial conditions because Model~2 is deterministic.

In contrast, for non-deterministic Models~1 and 3, we obtain forecasts by simulating future values using latent states that are harmonious with the most recent data. This is done by simulating forward from latent states drawn at the last observation time ($t_N$) from the filtering distribution $f_{\bm{X}_N|\bm{Y}_{1:N}}(\bm{x}_{N} | \bm{y}^*_{1:N} ; \hat\paramVec)$. The decision to obtain model forecasts from initial conditions partially explains the unsuccessful forecasts of Lee et al.~\cite{lee20}. Table~S7 in their supplement material, which contains results that were not discussed in their main article, shows that the subset of their simulations with zero cholera cases from 2019-2020 also correspond with its disappearance until 2022. These results support our argument that forecasts should be made by ensuring the starting point for the forecast is consistent with available data.

Uncertainty in just a single parameter can lead to drastically different forecasts \cite{saltelli20}.
Therefore, parameter uncertainty should also be considered when obtaining model forecasts to influence policy.
If a Bayesian technique is used for parameter estimation, a natural way to account for parameter uncertainty is to obtain simulations from the model where each simulation is obtained using parameters drawn from the estimated posterior distribution.
For frequentist inference, one possible approach is obtaining model forecasts from various values of $\paramVec$, where the values of $\paramVec$ are sampled proportionally according to their corresponding likelihoods \cite{king15} (\nameref{S_uncertain}).
Both of these approaches share the similarity that parameters are chosen for the forecast approximately in proportion to their corresponding value of the likelihood function, $f_{\bm{Y}_{1:N}}(\bm{y}_{1:N}^*; \paramVec)$.
In this analysis, we do not construct forecasts accounting for parameter uncertainty as our focus is on the estimation and diagnosis of mechanistic models, rather than providing forecasts intended to influence policy.
Furthermore, we use the projections from a single point estimate to highlight the deficiency of deterministic models that the only variability in model projections is a result of parameter and measurement uncertainty, which can lead to over-confidence in forecasts \cite{king15}.

The primary forecasting goal of Lee et al.~\cite{lee20} was to investigate the potential consequences of vaccination interventions on a system to inform policy.
One outcome of their study include estimates for the probability of cholera elimination under several possible vaccination scenarios.
Mimicking their approach, we define cholera elimination as an absence of cholera infections for at least 52 consecutive weeks, and we provide forecasts under the following vaccination scenarios:

\begin{itemize}
  \item[$V0$:] No additional vaccines are administered.
  \item[$V1$:] Vaccination limited to the departments of Centre and Artibonite, deployed over a two-year period.
  \item[$V2$:] Vaccination limited to three departments: Artibonite, Centre, and Ouest deployed over a two-year period.
  \item[$V3$:] Countrywide vaccination implemented over a five-year period.
  \item[$V4$:] Countrywide vaccination implemented over a two-year period.
\end{itemize}

\begin{figure}[!h]
\begin{knitrout}
\definecolor{shadecolor}{rgb}{0.969, 0.969, 0.969}\color{fgcolor}
\includegraphics[width=\maxwidth]{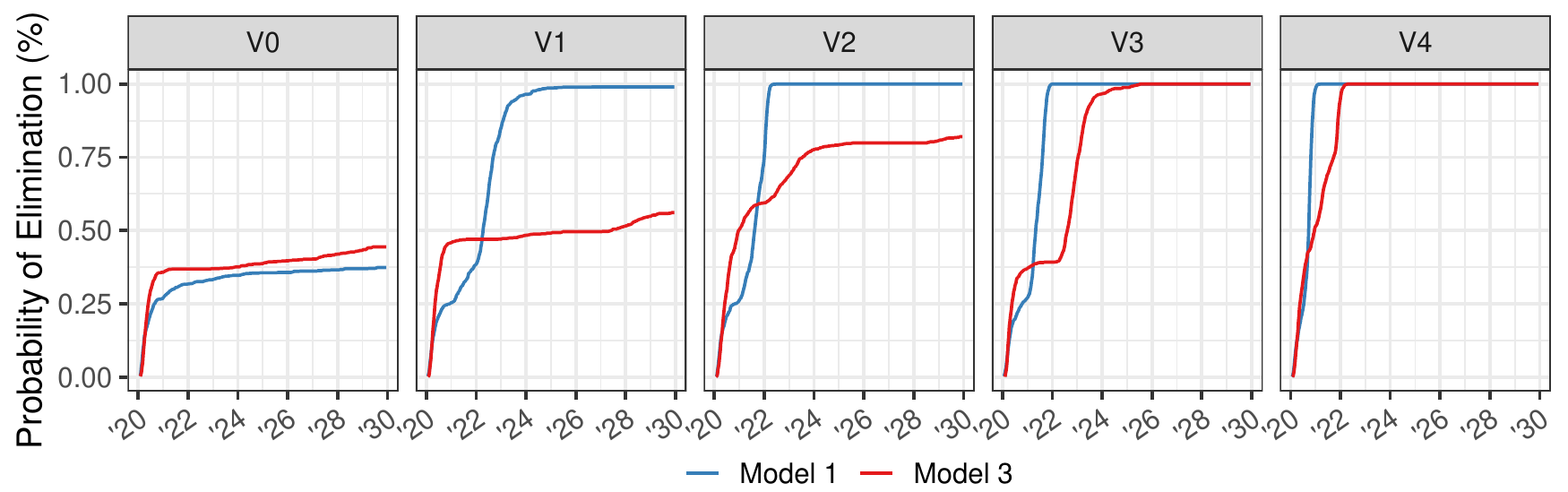} 
\end{knitrout}
\caption{\label{fig:elimProbs}
{\figTitle Simulated probability of elimination using Models~1 and 3.}
Probability of cholera elimination, defined as having zero cholera infectious for at least 52 consecutive weeks, based on 10 year simulations from calibrated versions of Models~1 and 3.
Compare to Fig.~3A of \cite{lee20}.}
\end{figure}

Simulations from probabilistic models (Models~1 and~3) represent possible trajectories of the dynamic system under the scientific assumptions of the models.
Because Model~1 only accounts for national level disease dynamics, the pre-determined department-specific vaccination campaigns are carried out by assuming the vaccines are administered in one week to the same number of individuals that would have obtained vaccines if explicitly administered to the specific departments. We refer readers to \cite{lee20} and the accompanying supplement material for more details.
Estimates of the probability of cholera elimination can therefore be obtained as the proportion of simulations from these models that result in cholera elimination.
The results of these projections are summarized in Figs. \ref{fig:elimProbs}.

Probability of elimination estimates of this form are not meaningful for deterministic models, as the trajectory of these models only represent the mean behavior of the system rather than individual potential outcomes.
We therefore do not provide probability of elimination estimates under Model~2, but show trajectories under the various vaccination scenarios using this model (Fig.~\ref{fig:mod2Traj}).

\section*{Discussion}\label{sec:discussion}

The ongoing global COVID-19 pandemic has demonstrated how government policy may be affected by the inferences drawn from mathematical modeling \cite{saltelli20}.
However, the development of credible models---which are supported by data and can provide quantitative insights into a dynamic system---remains a challenging task.
In this article, we demonstrated opportunities available for raising the current standards of statistical inference for mathematical models of biological systems.

We presented methodology consistent with existing guidelines \cite{behrend20} but going beyond standard practice.
In particular, we showed the value of comparing the likelihood of fitted mechanistic models versus non-mechanistic benchmarks, a practice that has been previously advocated for \cite{he10} but was not done by any of the studies in our literature review.
These comparisons, along with other likelihood based diagnostics, help identify specific limitations of proposed models.
Diagnostic tools include likelihood profile methods, which help to assess parameter identifiability and enable the construction of confidence intervals for parameter estimates \cite{ionides17, simpson23}.
When reaching conclusions, it is important to consider potential consequences of confounded variables and model misspecification.

Model diagnostics are a key tool for exposing unresolved model limitations and improving model fit. In our case study, we compared the three models from Lee et al.~\cite{lee20} to statistical benchmarks, revealing areas for improvement. For example, comparisons of Model~3 to a benchmark revealed its inadequacy in accounting for the post-hurricane increase in transmission, leading to a beneficial model refinement. When a mechanistic model is competitive with statistical benchmarks, we have a license to begin critical evaluation of its causal implications.
If a model falls far behind simple benchmarks, there is likely to be substantial limitation in the data analysis that should be identified and remedied.
In our case study, the re-calibrated version of Model~1 outperformed its benchmark, so we proceeded to examine causal implications. When doing so, we found that the fitted model provides a causal description of the dynamic system that is consistent with known features of the system, such as the importance of rainfall as a driver of cholera infection. The congruency between causal implications of the model and our belief about the dynamic system, coupled with a strong quantitative description of observed data relative to a benchmark, provides support for viewing the model as a plausible quantitative representation of the system under investigation.

When fitting a mechanistic model to a dynamic system, the complexity of the model warrants consideration. Mathematical models provide simplified representations of complex systems, with the simplicity serving both to facilitate scientific understanding and to enable statistical inference on unknown parameters.
In our case study, employing deterministic dynamics in Model~2 was found to be an over-simplification by comparing model fit with benchmarks.
Model~3 is distinct in that it is both stochastic and has a meta-population structure, making it challenging to draw likelihood-based inferences. In this paper, we demonstrated how this model class can be calibrated to incidence data using the innovative IBPF algorithm. One of only a few examples of fitting a nonlinear non-Gaussian meta-population model via maximum likelihood \cite{li23,ionides22}, this case study exemplifies the algorithm's potential benefits and provides an example for future researchers on a possible approach to fitting a high-dimensional non-linear model.

Likelihood-based methods aid in determining an appropriate level of model complexity. Models fit to the same data can be compared using a criteria such as AIC. Nested model variations are particularly useful as they enable formal statistical testing of the nested features via likelihood ratio tests. Our case study demonstrated the examination of nested model features for all three models. Model~1 investigated a time-varying transmission rate; Model~2 assessed a phase-shift parameter in seasonal cholera peaks; Model~3 incorporated hurricane-related parameters.

Unmodeled features of a dynamic system can lead to spurious or misleading parameter estimates if the features substantially impact observed data. 
In deterministic models, features that cannot be explained by measurement error must be accounted for by the choice of parameters. 
For our case study, some of the parameter estimates for the deterministic Model~2 are implausible, such as the infinite immunity discussed above, and this may be explained by compensation for model misspecification.
Incorporating demographic and environmental stochasticity into models can mitigate the impact of unmodeled features. Stochastic phenomena are not only arguably present in biological systems, but their inclusion in a model also allows observed data variations to be attributed to inherent uncertainty rather than to distorted parameter values.
Models~1 and 3 suggest the presence of extra-demographic stochasticity \cite{he10,stocks20,li23}, as evidenced by the confidence intervals for the corresponding parameter $\sigmaProc$ (\nameref{S_CI}).

If forecasts are an important component of a modeling task, the forecasts should be consistent with the available data, particularly at the most recently available time points. In our case study, we did this by simulating forward from the filtering distribution, as this procedure conditions latent variables on the available data. This type of forecasting, however, is not directly available using a deterministic model, where future dynamics are fully determined by initial conditions and parameter values. This can result in over-confident model forecasts \cite{king15}. Despite their limitations, deterministic models can offer valuable insights into dynamic systems \cite{may04}. In \cite{lee20}, the forecasts from the deterministic Model~2 were qualitatively more consistent with the observed disappearance of cholera than the stochastic models. In our case study, we found improvements to Models~1 and 3 that resulted in improved forecasts for these models.

In our case study, we found that additional attention to statistical details could have resulted in an enhanced statistical fit to the observed incidence data. This would have improved the accuracy of the policy guidance resulting from the study.
We used the same data, models, and much of the same code used by Lee et al.~\cite{lee20}, but we arrived at drastically different conclusions.
Specifically, each of the re-calibrated models predicted with moderate probability that cholera would disappear from Haiti.
Although there have been new cases of cholera in Haiti, this conclusion aligns more with the prolonged absence of cholera cases from 2019-2022.
We acknowledge the benefit of hindsight: our demonstration of a statistically principled route to obtain better-fitting models resulting in more robust insights does not rule out the possibility of discovering other models that fit well yet predict poorly.

Mechanistic models offer opportunities for understanding and controlling complex dynamic systems.
This case study has investigated issues requiring attention when applying powerful new statistical techniques that can enable statistically efficient inference for a general class of partially observed Markov process models.
Researchers should ensure that intensive numerical calculations are adequately executed.
Using benchmarks and alternative model specifications to assess statistical goodness-of-fit should also should be common practice.
Once a model has been adequately calibrated to data, care is required to assess what causal conclusions can properly be inferred given the possibility of alternative explanations consistent with the data.
Studies that combine model development with thoughtful data analysis, supported by a high standard of reproducibility, build knowledge about the system under investigation.
Cautionary warnings about the difficulties inherent in understanding complex systems \cite{saltelli20,ioannidis20,ganusov16} should motivate us to follow best practices in data analysis, rather than avoiding the challenge.

\subsection*{Reproducibility and Extendability}

Lee et al.~\cite{lee20} published their code and data online, and this reproducibility facilitated our work.
Robust data analysis requires not only reproducibility but also extendability: if one wishes to try new model variations, or new approaches to fitting the existing models, or plotting the results in a different way, this should not be excessively burdensome.
Scientific results are only trustworthy so far as they can be critically questioned, and an extendable analysis should facilitate such examination \cite{gentleman07}.

We provide a strong form of reproducibility, as well as extendability, by developing our analysis in the context of a software package, \code{haitipkg}, written in the R language \cite{r}.
Using a software package mechanism supports documentation, standardization and portability that promote extendability.
In the terminology of Gentleman and Temple Lang~\cite{gentleman07}, the source code for this article is a {\it dynamic document} combining code chunks with text.
In addition to reproducing the article, the code can be extended to examine alternative analysis to that presented.
The dynamic document, together with the R packages, form a {\it compendium}, defined by Gentleman and Temple Lang~\cite{gentleman07} as a distributable and executable unit which combines data, text and auxiliary software (the latter meaning code written to run in a general-purpose, portable programming environment, which in this case is R).

\section*{Supporting information}

\paragraph*{S1 Fig.}
\label{S_Mod1}
{\bf Model~1.} Flow chart representation of Model~1.

\paragraph*{S2 Fig.}
\label{S_Mod2}
{\bf Model~2.} Flow chart representation of Model~2.

\paragraph*{S3 Fig.}
\label{S_Mod3}
{\bf Model~3.} Flow chart representation of Model~3.

\paragraph*{S1 Table}
\label{S1_Table}
{\bf Notation conversion table.} Conversions between the notation used here and the notation of Lee et al. (2020) \cite{lee20}.

\paragraph*{S1 Text}
\label{S1_Text}
{\bf Model details.} In depth description of the Markov chain and differential equation interpretations of compartment flow rates, as well as details on the numeric implementation of these models.

\paragraph*{S2 Text}
\label{S_init}
{\bf Initialization models.} Additional details of the initialization model that were used for Models~1--3.

\paragraph*{S3 Text}
\label{S_meas}
{\bf Measurement models.} Additional details of the measurement model that were used for Models~1--3.

\paragraph*{S4 Text}
\label{S_lee20}
{\bf Likelihood of Lee et al. models.} Details of how the log-likelihood of the Lee et al. models were calculated.

\paragraph*{S5 Text}
\label{S_mod3cal}
{\bf Calibrating Model 3.} Additional details on the procedures for fitting and diagnosing Model~3.

\paragraph*{S6 Text}
\label{S_uncertain}
{\bf Accounting for parameter uncertainty} A description of a empirical-Bayes approach that can be used to account for parameter uncertainty.

\paragraph*{S7 Text}
\label{S_CI}
{\bf Confidence Intervals} Tables and figures describing 95\% confidence intervals for estimated parameters.

\section*{Acknowledgments}

This work was supported by National Science Foundation grants DMS-1761603 and DMS-1646108.
The authors would like to thank Mercedes Pascual and Betz Halloran for helpful discussions. Laura Matrajt provided additional data for the Model~2 analysis.

s

\end{document}


\date{\today}
\title{Supplement to ``{\mytitle}''}
  \author{Jesse Wheeler, AnnaElaine L. Rosengart, Zhuoxun Jiang, \\ Kevin Tan, Noah Treutle and Edward L. Ionides
 \\ \hspace{.2cm}\\
    Department of Statistics, University of Michigan\\
}

\newcommand{\blind}{1}

\if1\blind
{
\maketitle
}
\fi

\if0\blind
{
  \bigskip
  \bigskip
  \bigskip
  \begin{center}
    {\LARGE\bf \mytitle}
\end{center}
  \bigskip
  \bigskip
}\fi

\tableofcontents

\newpage

\section{Model Diagrams}

Each of the dynamic models considered in this manuscript can be fully described using the model descriptions in the manuscript, coupled with the additional information described in Sections 2 and 3 of this supplement.
Despite this, diagrams of dynamic systems are often helpful to understand the equations.
In this section, we give three diagrams representing Models~1--3, respectively.
Because the models are defined by their mathematical equations and numeric implementation, these diagrams are not unique visual representations of the model.
Alternative representations that may be helpful in understanding the models explored in this paper were made by \citet{lee20sup}.

\subsection{Model~1}

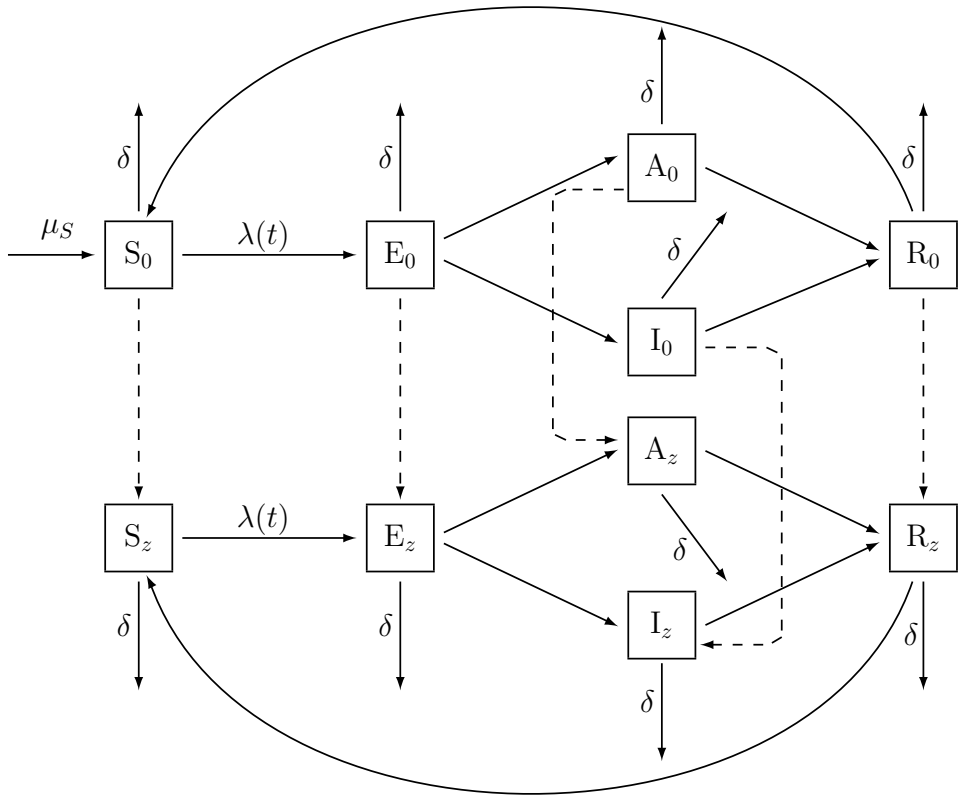
\begin{figure}[!ht]
\begin{center}
  \resizebox{0.9\textwidth}{!}{
    \Large
    \setlength{\unitlength}{5pt}
    \begin{picture}(100,95)(0,15)
      \thicklines

      \put(10,63){\framebox(6,6){$\mathrm{S}_0$}}
      \put(34,63){\framebox(6,6){$\mathrm{E}_0$}}
      \put(58,71){\framebox(6,6){$\mathrm{A}_0$}}
      \put(82,63){\framebox(6,6){$\mathrm{R}_0$}}
      \put(58,55){\framebox(6,6){$\mathrm{I}_0$}}

      \put(1, 66){\vector(1, 0){8}}
      \put(4, 68){$\muBirth$}

      \put(17,66){\vector(1,0){16}}
      \put(22,67){$\lambda(t)$}

      \put(41,67.5){\vector(2.1,1){16}}

      \put(41,65.5){\vector(2.1,-1){16}}

      \put(65,74){\vector(2.1,-1){16}}

      \put(65,59){\vector(2.4,1){16}}

      \cbezier(84,70)(75, 95)(23,95)(14,70)
      \put(14,70){\vector(-1, -2){0.3}}

      \put(13, 70){\vector(0, 1){10}}
      \put(11, 74){$\muDeath$}

      \put(37, 70){\vector(0, 1){10}}
      \put(35, 74){$\muDeath$}

      \put(61, 78){\vector(0, 1){9}}
      \put(59, 80){$\muDeath$}

      \put(61, 62){\vector(1.5, 2){6}}
      \put(61.5, 65.5){$\muDeath$}

      \put(85, 70){\vector(0, 1){10}}
      \put(83, 74){$\muDeath$}

      \put(10,37){\framebox(6,6){$\mathrm{S}_\vaccCounter$}}
      \put(34,37){\framebox(6,6){$\mathrm{E}_\vaccCounter$}}
      \put(58,45){\framebox(6,6){$\mathrm{A}_\vaccCounter$}}
      \put(82,37){\framebox(6,6){$\mathrm{R}_\vaccCounter$}}
      \put(58,29){\framebox(6,6){$\mathrm{I}_\vaccCounter$}}

      \put(17,40){\vector(1,0){16}}
      \put(22,41){$\lambda(t)$}

      \put(41,40.5){\vector(2.1,1){16}}

      \put(41,39.5){\vector(2.1,-1){16}}

      \put(65,48){\vector(2.1,-1){16}}

      \put(65,32){\vector(2.1,1){16}}

      \cbezier(84, 36)(75, 10)(23, 10)(14, 36)
      \put(14,36){\vector(-1, 2){0.3}}

      \put(13, 36){\vector(0, -1){10}}
      \put(11, 31){$\muDeath$}

      \put(37, 36){\vector(0, -1){10}}
      \put(35, 31){$\muDeath$}

      \put(61, 28.5){\vector(0, -1){9}}
      \put(59, 24){$\muDeath$}

      \put(61, 44){\vector(1.5, -2){6}}
      \put(62, 38){$\muDeath$}

      \put(85, 36){\vector(0, -1){10}}
      \put(83.2, 30.2){$\muDeath$}

      \multiput(13,62)(0,-2){9}{\line(0,-1){1}}
      \put(13, 43.5){\vector(0, -1){0}}

      \multiput(37,62)(0,-2){9}{\line(0,-1){1}}
      \put(37, 43.5){\vector(0, -1){0}}

      \multiput(85,62)(0,-2){9}{\line(0,-1){1}}
      \put(85, 43.5){\vector(0, -1){0}}

      \multiput(57.5,72)(-1.9, 0){3}{\line(-1, 0){0.9}}
      \put(52.2, 72){\line(-2.1, -1){1}}
      \multiput(51,71)(0, -2){11}{\line(0, -1){1}}
      \put(52.2, 49){\line(-2.1, 1){1}}
      \multiput(54,49)(1.9, 0){2}{\line(-1, 0){0.9}}
      \put(57, 49){\vector(1, 0){0}}

      \multiput(65, 57.5)(1.9, 0){3}{\line(1, 0){0.9}}
      \put(70.6, 57.5){\line(2.1, -1){1}}
      \multiput(72, 56.5)(0, -2){13}{\line(0, -1){1}}
      \put(72, 30.9){\line(-2, -1.2){0.9}}
      \multiput(65.6, 30.2)(1.9, 0){3}{\line(1, 0){0.9}}
      \put(64.5, 30.2){\vector(-1, 0){0}}

    \end{picture}
  }
\end{center}
\caption{A flow diagram for the SEAIR model.}\label{fig:flow_diagram}
\end{figure}

\subsection{Model~2}

\begin{figure}[!h]
\begin{center}
  \resizebox{\textwidth}{!}{
    \Large
    \setlength{\unitlength}{5pt}
    \begin{picture}(100,95)(0,15)
      \thicklines

      \put(10,63){\framebox(6,6){$\mathrm{S}_{u0}$}}
      \put(34,69){\framebox(6,6){$\mathrm{E}_{u0}$}}
      \put(58,77){\framebox(6,6){$\mathrm{A}_{u0}$}}
      \put(82,77){\framebox(6,6){$\mathrm{R^A}_{u0}$}}
      \put(82,61){\framebox(6,6){$\mathrm{R^I}_{u0}$}}
      \put(58,61){\framebox(6,6){$\mathrm{I}_{u0}$}}

      \put(17,66){\vector(3,1.1){16}}

      \put(41,72.5){\vector(2.1,1){16}}

      \put(41,71.5){\vector(2.1,-1){16}}

      \put(65,80.5){\vector(1,0){16}}

      \put(58,39){\framebox(6,6){$\mathrm{A}_{u\vaccCounter}$}}
      \qbezier[25](60.5, 46)(51, 52)(42, 53)
      \put(42, 53){\vector(-1, 0.1){0.5}}

      \qbezier[25](60.5, 60)(51, 54)(42, 54)
      \put(42, 54){\vector(-1, -0.1){0.5}}

      \qbezier[30](60.5,30)(48, 45)(40, 50)
      \put(40, 50){\vector(-1, 0.8){0.5}}

      \qbezier[29](60.5,76)(48, 61)(40.6, 57)
      \put(40.8, 57.1){\vector(-1, -0.7){0.5}}

      \put(65,63.5){\vector(1,0){16}}

      \put(10,37){\framebox(6,6){$\mathrm{S}_{u\vaccCounter}$}}
      \put(34,31){\framebox(6,6){$\mathrm{E}_{u\vaccCounter}$}}
      \put(58,39){\framebox(6,6){$\mathrm{A}_{u\vaccCounter}$}}
      \put(82,39){\framebox(6,6){$\mathrm{R^A}_{u\vaccCounter}$}}
      \put(58,23){\framebox(6,6){$\mathrm{I}_{u\vaccCounter}$}}
      \put(82,23){\framebox(6,6){$\mathrm{R^I}_{u\vaccCounter}$}}

      \cbezier(85,22)(80, 8)(18,8)(12,36)
      \put(12.1,35.5){\vector(-1, 3.5){0.3}}

      \cbezier(85,38)(80, 12)(18,15)(14,36)
      \put(14.1,35.5){\vector(-1, 3.5){0.3}}

      \cbezier(85,84)(80, 100)(18,100)(12,70)
      \put(12.1,70.5){\vector(-1, -3.5){0.3}}

      \cbezier(85,68)(80, 95)(18, 92)(14,70)
      \put(14.1,70.5){\vector(-1, -3.5){0.3}}

      \put(17,40){\vector(3,-1.1){16}}

      \put(41,34.5){\vector(2.1,1){16}}

      \put(41,33.5){\vector(2.1,-1){16}}

      \put(65,42.5){\vector(1,0){16}}

      \put(65,25.5){\vector(1,0){16}}

      \multiput(12,62)(0,-2){9}{\line(0,-1){1}}
      \put(12, 44){\vector(0, -1){0}}

      \multiput(14,44)(0,2){9}{\line(0,1){1}}
      \put(14, 62){\vector(0, 1){0}}

      \put(37, 54){\circle{7}}
      \put(35, 53){$\mathrm{W_u}$}

      \qbezier[10](32.5, 54)(29.5, 54)(26.5, 54)
      \put(26, 54){\vector(-1, 0){0.1}}
      \put(28, 55){$\Wremoval$}

    \end{picture}
  }
\end{center}
\caption{A flow diagram for the SEAIR model 2. This is a constant population model, there are no births/deaths. Vaccinations are assumed to only be given to susceptible individuals, and vaccine immunity wanes only with susceptible vaccinated individuals.}\label{fig:flow_diagram2}
\end{figure}
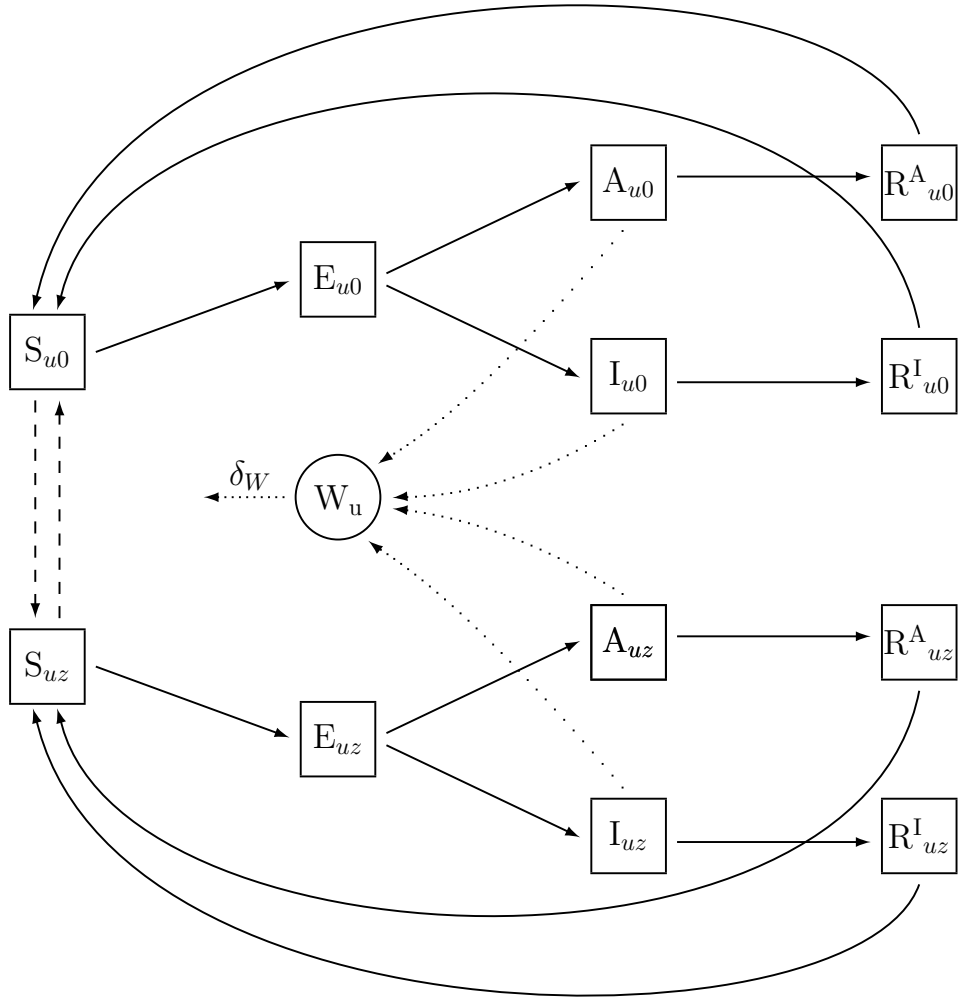

\newpage

\subsection{Model~3}

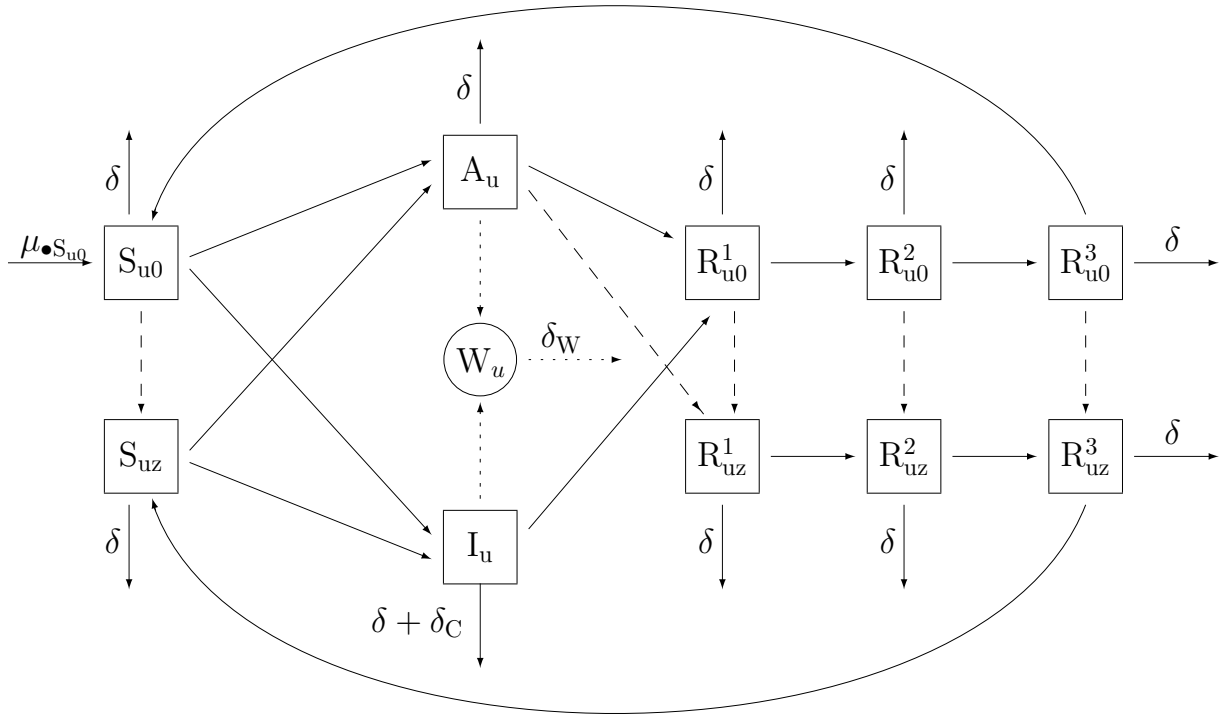
\begin{figure}[!h]
\begin{center}
  \resizebox{\textwidth}{!}{
    \Large
    \setlength{\unitlength}{5pt}
    \begin{picture}(100,85)(0,15)

    \put(39, 50){\circle{6}}
    \put(37, 49){$\mathrm{W}_{u}$}

    \put(8, 55){\framebox(6, 6){$\mathrm{S_{u0}}$}}
    \put(8, 39){\framebox(6, 6){$\mathrm{S_{u\vaccCounter}}$}}
    \put(36, 62.5){\framebox(6, 6){$\mathrm{A_{u}}$}}
    \put(36, 31.5){\framebox(6, 6){$\mathrm{I_{u}}$}}
    \put(56, 55){\framebox(6, 6){$\mathrm{R^1_{u0}}$}}
    \put(71, 55){\framebox(6, 6){$\mathrm{R^2_{u0}}$}}
    \put(86, 55){\framebox(6, 6){$\mathrm{R^3_{u0}}$}}
    \put(56, 39){\framebox(6, 6){$\mathrm{R^1_{u\vaccCounter}}$}}
    \put(71, 39){\framebox(6, 6){$\mathrm{R^2_{u\vaccCounter}}$}}
    \put(86, 39){\framebox(6, 6){$\mathrm{R^3_{u\vaccCounter}}$}}

    \put(15, 58.5){\vector(3, 1.2){20}}

    \put(15, 57.5){\vector(1, -1.1){20}}

    \put(15, 41.5){\vector(3, -1.2){20}}

    \put(15, 42.5){\vector(1, 1.1){20}}

    \put(43, 66){\vector(2, -1){12}}

    \multiput(43, 64)(1.25, -1.6){12}{\line(1, -1.3){0.7}}
    \put(57, 46){\vector(1, -1.3){0.2}}

    \put(43, 36){\vector(1.7, 2){15}}

    \multiput(60, 54)(0, -2){4}{\line(0, -1){1}}
    \put(60, 46){\vector(0, -1){0.5}}

    \multiput(74, 54)(0, -2){4}{\line(0, -1){1}}
    \put(74, 46){\vector(0, -1){0.5}}

    \multiput(89, 54)(0, -2){4}{\line(0, -1){1}}
    \put(89, 46){\vector(0, -1){0.5}}

    \put(63, 58){\vector(1, 0){7}}
    \put(78, 58){\vector(1, 0){7}}

    \put(63, 42){\vector(1, 0){7}}
    \put(78, 42){\vector(1, 0){7}}

    \multiput(39, 61.5)(0, -1){8}{\line(0, -1){0.3}}
    \put(39,54){\vector(0, -1){0.5}}

    \multiput(39, 38.5)(0, 1){8}{\line(0, 1){0.3}}
    \put(39,46){\vector(0, 1){0.5}}

    \cbezier(89,62)(80, 85)(20, 85)(12,62)
    \put(12.2,62.5){\vector(-1, -3.5){0.3}}

    \cbezier(89,38)(80, 15)(20, 15)(12,38)
    \put(12.2,37.5){\vector(-1, 3.5){0.3}}

    \multiput(11, 54)(0, -2){4}{\line(0, -1){1}}
    \put(11, 46){\vector(0, -1){0.5}}

    \put(0, 58){\vector(1, 0){7}}
    \put(1, 59){$\mathrm{\mu_{\demography S_{u0}}}$}

    \put(10, 62){\vector(0, 1){7}}
    \put(8, 64){$\mathrm{\muDeath}$}
    \put(10, 38){\vector(0, -1){7}}
    \put(8, 34){$\mathrm{\muDeath}$}

    \put(39, 69.5){\vector(0, 1){7}}
    \put(37, 71.5){$\mathrm{\muDeath}$}

    \put(39, 31.5){\vector(0, -1){7}}
    \put(30, 27.5){$\mathrm{\muDeath + \choleraDeath}$}

    \put(59, 62){\vector(0, 1){7}}
    \put(57, 64){$\mathrm{\muDeath}$}

    \put(74, 62){\vector(0, 1){7}}
    \put(72, 64){$\mathrm{\muDeath}$}

    \put(59, 38){\vector(0, -1){7}}
    \put(57, 34){$\mathrm{\muDeath}$}

    \put(74, 38){\vector(0, -1){7}}
    \put(72, 34){$\mathrm{\muDeath}$}

    \put(93, 42){\vector(1, 0){7}}
    \put(95.5, 43){$\mathrm{\muDeath}$}

    \put(93, 58){\vector(1, 0){7}}
    \put(95.5, 59){$\mathrm{\muDeath}$}

    \multiput(43, 50)(1, 0){7}{\line(1, 0){0.2}}
    \put(50.2, 50){\vector(1, 0){0.5}}
    \put(44, 51){$\mathrm{\Wremoval}$}

    \end{picture}
  }
\end{center}
\caption{A flow diagram for the SAIR model 3. This model assumes a constant population while also including a mechanism for births/deaths; all deaths are balanced by births into the unvaccinated susceptible compartment, so the birth rate $\mu_{\demography S_{u0}}$ corresponds to the sum total deaths from the remaining compartments. The model assumes that symptomatic individuals will not be vaccinated, hence no vaccination arrow exiting the $I_u0$ compartment.}\label{fig:flow_diagram3}
\end{figure}

\section{Markov chain and differential equation interpretations of compartment flow rates}

In Sections 2.1, 2.2 and 2.3 of the main article, we define compartment models in terms of their flow rates.
For a discrete population model, these rates define a Markov chain.
For a continuous and deterministic model, the rates define a system of ordinary differential equations.
Here, we add additional details to clarify the mapping from a collection of rate functions to a fully specified process.
Our treatment follows \citet{breto09}.

A general compartment model is a vector-valued process $X(t)=(X_1(t),\dots,X_c(t))$ denoting the (integer or real-valued) counts in each of $c$ compartments, where $t$ is any continuous value in the interval  $\left[t_0, \infty\right)$ for some real valued starting time $t_0$.
The compartments may also have names, but to set up general notation we simply refer to them by their numerical index.
The basic characteristic of a compartment model is that $X(t)$ can be written in terms of the flows $N_{ij}(t)$ from $i$ to $j$.
A flow into compartment $i$ from outside the system is denoted by $N_{\demography i}$, and a flow out of the system from compartment $i$ is denoted by $N_{i\demography}$.
We call $\demography$ a source/sink compartment, though it is an irregular compartment since $X_{\demography}(t)$ is not defined.
These flows are required to satisfy a ``conservation of mass'' identity:
\begin{equation}
X_i(t)=X_i(t_0)+N_{\demography i}(t) - N_{i\demography}(t) + \sum_{j\neq
i}N_{ji}(t)-\sum_{j\neq i}N_{ij}(t). \label{eq:conservation}
\end{equation}
Each {\em flow} $N_{ij}(t)$ is associated with a {\em rate} function
$\mu_{ij}=\mu_{ij}(t,X(t))$, where we include the possibility that $i$ or $j$ takes value $\demography$.

There are different ways to use a collection of rate functions to build a fully specified model.
We proceed to describe the ones we use in this paper: via a system of ordinary differential equations (Sec.~\ref{subsec:ode}), a simple Markov counting system (Sec.~\ref{subsec:smcs}), and an over-dispersed Markov counting system (Sec.~\ref{subsec:odmcs}). Other representations include stochastic differential equations driven by Gaussian noise or Gamma noise \citep{bhadra11}.

\subsection{Ordinary differential equation (ODE) interpretation}
\label{subsec:ode}

A basic deterministic specification is
\begin{equation}
\label{eq:ode1}
dN_{ij}/dt = \mu_{ij}\big(t,X(t) \big) X_i(t), \hspace{3mm} i\in \seq{1}{c}, \hspace{3mm} j\in \seq{1}{c} \cup\{\demography\}, \hspace{3mm} i\neq j,
\end{equation}
where $\mu_{ij}\big(t,X(t)\big)$ is called a per-capita rate or a unit rate.
Flows into the system require special treatment since $X_i(t)$ in \myeqref{eq:ode1} is not defined for $i=\demography$.
Instead, we specify
\begin{equation}
\label{eq:ode2}
dN_{\demography i}/dt = \mu_{\demography i}\big(t,X(t) \big).
\end{equation}
This is the the interpretation and implementation used for Model~2 in our study.

\subsection{Simple Markov counting system interpretation}
\label{subsec:smcs}
A continuous time Markov chain can be specified via its infinitesimal transition probabilities.
A basic approach to this is to define
\begin{eqnarray}
\label{eq:smcs1}
\prob\big[ N_{ij}(t+\delta)-N_{ij}(t)=0 \given X(t)\big]
 &=& 1-\delta \mu_{ij}\big(t,X(t)\big)X_i(t) + o(\delta),
\\
\label{eq:smcs2}
\prob\big[ N_{ij}(t+\delta)-N_{ij}(t)=1 \given X(t)\big]
 &=& \delta \mu_{ij}\big(t,X(t)\big)X_i(t) + o(\delta),
\end{eqnarray}
for $i\in \seq{1}{c}$ and $j\in\seq{1}{c}\cup\{\demography\}$ with $i\neq j$.
As with the ODE case, we need special attention for flows into the system, and we define
\begin{eqnarray}
\label{eq:smcs3}
\prob\big[ N_{\demography i}(t+\delta)-N_{\demography i}(t)=0 \given X(t)\big]
 &=& 1-\delta \mu_{\demography i}\big(t,X(t)\big) + o(\delta),
\\
\label{eq:smcs4}
\prob\big[ N_{\demography i}(t+\delta)-N_{\demography i}(t)=1 \given X(t)\big]
 &=& \delta \mu_{\demography i}\big(t,X(t)\big) + o(\delta).
\end{eqnarray}
Together with the initial conditions $X(0)$, equations \myeqref{eq:smcs1}--\myeqref{eq:smcs4} define a Markov chain.
Each flow is a simple counting process, meaning a non-decreasing integer-valued process that only has jumps of size one.
We therefore call the Markov chain a simple Markov counting system (SMCS).
The infinitesimal mean of every flow is equal to its infinitesimal variance \citep{breto11} and so an SMCS is called equidispersed.
We note that the special case of Model~1 used by \cite{lee20} (with $\sigmaProc = 0$) is an SMCS.
To permit more general mean-variance relationships for a Markov counting system, we must permit jumps of size greater than one.
The utility of over-dispersed models, where the infinitesimal variance of the flow exceeds the infinitesimal mean, has become widely recognized \citep{stocks20,he10}.

\subsection{Overdispersed Markov counting system interpretation}
\label{subsec:odmcs}

Including white noise in the rate function enables the possibility of an over-dispersed Markov counting system \citep{breto11,breto09,he10}.
Since rates should be non-negative, Gaussian noise is not appropriate and gamma noise is a convenient option that has found various applications \citep{romero-severson15, subramanian20}.
Specifically, we consider a model given by
\begin{equation}
\label{eq:odmcs1}
\mu_{ij}\big(t,X(t)\big) = \bar\mu_{ij}\big(t,X(t)\big) \, d\Gamma_{ij}(t)/dt,
\end{equation}
where $\Gamma_{ij}(t)$ is a stochastic process having independent gamma distributed increments, with
\begin{equation}
\label{eq:odmcs2}
\E\big[\Gamma_{ij}(t)\big] = t, \quad \var\big[\Gamma_{ij}(t)\big] = \sigma_{ij}^2 t.
\end{equation}
Formally interpreting the meaning of \myeqref{eq:odmcs1} is not trivial, and we do so by constructing a Markov process $X(t)$ as the limit of the Euler scheme described in Section~\ref{sec:numerics}, below.
Therefore, the numerical scheme in Sec.~\ref{sec:numerics} can be taken as a definition of the meaning of \myeqref{eq:odmcs1}.
The Markov chain defined by the limit of this Euler scheme as the step size decreases is an over-dispersed Markov counting system, with the possibility of instantaneous jumps of size greater than one \citep{breto11}.

\section{Numerical solutions to compartment models}
\label{sec:numerics}

Models may be fitted and their implications assessed via numerical solutions (i.e., simulations) from the model equations.
All the analyses we consider have this simulation-based property, known as plug-and-play or equation-free or likelihood-free.
The numerical solutions to the model are arguably of more direct scientific interest than the exact solutions to the postulated equations.
For ODE models, numerical methods are well studied and a standard numerical solution package such as \code{deSolve} in \code{R} is adequate for many purposes.
For SMCS and ODMCS models, exact schemes are feasible when the number of events is small, which may be the case for small populations.
However, for applicability to larger populations, we use instead the following Euler scheme.
Write $\delta$ for an Euler time step, and $\Delta N_{ij}$ for the numerical approximation to $N_{ij}(t+\delta)-N_{ij}(t)$ given $X(t)$.
For each $i$ and $j$ in $\seq{1}{c} \cup \{\demography\}$ with $i \neq j$, we draw independent Gamma distributed noise increments with mean $\delta$ and variance $\sigma_{ij}^2 \delta$, denoted using a mean-variance parameterization of the gamma distribution as
\begin{equation}
\label{eq:numerics1}
\Delta\Gamma_{ij} \sim \mathrm{gamma}(\delta, \sigma_{ij}^2 \delta).
\end{equation}
In the case of an SMCS model, $\sigma_{ij}=0$ for all $i$ and $j$, so we have $\Delta\Gamma_{ij}=\delta$.
Then, for $i\neq \demography$ and $j\neq i$, and writing
\begin{equation}
\label{eq:numerics2}
\mu_{ij}=\bar\mu_{ij}\big(t,X(t)\big) \Delta\Gamma_{ij} / \delta,
\end{equation}
we calculate transition probabilities
\begin{eqnarray}
\label{eq:numerics3}
p_{ij} &=& \exp\left\{-\sum_{k\in 1:c \, \cup \{\demography\}} \mu_{ik} \, \delta \right\}
\frac{\mu_{ij}}{\sum_{k\in 1:c\cup \{\demography\}} \mu_{ik}},
\\
\label{eq:numerics4}
p_{ii} &=& 1 - \sum_{j\neq i} p_{ij}.
\end{eqnarray}
These probabilities correspond to competing hazards for every individual in compartment $i$ to transition to some compartment $j$, interpreting $j=i$ to mean that the individual remains in $i$.
Then, $\big(\Delta N_{i1},\dots,\Delta N_{ic},\Delta N_{i\demography}\big)$ has the multinomial distribution where $X_i(t)$ individuals are allocated independently to $\seq{1}{c}\cup\{\demography\}$ with probabilities given by \eqref{eq:numerics3} and \eqref{eq:numerics4}.
We use the \code{reulermultinom} function in the \code{pomp} package to draw from this multinomial distribution.

Different treatments of demographic flows---such as birth, death, immigration and emigration---are possible.
For the case $i=\demography$, the treatment used by Model~1 is to set
\begin{equation}
\label{eq:numerics6}
\Delta N_{\demography j} \sim \mathrm{poisson}( \mu_{\demography j} \delta),
\end{equation}
an independent Poisson random variable with mean $\mu_{\demography j} \delta$.

Models~2 and 3 used an alternative approach, balancing the total number of flows in and out of the compartment, i.e., $\sum_{i}N_{\demography i}(t) = \sum_{i}N_{i \demography}(t)$, in order to make the model consistent with the known total population.
In this case, we formally model the death rate as a rate of returning to the susceptible class $S$, and use external transitions from $\demography$ into $S$ to describe only net population increase.

\section{Measurement Models}

Each POMP requires specification of a measurement model, which is a statistical description of how observations on the system are obtained.
In general, we used the same measurement models that were reported by \cite{lee20} unless specifically noted in the following subsections.

\subsection{Model~1}

In this model, the advantage afforded by vaccination is an increased probability that an infection is asymptomatic.
Therefore, under the assumptions of this model, all reported cases are assumed to be a fraction of individuals that transition from the exposed to the infected compartment, as noted in Eq.~\myeqref{model1:meas}:
\begin{equation}
  \label{model1:meas}
  Y_{n} \mid \Delta N_{E_{\cdot}I_{\cdot}}(n) = \reportChange \sim \text{NB}\left(\reportRate \reportChange, \obsOverdispersion \right),
\end{equation}
where $Y_n$ is the reported cholera cases at time $t_n \in t_1:t_N$ and $\Delta N_{E_{\cdot}I_{\cdot}}(n)$ is the sum total of individuals across each vaccination compartment $\vaccCounter \in 1:\vaccClass$ who moved from compartment $E_\vaccCounter$ to $I_\vaccCounter$ since observation $t_{n-1}$.
Here, $\text{NB}\left(\reportRate \reportChange, \obsOverdispersion \right)$ denotes a negative binomial distribution with mean $\reportRate \reportChange$ and variance $\reportRate \reportChange \big(1 + \frac{\reportRate \reportChange}{\obsOverdispersion}\big)$.

\subsection{Model~2}\label{sec:mod2Meas}

Model~2 was fit using reported case counts that were transformed using the natural logarithm.
We fit Model~2 using the subplex algorithm in the \code{subplex} package, using
a Gaussian measurement model (Eq.~\myeqref{model2:meas}) on the log transformed cases within each unit.
The final loss function that is maximized is the product of the likelihoods of the individual units, or the sum of the log-likelihood of the individual units.
The measurement model for individual units is given in Eq.~\myeqref{model2:meas}.
\begin{equation}
  \label{model2:meas}
  \log\big(Y_{u, n} + 1\big) \mid \Delta N_{E_{u\cdot}I_{u\cdot}}(n) = \reportChange_u \sim \text{N}\left(\log \big(\reportRate \reportChange_u + 1\big), \obsOverdispersion^2 \right),
\end{equation}
where $\Delta N_{E_{u\cdot}I_{u\cdot}}(n)$ is the sum total of individuals across vaccination compartment $\vaccCounter \in 0:4$ within unit $u$ who moved from compartment $E_{u\vaccCounter}$ to $I_{u\vaccCounter}$ since observation $t_{n-1}$.
Therefore, because the natural logarithm of observed case counts (plus one, to avoid taking the logarithm of zero) has a normal distribution, $Y_{u, n} + 1$ is assumed to follow a log-normal distribution with log-mean parameter $\log\big(\reportRate \Delta N_{E_{u\cdot}I_{u\cdot}}(n) + 1\big)$ and log-variance $\obsOverdispersion^2$.
We note that fitting a model with this measurement model is equivalent to fitting using least squares, with $\log(Y_{u, n} + 1)$ as the response variable.

This measurement model differs from that used by \citet{lee20}, who fit the model in two stages: epidemic and endemic phases.
Although their text and supplement material do not explicitly describe the measurement model used, inspection of the code provided with their submission suggests a change in measurement model between the epidemic and endemic phases.
The measurement model they used for the epidemic phase is
\begin{equation}
  \label{model2:measEpi}
  Y_{u, n} \mid \Delta N_{E_{u\cdot}I_{u\cdot}}(n) = \reportChange_u \sim \text{N}\left(\reportRate \reportChange_u, \obsOverdispersion^2 \right),
\end{equation}
The measurement model they used for the endemic phase modifies the epidemic model by counting both asymptomatically infected (A) and symptomatically infected (I) individuals in the case counts:
\begin{equation}
  \label{model2:measEnd}
  Y_{u, n} \mid \Delta N_{E_{u\cdot}I_{u\cdot}}(n) = \reportChange_{u1}, \Delta N_{E_{u\cdot}A_{u\cdot}}(n) = \reportChange_{u2} \sim \text{N}\left(\reportRate \big(\reportChange_{u1} + \reportChange_{u2}\big), \obsOverdispersion^2 \right),
\end{equation}
where the notation for $\Delta N_{E_{u\cdot}A_{u\cdot}}(n)$ is similar to $\Delta N_{E_{u\cdot}I_{u\cdot}}(n)$, described above.

\subsection{Model~3}

In this model, reported cholera cases are assumed to stem from individuals who develop symptoms and seek healthcare.
Therefore reported cases are assumed to come from an over-dispersed negative binomial model, given the increase in infected individuals:
\begin{equation}
  \label{model3:meas}
  Y_{u, n} \mid \Delta N_{S_{u \cdot}I_{u}}(t) = \reportChange_u \sim \text{NB}\left(\reportRate \reportChange_{u}, \obsOverdispersion \right),
\end{equation}
where $\Delta N_{S_{u \cdot}I_{u \vaccCounter}}(n)$ is the number of individuals who moved from compartment $S_{u \vaccCounter}$ to $I_{u}$ since observation $t_{n-1}$.

This measurement model is a minor change from that used by \cite{lee20}, which allowed for a change in the reporting rate on January 1st, 2018.
The fitted values of the reporting rate---before and after January 2018---were $0.97$ and $0.097$, respectively.
An instantaneous change from near perfect to almost non-existent reporting can be problematic, as it forces the model to explain the observed reduction in reported cases as a decrease in the reporting of cases, rather than a decrease in the prevalence of cholera.
This shift was justified by a ``change of the case definition that occurred on January 1st, 2018";
this claim was not cited, and we could find no evidence that such a drastic change in the reporting rate would be warranted.
We therefore do not allow a change in reporting rate when fitting Model~3.

\section{Initial Values}

To perform inference on POMP models, it is necessary to propose an initial probability density for the latent process $f_{X_0}(x_0;\theta)$, including the possibility that the initial values of the latent states are known, or are a non-random function of the unknown parameter vector, $\theta$.
This density is used to obtain initial values of the latent state when fitting and evaluating the model.
For each of the models considered in this analysis, the initial conditions are derived by enforcing the model dynamics on reported cholera cases.
It is also sometimes necessary to fit some initial value parameters in order to help determine initial values for weakly identifiable compartments.
In the following subsections, we mention initial value parameters that were fit for each model.

\subsection{Model~1}

For this model, the number of individuals in the Recovered and Asymptomatic compartments are set to zero, but the initial proportion of Infected and Exposed individuals is estimated as initial value parameters ($\Iinit$ and $\Einit$, respectively) using the IF2 algorithm, implemented as \texttt{mif2} in the \texttt{pomp} package.
Finally, the initial proportion of Susceptible individuals $S_{0, 0}$ is calculated as $S_{0, 0} = 1 - \Iinit - \Einit$.
This model for the initial values of the latent states matches that which was used by \citet{lee20}.

\subsection{Model~2}

Model~2 assumes that the initial values are a deterministic function of the reporting rate and the initial case reports, and so no initial value parameters need to be estimated.
Initial values for latent state compartments are chosen so as to enforce the model dynamics on the observed number of cases.
Specifically, the model sets $I_{u0}(0) = y^*_{1u} / \reportRate$ for each unit $u \in 1:10$, where $I_{u0}(t)$ is the number of infected individuals in unit $u$ at time $t$, vaccination scenario $z = 0$, $y^*_{tu}$ is the reported number of cases, and $\reportRate$ is the reporting rate.
It is further assumed that there are no individuals in the recovered compartment, as the epidemic has just begun. 
This model for the initial values of the latent states matches that which was used by \citet{lee20}.

The decision to fix initial values so that they satisfy the dynamics of the model has the benefit of reducing the number of estimated parameters and enforcing latent states at time $t_0$ to be consistent with the calibrated model. 
The risk of fixing initial values rather than estimating them is doing so may have substantial effects on the model dynamics, and hence on the consequences of the analysis. 
To consider the impact of the chosen model approach, we consider an alternative initialization model that enables flexible estimation of certain latent states. Specifically, we initialize $I_{u0}(0) = \tilde{I}_{u0}$ for $u \in 1:10$, where $\{\tilde{I}_{u0}\}_{u = 1}^{10}$ is a set of additional model parameters. 
We then fix $S_{u0}(0) = \text{pop}_u - I_{u0}(0)$, and all other starting values are set to zero, as with the fixed value approach.

\begin{figure}[!ht]
\begin{knitrout}
\definecolor{shadecolor}{rgb}{0.969, 0.969, 0.969}\color{fgcolor}

{\centering \includegraphics[width=\maxwidth]{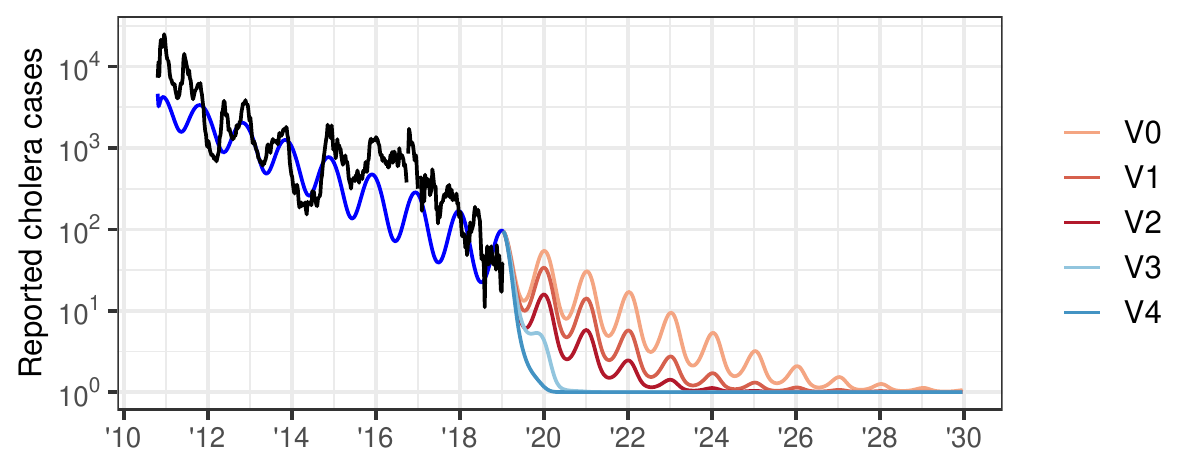} 

}

\end{knitrout}
\caption{\label{fig:h2InitialTraj}
{\figTitle Simulated trajectory of alternate initialization of Model~2.}
The black line shows the nationally aggregated weekly cholera incidence data.
The blue curve from 2012-2019 is the trajectory of the calibrated version of Model~2.
Compare to Fig.~4 of the article.}
\end{figure}

The AIC of this alternative approach is $43854.0$, compared to the fixed approach with an AIC of $43926.5$. 
This alternative initialization approach results in quantitative improvement to the model-fit, but does not result in qualitative differences in the conclusions made using this model. Figure~\ref{fig:h2InitialTraj} displays the trajectory of the model with this alternate initialization model. 
Table~\ref{tab:InitValues} gives the estimated initial values. 
The estimated value of the latent state is similar to the fixed value in Artibonite and Centre, where the largest number of cases are present at the start of the epidemic. 
Because of this, the qualitative dynamics do not differ by much when the parameters are estimated versus held constant, despite the improvement in model fit measured by AIC.

\begin{table}[!h]
\centering
\caption{\label{tab:InitValues}Initial values estimated for alternative initialization model for Model~2 compared to the fixed-value initialization model.}
\vspace{2mm}
\begin{tabular}{|c|c|c|c|}
\hline
Latent State & Department & Calibrated Model Value & Fixed Value
\\
\hline
\hline

$I_{1,0}(0)$ & Artibonite &
  $27125$ &
  $31170$ 
\\
\hline

$I_{2,0}(0)$ & Centre &
  $2095$ &
  $2535$ 
\\
\hline

$I_{3,0}(0)$ & Grand'Anse &
  $64$ &
  $0$ 
\\
\hline

$I_{4,0}(0)$ & Nippes &
  $35$ &
  $0$ 
\\
\hline

$I_{5,0}(0)$ & Nord &
  $0$ &
  $85$ 
\\
\hline

$I_{6,0}(0)$ & Nord-Est &
  $0$ &
  $0$ 
\\
\hline

$I_{7,0}(0)$ & Nord-Ouest &
  $3233$ &
  $10$ 
\\
\hline

$I_{8,0}(0)$ & Ouest &
  $878$ &
  $2700$ 
\\
\hline

$I_{9,0}(0)$ & Sud &
  $50$ &
  $0$ 
\\
\hline

$I_{10,0}(0)$ & Sud-Est &
  $0$ &
  $0$ 
\\
\hline

\end{tabular}
\end{table}

\subsection{Model~3}

The latent states of this model are initialized by enforcing the model dynamics on the incidence data from the start of the recorded cases until time $t_0$, requiring that some of the available data be used to determine the initial values of the latent states.
This is the same approach that was taken by \citet{lee20}, who used the value $t_0 = \text{2014-02-22}$; this choice of $t_0$ results in modeling roughly only $60\%$ of the available data, some of which is later discarded for alternative reasons \citep{lee20sup}.

We do not see any immediate reason that this model could not be extended to cover a larger range of the data, and chose the value $t_0 = \text{2010-11-13}$.
This choice of $t_0$ corresponds to using approximately $1\%$ of the available data to determine initial values of the latent states.
In addition to modeling a larger portion of the available data, this choice of $t_0$ corresponds to an important real-world event, as daily reports from each of the departments were not being sent to the health ministry until November 10, 2010 \citep{barzilay13};
this choice of $t_0$ therefore makes $\bm{Y}\big(t_1\big)$ the first week of data once daily reports are being sent to the health ministry.
The few observation times that exist before $t_0$ are used to initialize the model by enforcing model dynamics on these preliminary observations.
For convenience, we denote these observations as $t_{-3}, t_{-2}$ and $t_{-1}$; as before, we let $y_{u,-k} = Y_{u}\big(t_{-k}\big)$ denote the observed case count for unit $u$ at time point $t_{-k}$, where $k \in 1:3$.
Equations for the initial values of non-zero latent states are provided in Eqs.~\myeqref{eq:model3:I0}--\myeqref{eq:model3:B0}; these equations match those that were used by \citet{lee20}, the primary change being a change to the value of $t_0$.

\begin{eqnarray}
\label{eq:model3:I0}
I_{u}\big(t_0\big) &=& \frac{y^*_{u, -1}}{7\reportRate\left(\muIR + \left(\muDeath + \choleraDeath\right)/365\right)},
\\
\label{eq:model3:A0}
A_{u}\big(t_0\big) &=& \frac{I_{u0}\big(t_0\big)\left(1 - \symptomFrac\right)}{\symptomFrac},
\\
\label{eq:model3:R0}
R_{u01}\big(t_0\big) &=& R_{u02}\big(t_0\big) = R_{u03}\big(t_0\big) = \left(\frac{\sum_{k = -3}^0 y^*_{u, k}}{\reportRate \symptomFrac} - \left(I_{u0}\big(t_0\big) + A_{u0}\big(t_0\big)\right)\right) / 3
\\
\label{eq:model3:S0}
S_{u0}\big(t_0\big) &=& \mathrm{Pop}_u - I_{u}\big(t_0\big) - A_{u}\big(t_0\big) - \sum_{k=1}^3 R_{u0k}\big(t_0\big)
\\
\label{eq:model3:B0}
B_{u}\big(t_0\big) &=& \big[1 + \seasAmplitude \tilde{J}^r \big] \mathrm{Den}_u \, \Wshed \big[ I_{u}(t_0)+ \asymptomRelativeShed A_{u_0}(t) \big]/\Wshed.
\end{eqnarray}

In Eq.~\myeqref{eq:model3:B0}, $\tilde{J} = 0.002376$ is the median adjusted rainfall over the observation period.
One important consideration to make with this parameter initialization model is when $y_{u, -1}^* = 0$, which occurs for units $u \in \{3, 4\}$, which correspond to the Grand'Anse and Nippes departments, respectively.
When this is the case, each of the infectious $I_u(t_0)$, asymptomatic $A_u(t_0)$, and bacterial reservoir $W_u(t_0)$ compartments have a value of zero.
Because Model~3 models cholera transmission primarily by means of the bacterial reservoir, this makes it nearly impossible for an outbreak to occur.
Therefore for units $u \in \{3, 4\}$, we introduce initial value parameters $I_{0,0}^3$ and $I_{0,0}^4$, and calibrate these parameter values using the data.
The resulting parameter estimates are used to obtain the remaining non-zero initial values of the latent states using Eqs.~\myeqref{eq:model3:A0}--\myeqref{eq:model3:B0}.

\section{Calibrating Model~3 to observed cases}\label{sec:mod3Cal}

In this section, we provide more detail on the process that was used to estimate the coefficients of Model~3.
In particular, we discuss why we decided to include additional model parameters---those that are associated with the behavior of the system during Hurricane Matthew---that were not considered by \cite{lee20}.
To calibrate this model, we used the iterated block particle filter (IBPF) method of \cite{ionides22}.
Due to the novelty of this algorithm, there exists only a few examples where the IBPF algorithm is used for data analysis \citep{li23,ionides22}, which is one motivation of the inclusion additional details related to fitting and diagnosing the model fit provided here.

\cite{lee20} only estimated model parameters to a simplified version of Model~3 on a subset of the available data, as no method existed at the time of their publication to fit a fully coupled meta-population model to disease incidence data.
Building on their results, we fit the fully coupled version of Model~3 to (nearly) all available data, reserving only a few observations to use to calibrate the initial conditions of the model (see the supplement for initialization models for more details).
To maximize model likelihoods, we relied on parameter estimates obtained while calculating profile-likelihood confidence intervals, as this calculation requires many replicated IBPF searches.
In our preliminary investigations that were done prior to conducting a profile likelihood search, we found that it was necessary to use multiple searches for the MLE, periodically pruning away less successful searches.
To do this, the first collection of searches is performed by obtaining initial values for the parameters by uniformly sampling values from a predefined hypercube.
A subsequent refinement search used parameter values corresponding to the largest model likelihoods as starting parameter values.
The need for multiple searches does not appear to be uncommon, as a similar approach was used by \cite{ionides22}.
While computationally intensive, profile likelihoods proved to be an effective alternative to maximizing model likelihoods without the need to apply this multistage heuristic. 

\begin{figure}[!ht]
\begin{knitrout}
\definecolor{shadecolor}{rgb}{0.969, 0.969, 0.969}\color{fgcolor}

{\centering \includegraphics[width=\maxwidth]{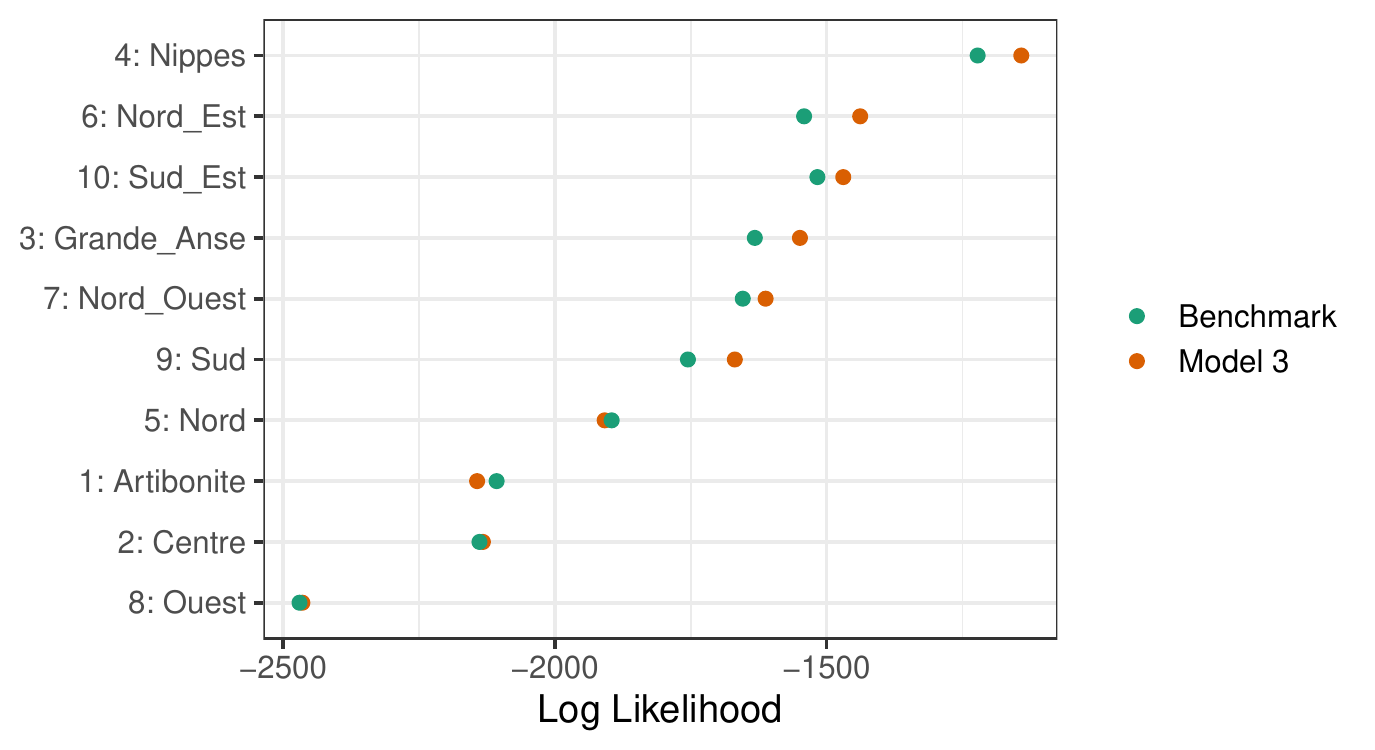} 

}

\end{knitrout}
\caption{\label{fig:h3UnitLikes}Log-likelihoods of Model~3 for each department compared to the corresponding benchmark model prior to the inclusion of parameters that account for Hurricane Matthew.}
\end{figure}

We use the iterative fitting / pruning technique described above to fit the fully coupled version of Model~3 proposed by \cite{lee20}.
The maximum likelihood we obtained after two rounds of searching was $-17549$, which is higher than the benchmark model ($-17933$).
While beating a simple associative benchmark is promising, this does not immediately imply that the model is a good description of the system.
Additional investigation of parameters estimates and their corresponding implications on model based conclusions should always be conducted.
For meta-population models, it is worth considering how well the model fits the data to each spatial unit.
The likelihoods for each department, compared to the corresponding benchmark model, are displayed in Fig.~\ref{fig:h3UnitLikes}.
The figure demonstrates that while the log-likelihood of the fitted model outperforms the auto-regressive negative binomial benchmark model at the aggregate level, Model~3 has lower likelihoods for some departments.

\begin{figure}[!ht]
\begin{knitrout}
\definecolor{shadecolor}{rgb}{0.969, 0.969, 0.969}\color{fgcolor}

{\centering \includegraphics[width=\maxwidth]{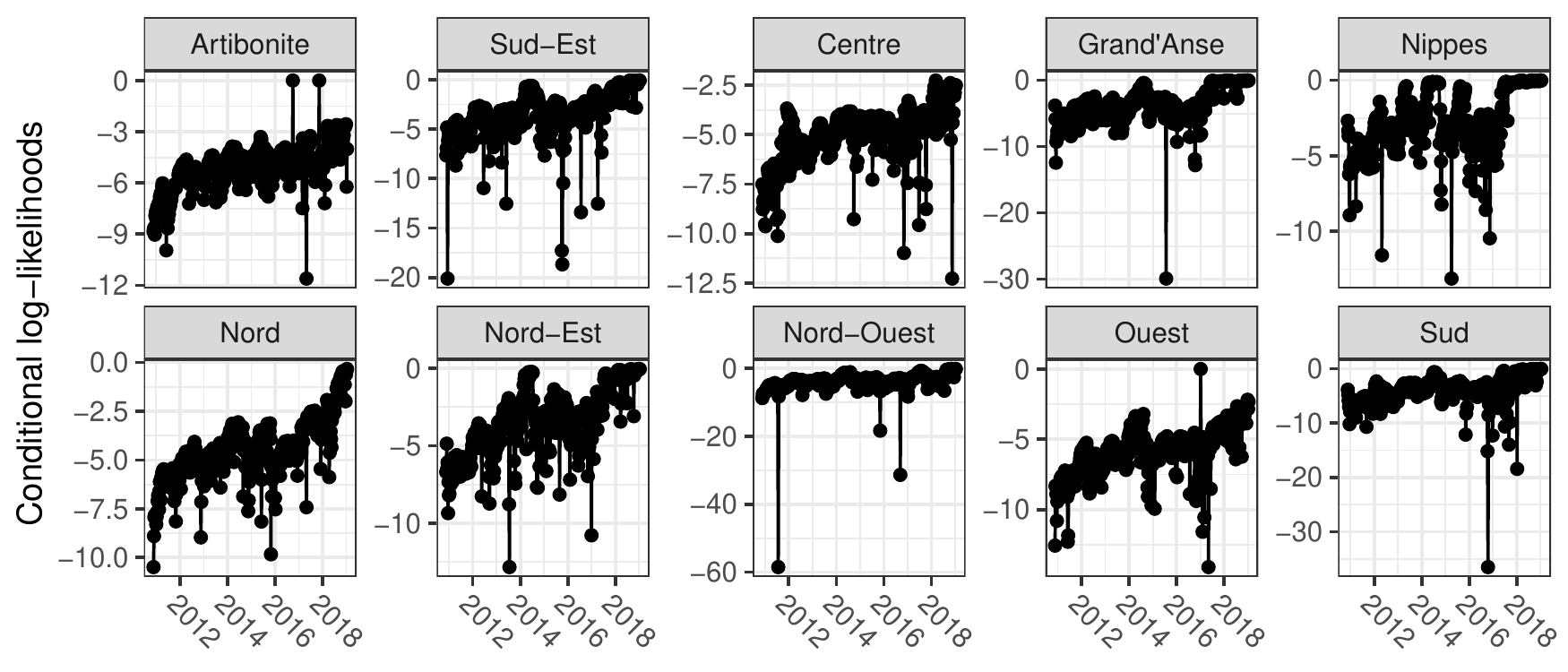} 

}

\end{knitrout}
\caption{\label{fig:condLL}Conditional log-likelihoods of Model~3 prior to the inclusion of the Hurricane Matthew related parameters.}
\end{figure}

In addition to considering the conditional log-likelihoods of each unit, one can consider conditional log-likelihoods of each observation.
When compared to a benchmark, this level of detail can provide useful information about which observations are well described by the model and which are not.
In Fig.~\ref{fig:condLL}, we plot the conditional log-likelihoods of Model~3 for each observation.
Typically it is most useful to compare the conditional log-likelihoods of the model under consideration to a benchmark, as plotting only conditional log-likelihoods without a comparison may not be helpful.
In this case, however, the same insight can be drawn using a figure without a benchmark comparison, so we do not include the benchmark in order to avoid the issue of over-plotting.

Fig.~\ref{fig:condLL} reveals that the fitted model poorly describes certain features of the data.
For example, many departments (in particular Sud) have observations with lower conditional log-likelihoods near October 2016 than at other time points.
Further investigation of the model output reveals that the model is struggling to explain the sudden surge in cholera cases that occurred at this time, which coincides with the time that Hurricane Matthew struck Haiti.
While the model does include a mechanism to account for increased cholera transmission due to large rainfall events, the mechanism does not appear to be sufficient to capture the damaging effects of the hurricane, which had the greatest impact in the the Sud and Grand'Anse departments \citep{ferreirai16}.
This result led us to include parameters $\Whur{u}$ and $\hHur{u}$ in Eq.~23 of the main text, which allows for an increase in the transmission rate between environmental cholera and humans for in Sud and Grand'Anse during and after the hurricane.
The effect of the hurricane on cholera transmission is assumed to have an exponential decay, where the magnitude is determined by $\Whur{u}$ and the duration of the effect determined by $\hHur{u}$.

\begin{figure}[!ht]
\begin{knitrout}
\definecolor{shadecolor}{rgb}{0.969, 0.969, 0.969}\color{fgcolor}

{\centering \includegraphics[width=\maxwidth]{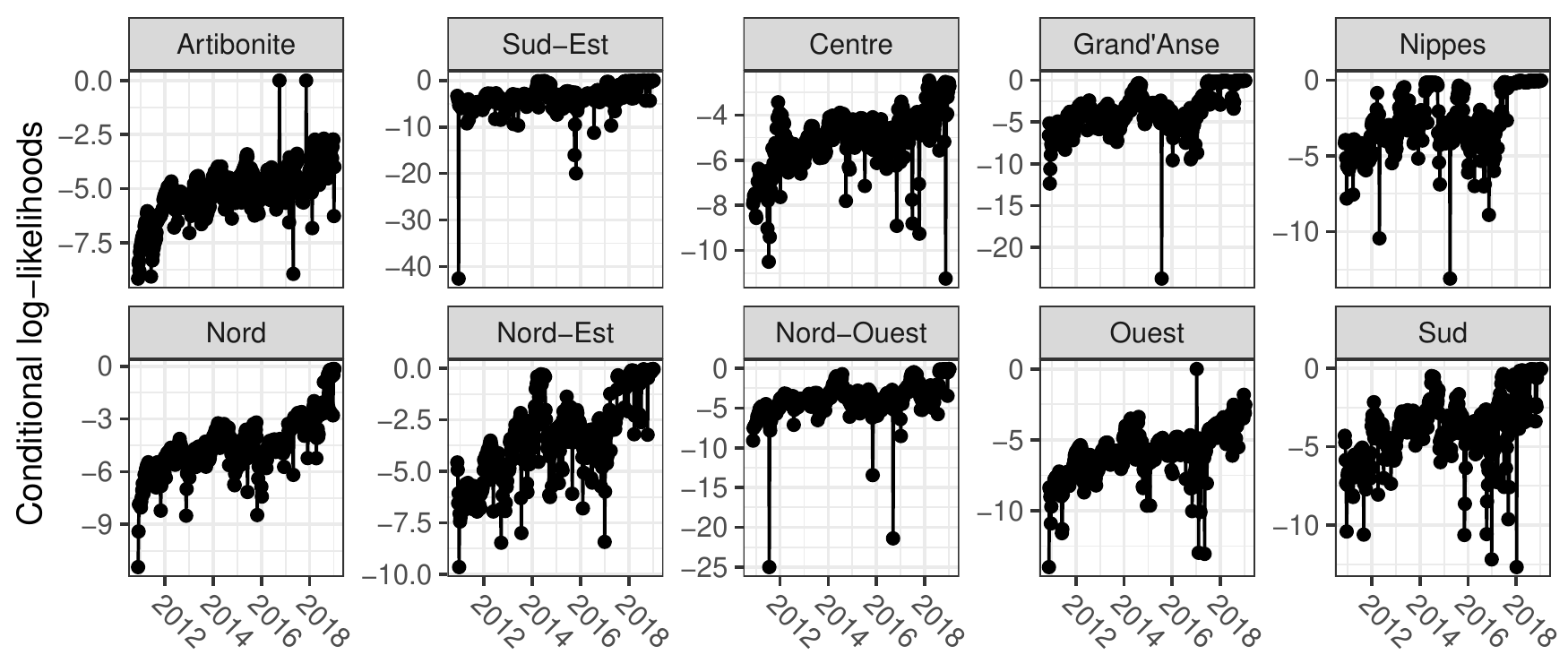} 

}

\end{knitrout}
\caption{\label{fig:finalCondLL}Conditional log-likelihoods of Model~3 after adding and estimating the parameters related to Hurricane Matthew.}
\end{figure}

We refit Model~3 after introducing these hurricane-related parameters.
The resulting model has a log-likelihood value of $-17332.9$.
The inclusion of these parameters resulted in an overall increase of $216.4$ log-likelihood units.
Such a large difference in log-likelihoods is well beyond the threshold of statistical uncertainty determined by Wilks' theorem, suggesting that the data highly favor the inclusion of the additional parameters.
The addition of the Hurricane parameters also increases in conditional likelihoods for each observation, particularly around October 2016 (Fig.~\ref{fig:finalCondLL}).

\begin{figure}[!ht]
\begin{knitrout}
\definecolor{shadecolor}{rgb}{0.969, 0.969, 0.969}\color{fgcolor}

{\centering \includegraphics[width=\maxwidth]{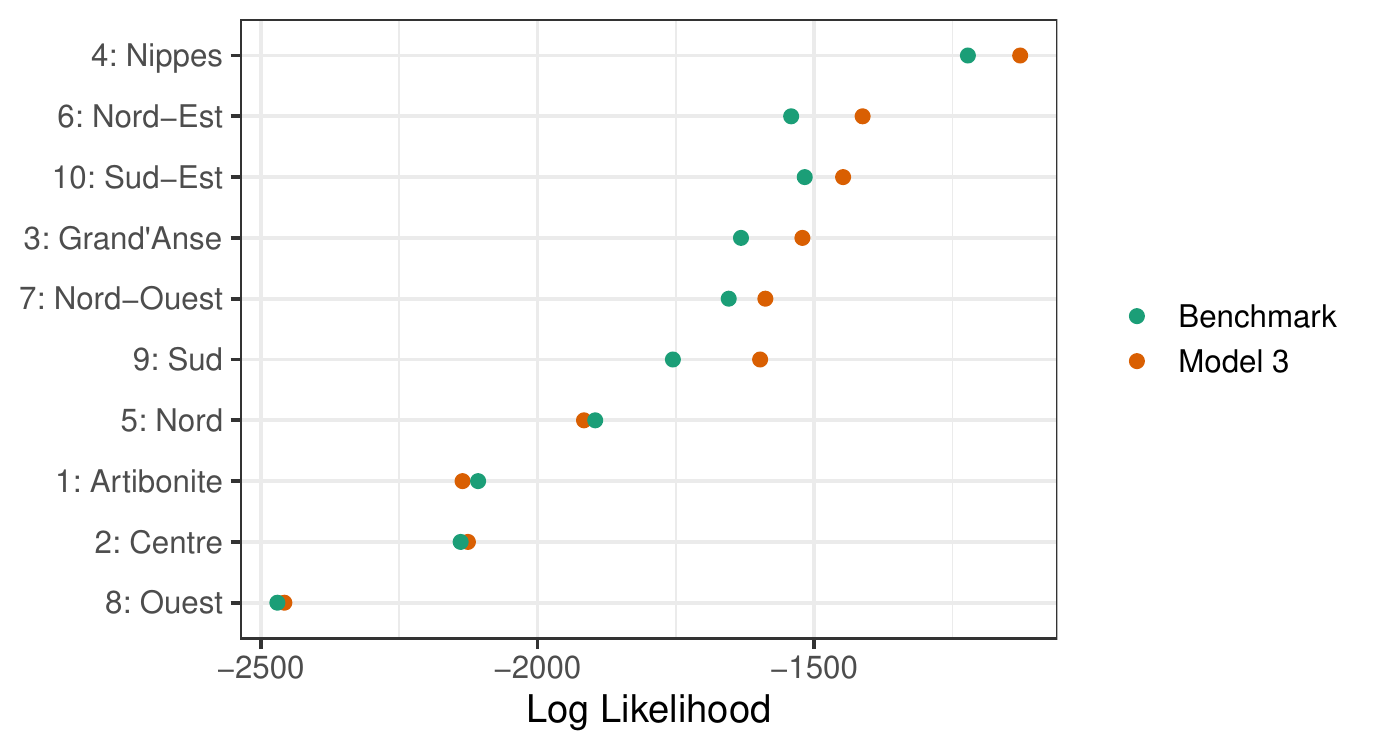} 

}

\end{knitrout}
\caption{\label{fig:finalUnitLL}Log-likelihoods of Model~3 for each department compared to the corresponding benchmark model after adding and estimating parameters related to Hurricane Matthew.}
\end{figure}

Now that the model with additional parameters has been calibrated to the incidence data, we plot the conditional log-likelihood of each department compared to a benchmark model in Fig.~\ref{fig:finalUnitLL}.
The difference in log-likelihoods between Model~3 and the benchmark model is smallest in the departments Artibonite, Nord, Ouest and Centre.
Each of these departments also exhibited the most sustained cholera transmission, defined by having the fewest number of weeks with no recorded cholera cases.
Specifically, these four departments have zero cholera cases recorded in less than $4\%$ of the available data, and all remaining departments---except for Nord-Ouest, which has approximately $9.5\%$ of cases that are zeros and also exhibits the next smallest difference in log-likelihoods---have zero cases recorded in at least $14\%$ of the available weekly data.
This result suggests that the quantitative advantage Model~3 has over its respective benchmark is primarily due to the model's ability to describe a resurgence of cases after a department records a week of zero cholera cases.
This result may be unsurprising in the context of the models that we are comparing.
Because the cholera transmission in individual departments likely depends on the national prevalence of cholera and the Vibrio cholerae bacteria, our spatially-independent benchmark model that relies exclusively on the previous number of case within any given unit has a difficult time predicting a resurgence of cases.

The difference in log-likelihoods between Model~3 and its benchmark model for each individual units suggests that Model~3 has a relatively poor fit for the four units with the most sustained cholera transmission.
The simple four parameter benchmark has a higher likelihood than Model~3 for the Artibonite and Nord departments, and also has log-likelihoods only a few units smaller than Model~3 for the departments Ouest and Centre.
This is particularly worrisome given that these four departments account for more than $77\%$ of the total number of reported cholera cases.

\subsection{Examining the Hidden States of the Calibrated Model}

For mechanistic models, beating a suitable statistical benchmark does not alone guarantee that the model provides an accurate description of a dynamic process.
Indeed, a good statistical fit does not require the model to assert a causal explanation.
For example, reconstructed latent variables should make sense in the context of alternative measurements of these variables \citep{grad12}.
We demonstrate this principle by examining the latent state of the calibrated model.
In particular, we examine the compartment of susceptible individuals under various scenarios.
This analysis can also provide insight into why the calibrated model fails to outperform the benchmark model on the four departments with the most sustained cholera transmission. 

\begin{figure}[!ht]
\begin{knitrout}
\definecolor{shadecolor}{rgb}{0.969, 0.969, 0.969}\color{fgcolor}

{\centering \includegraphics[width=\maxwidth]{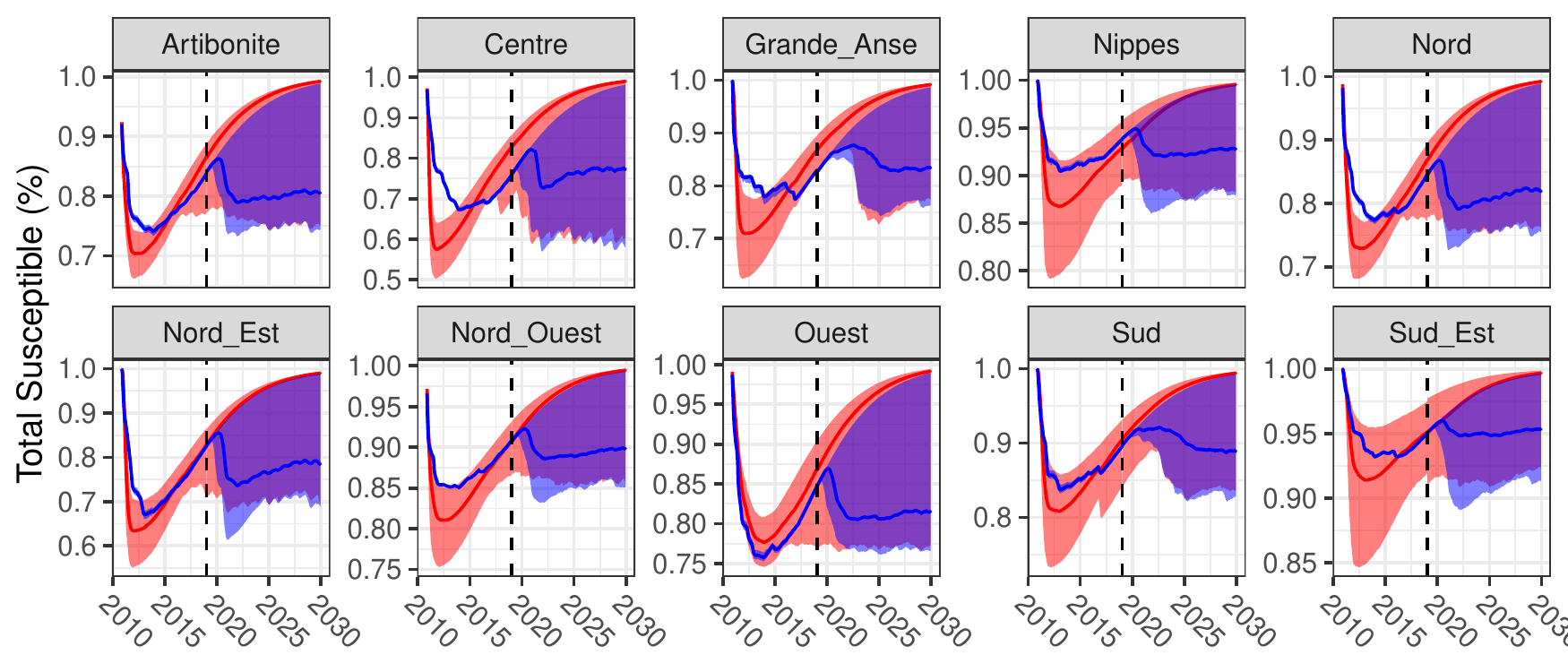} 

}

\end{knitrout}
\caption{\label{fig:h3Sus}Percentage of individuals that are in the susceptible compartment.
Simulations from Model~3 under initial conditions are displayed in red; simulations from the filtering distribution of model are displayed in blue.
The dashed line represents the end of the observed data.}
\end{figure}

Recall that the filtering distribution for the calibrated version of Model~3 at time $t_k$ is defined as the distribution of the latent state at time $t_k$ given the data from times $t_{1}:t_{k}$, i.e. $f^{(3)}_{\bm{X}_k|\bm{Y}_{1:k}}(\bm{x}_{k} | \bm{y}^*_{1:k} ; \hat\theta)$.
In general, one may expect simulations from the filtering distribution of a model with a good statistical fit to result in hidden states that are highly consistent with the observed data because the filtering distribution is conditioned on the observed data.
Fig.~\ref{fig:h3Sus} shows the percentage of individuals that are in the susceptible compartment from various simulations of the model:
simulations from Model~3 under initial conditions are displayed in red; simulations from the filtering distribution of model are displayed in blue.
This figure shows that simulations from initial conditions tends to result in a much more rapid depletion of susceptible individuals at the start of the epidemic than simulations from the filtering distribution, suggesting the calibrated model has a propensity to predict larger outbreaks than what is typically seen in the data.
This result demonstrates that the calibrated model favors a more rapid growth in cholera cases than what is typically seen in the observed data, providing a possible explanation as to why the model fails to beat the simple benchmark for each spatial unit. 
This results hints at the possibility of model mispecification, and warrants a degree of caution in interpreting the model's outputs.

\clearpage

\section{Replication of \cite{lee20}}

In this article we claimed that we were able to obtain better fits to the observed data using the same models that were proposed by \cite{lee20}.
Along with visual comparisons to the data, this claim was supported by comparing likelihoods and AIC values in Table~2 in the manuscript.
Because model likelihoods were not provided by \cite{lee20}, it is necessary to replicate these models in order to obtain likelihood estimates.
Here we would like to thank the authors of \cite{lee20}, who provided detailed descriptions of their models, which enabled us to build on their work.
In the following subsections, we use our \code{R} package \code{haitipkg} to reproduce some of the results of \cite{lee20}.
This reproduction allows us to estimate the likelihoods of the \cite{lee20} version of Models~1--3, and also provides a demonstration of the importance and usefulness of reproducible research.

\subsection{Model~1 Replication}

The model was implemented by a team at Johns Hopkins Bloomberg School of Public Health (hereafter referred to as the Model~1 authors) in the \code{R} programming language using the \code{pomp} package \citep{king16}.
Original source code is publicly available with DOI: 10.5281/zenodo.3360991.
The final results reported by the Model~1 authors were obtained by using several different parameter sets rather than a single point estimate.
According to the supplement materials, this was because model realizations from a single parameter set retained substantial variability, but multiple realizations from a collection of parameter sets resulted in a reasonable visual fit to the data.
We are also inclined to believe that the use of multiple parameter values was in part intended to account for parameter uncertainty---the importance of which was discussed in the main text---an effort by the Model~1 authors that we applaud.
Simulations from each of the parameter sets, however, were treated with equal importance when being used to diagnose the model fit and make inference on the system.
This is problematic given Figures S8 and S9 of the supplement material, which suggest that some parameter sets that were used for inference may have been several hundred units of log-likelihood lower than other parameter sets that were simultaneously used to make forecasts.
Such a large difference in log-likelihoods is well beyond the threshold of statistical uncertainty determined by Wilks' theorem, resulting in the equal use of statistically inferior parameter sets in order to make forecasts and conduct inference on the system.

To fully reproduce the results of the Model~1 authors, it is necessary to use the exact same set of model parameters that were originally used to obtain the results presented by \cite{lee20}.
Because these parameter sets were not made publicly available, we relied on the source code provided by the Model~1 authors to approximately recreate the parameter set.
Due to software updates since the publication of the source code, we were unable to produce the exact same set of parameters.
Running the publicly available source code, however, resulted in a set of parameters that are visually similar to those used by the Model~1 authors (See Figures~\ref{fig:PlotEpiDist} and \ref{fig:plotEndParams}).
Furthermore, simulations using the set of parameters produced by the source code appear practically equivalent to those displayed by \cite{lee20} (See Figure~\ref{fig:plotMod1Sims}).

\begin{knitrout}
\definecolor{shadecolor}{rgb}{0.969, 0.969, 0.969}\color{fgcolor}\begin{figure}

{\centering \includegraphics[width=\maxwidth]{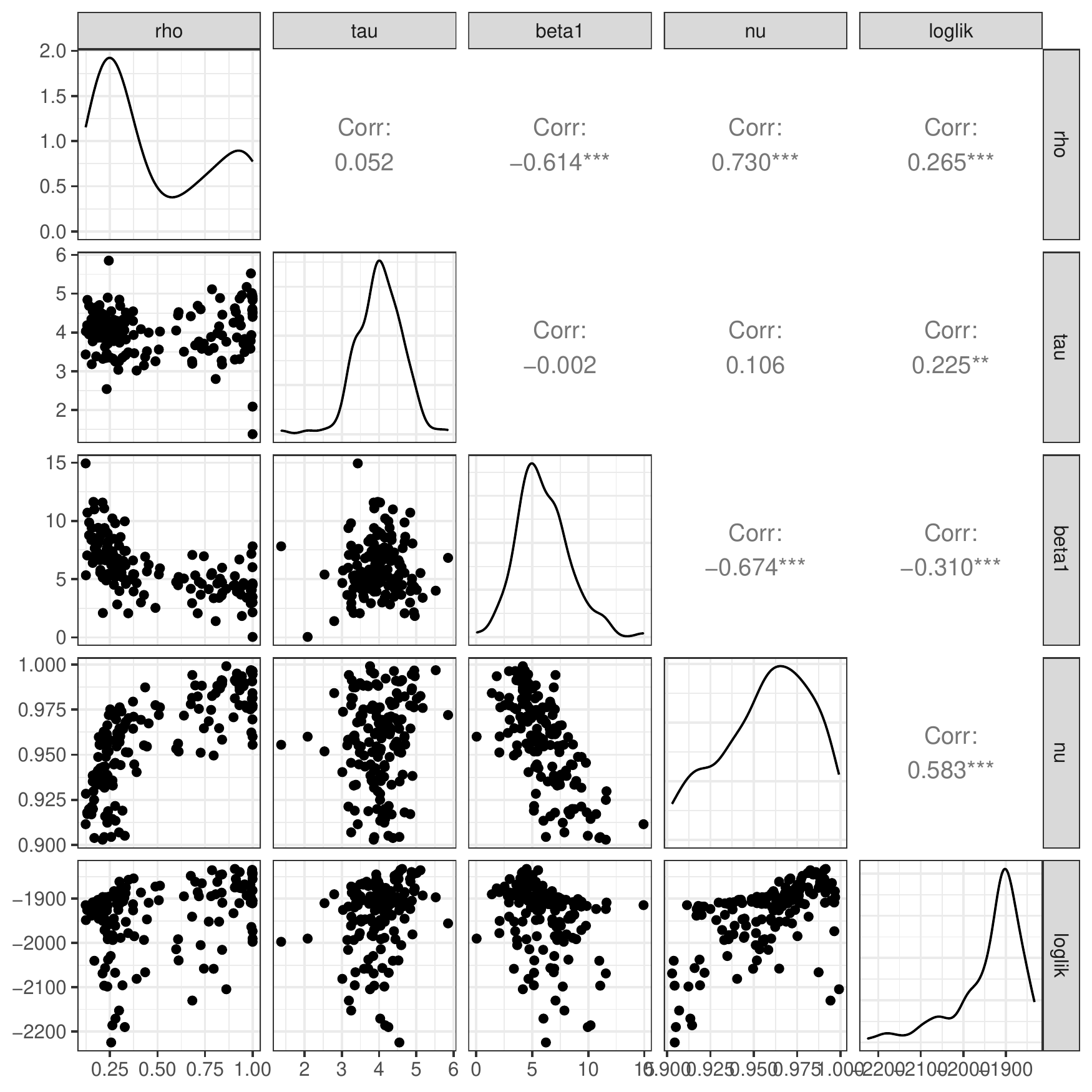} 

}

\caption[Bivariate distributions of parameter estimates after fitting epidemic phase of the Model~1 following the procedure described by Lee et al]{Bivariate distributions of parameter estimates after fitting epidemic phase of the Model~1 following the procedure described by Lee et al. (2020). Compare to Figure S8 in the supplement of Lee et al. (2020).}\label{fig:PlotEpiDist}
\end{figure}

\end{knitrout}

\begin{knitrout}
\definecolor{shadecolor}{rgb}{0.969, 0.969, 0.969}\color{fgcolor}\begin{figure}

{\centering \includegraphics[width=\maxwidth]{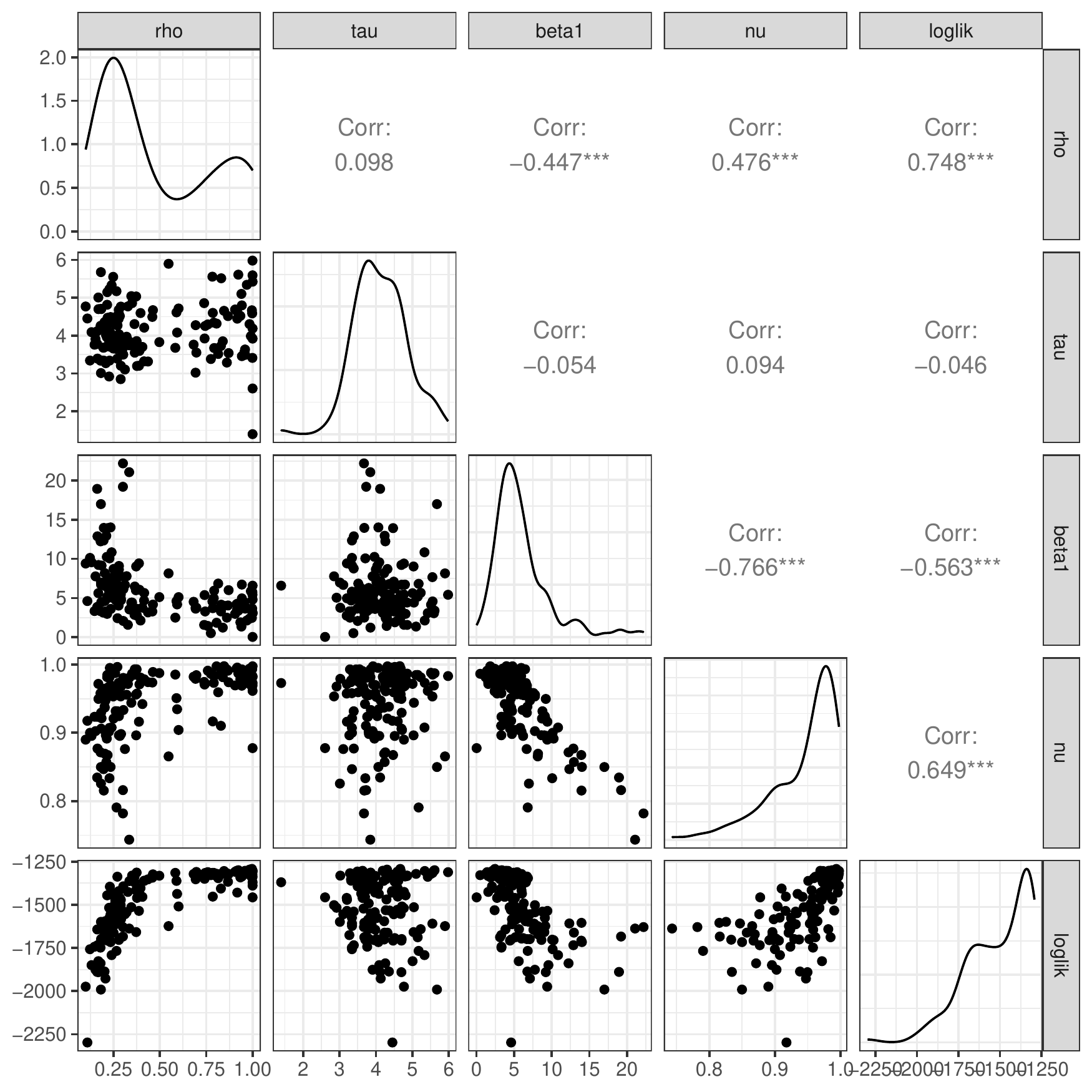} 

}

\caption[Bivariate distributions of parameter estimates after fitting endemic phase of the Model~1 following the procedure described by Lee et al]{Bivariate distributions of parameter estimates after fitting endemic phase of the Model~1 following the procedure described by Lee et al. (2020). Compare to Figure S9 in the supplement of Lee et al. (2020).}\label{fig:plotEndParams}
\end{figure}

\end{knitrout}

\begin{knitrout}
\definecolor{shadecolor}{rgb}{0.969, 0.969, 0.969}\color{fgcolor}\begin{figure}

{\centering \includegraphics[width=\maxwidth]{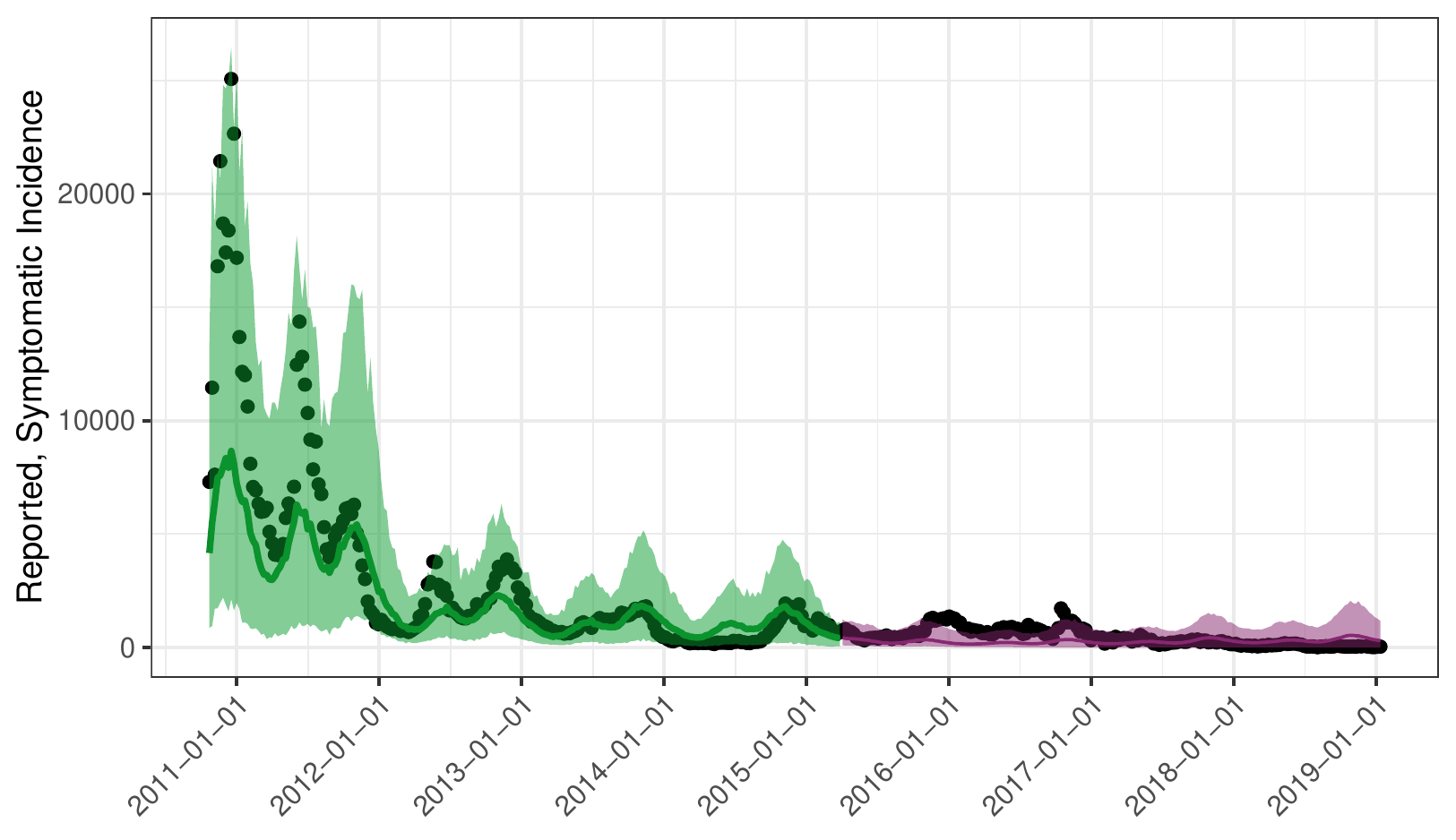} 

}

\caption[Simulations from Model~1 using parameter sets that were generated by running source code provided by Lee et al (2020)]{Simulations from Model~1 using parameter sets that were generated by running source code provided by Lee et al (2020). Compare to Figure S7 in the supplement of Lee et al. (2020). The upper bound for the likelihood of this model is -3031.}\label{fig:plotMod1Sims}
\end{figure}

\end{knitrout}

Because the model forecasts provided by \cite{lee20} come from various sets of parameters---which each correspond to a unique log-likelihood value---it is not obvious how one would obtain an estimate for the log-likelihood of the model that was used for simulations by the Model~1 authors.
One approach could be to calculate the logarithm of the weighted mean of the likelihoods for each parameter sets used to obtain the forecasts, where the weights are proportional to the number of times the parameter set was used.
However, in an effort to not underestimate the likelihood of the model of the Model~1 authors, we report the estimated log-likelihood as the log-likelihood value corresponding to the parameter set with the largest likelihood value, even though the majority of simulations were obtained using parameter sets with lower likelihood values.
In this sense, we consider the log-likelihood reported in Table~1 of the main text to be an upper-bound of the log-likelihood of the model used by \cite{lee20}.
For each parameter set, the log-likelihood was estimated using a particle filter, implemented as the \texttt{pfilter} function in the \texttt{pomp} package.

\subsection{Model~2 Replication}\label{sec:mod2rep}

Model~2 was developed by a team that consisted of members from the Fred Hutchinson Cancer Research Center and the University of Florida (hereafter referred to as the Model~2 authors).
While Model~2 is the only deterministic model we considered in our analysis, it contains perhaps the most complex descriptions of cholera in Haiti: Model~2 accounts for movement between spatial units; human-to-human and environment-to-human cholera infections; and transfer of water between spatial units based on elevation charts and river flows.

\begin{figure}[!h]
\begin{knitrout}
\definecolor{shadecolor}{rgb}{0.969, 0.969, 0.969}\color{fgcolor}

{\centering \includegraphics[width=\maxwidth]{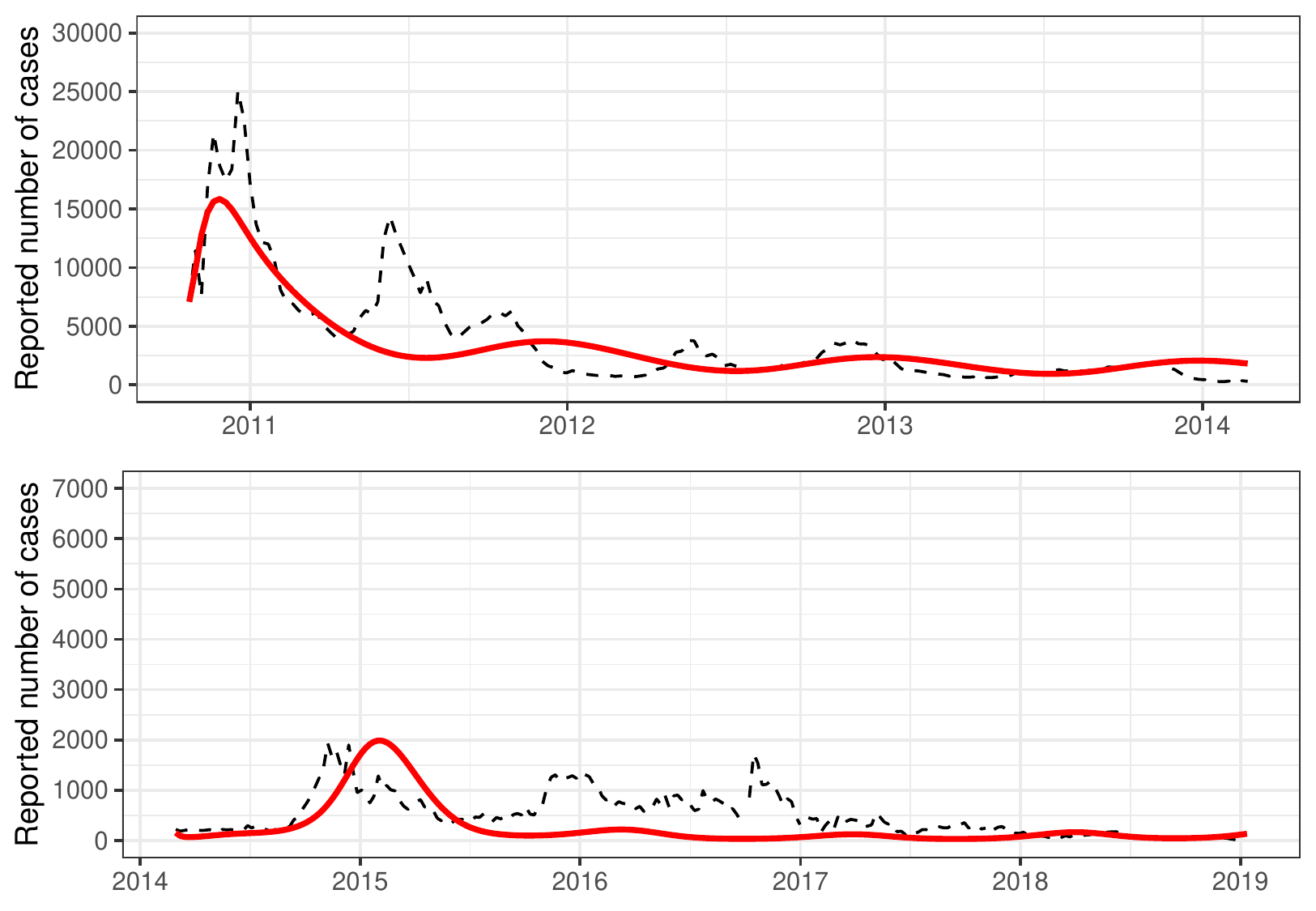} 

}

\end{knitrout}
\caption{\label{fig:mod2rep}
Model 2 trajectories using the \code{haitipkg}. Compare to Figure S15 in the supplement of \cite{lee20sup}.
}
\end{figure}

The source code that the Model~2 authors used to generate their results was written in the \code{Python} programming language, and is publicly available at \url{10.5281/zenodo.3360857} and its accompanying GitHub repository \url{https://github.com/lulelita/HaitiCholeraMultiModelingProject}.
In order to perform our analysis in a unified framework, we re-implemented this model in the \code{R} programming language using the \code{spatPomp} package \citep{asfaw23arxiv}, which facilitates the creation of meta-population models.
We note that the travel and water matrices used to implement the complex dynamics in Model~2 \citep{lee20sup} are not available in either the Zenodo archive or the GitHub repository;
instead, we obtained these matrices via personal correspondence with the Model~2 authors.
Using these matrices, and the point estimates for model parameters provided by \citep{lee20sup}, we created trajectories of the cholera dynamics and compared this to available data.
These trajectories, shown in Figure~\ref{fig:mod2rep}, are very similar to the trajectories shown in Figure~S15 of \cite{lee20sup}.

There are minor differences between Figure~\ref{fig:mod2rep} and Figure~S15 of \cite{lee20sup}.
While the discrepancy appears minor, the deterministic nature of Model~2 implies that an exact replication of model trajectories should be possible.
In this case, these discrepancies may possibly be attributed to implementing the model and plotting the model trajectory in two different programming languages.
Another potential explanation for the discrepancy is that the parameters that we used are only approximately the same as those used by \citet{lee20sup}.
For example, the parameters $\transmission$, $\Wbeta{}$ had reported values of $9.9 \times 10^{-7}$ and $4.03 \times 10^{-2}$, respectively (Table~S13 of \cite{lee20sup}), but were actually fit to data and therefore likely these values have been rounded.
Additionally, our implementation of Model~2 used a time scale of years and many of the parameters were reported on a weekly scale, so small differences may result due to unit conversions.
The collective effect of these small differences in model parameters likely will result in small differences in model trajectories.

Some additional concerns about being able to accurately replicate the results of \cite{lee20} are valid.
Details about the measurement models and how latent states were initialized for the epidemic model were not provided by \citet{lee20sup} and therefore these details must be inferred by looking at the provided source code.
According to repository comments, the files \code{fit\allowbreak In\allowbreak Pieces\allowbreak 3params\allowbreak Clean\allowbreak May2019\allowbreak Public.py} and \code{fit\allowbreak In\allowbreak Pieces\allowbreak Mu\allowbreak With\allowbreak Frac\allowbreak Sus\allowbreak Fixed\allowbreak All\allowbreak Infections\allowbreak Public.py} were used to fit the epidemic and endemic phases of the model respectively, although it is apparent that these exact files were not used to obtain the reported results since the files contain some variable-naming errors that make it impossible to run the files without making modifications \footnote{One example of why the code cannot be run that the file loads functions from a non-extant file named \code{choleraEqs.py} in line 13 rather than \code{cholera\allowbreak Eqs\allowbreak Public.py}.}.
The inability to replicate the results by \citet{lee20} by running the provided source code makes checking whether or not a our numeric implementation faithfully represents their results very difficult.
Additionally, the script that was said to been used to obtain the results reported by \cite{lee20} appears to use a different measurement model than what was described in the supplemental material, again making it difficult to fully replicate the result of \cite{lee20} without being able to easily run the provided source code.
In this case, we chose to use measurement model that considers only symptomatic individuals for both phases of the epidemic, as this seemed to visually match the results of \cite{lee20} most closely.

\subsection{Model~3 Replication}

\begin{figure}[!h]
\begin{knitrout}
\definecolor{shadecolor}{rgb}{0.969, 0.969, 0.969}\color{fgcolor}

{\centering \includegraphics[width=\maxwidth]{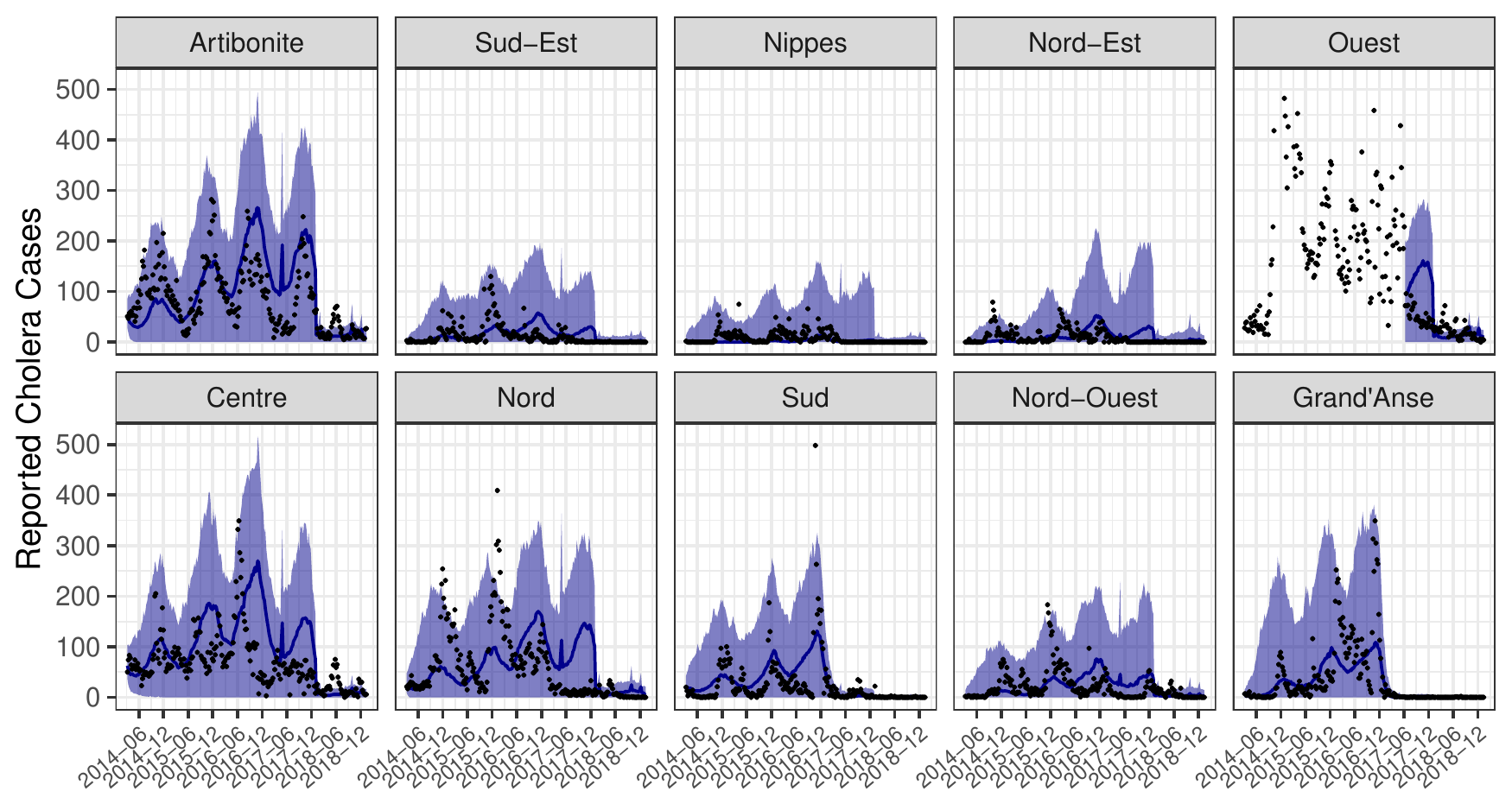} 

}

\end{knitrout}
\caption{\label{fig:mod3rep}
Simulations from Model~3.
Compare to Figure S18 in the supplement of \cite{lee20}.
}
\end{figure}

Model~3 was developed by a team of researchers at the Laboratory of the Swiss Federal Institute of Technology in Lausanne, hereafter referred to as the Model~3 authors.
The code that was originally used to implement Model~3 is archived with the DOI: \url{10.5281/zenodo.3360723}, and also available in the public GitHub repository: \code{jcblemai/haiti-mass-ocv-campaign}.
Because the code was made publicly available, and final model parameters were reported in the supplementary material of \cite{lee20}, we were able to reproduce Model~3 by directly using the source code.
In Fig.~\ref{fig:mod3rep}, we plot simulations from this model.
This figure can be compared to Figure S18 of \cite{lee20}.
We note that slight differences may be accounted for due to variance in the model simulations and the difference in programming language used to produce the figure.
Overall, the high standard of reproducibility that was achieved by the Model~3 authors facilitated the ability to readily replicate their model and results.

\section{Confidence Intervals for Model Parameters}\label{sec:ci}

In this section we provide confidence intervals for all model parameters, excluding those that take unique values for each spatial unit.
For each model and parameter, we use principles of profile likelihood to obtain confidence intervals \cite{pawitan01}.
Due to the non-linear and stochastic nature of Models~1 and 3, exact evaluations of the profile log-likelihood are difficult to obtain.
Instead, the log-likelihood at each point of the profile is estimated using via Monte-Carlo based particle filter methods.
We therefore obtain confidence intervals for the parameters of Model~1 and Model~3 using the Monte Carlo adjust profile (MCAP) algorithm \cite{ionides17}.

In each subsection, we provide figures that show the curvature of the profile log-likelihood near the MLE (Figures~\ref{fig:m1Profs}--\ref{fig:m3Profs}). 
In these figures, the parameter values are shown on the transformed scale in which the profile was calculated. 

\subsection{Model~1 parameters}

\begin{figure}[ht]
\begin{knitrout}
\definecolor{shadecolor}{rgb}{0.969, 0.969, 0.969}\color{fgcolor}

{\centering \includegraphics[width=\maxwidth]{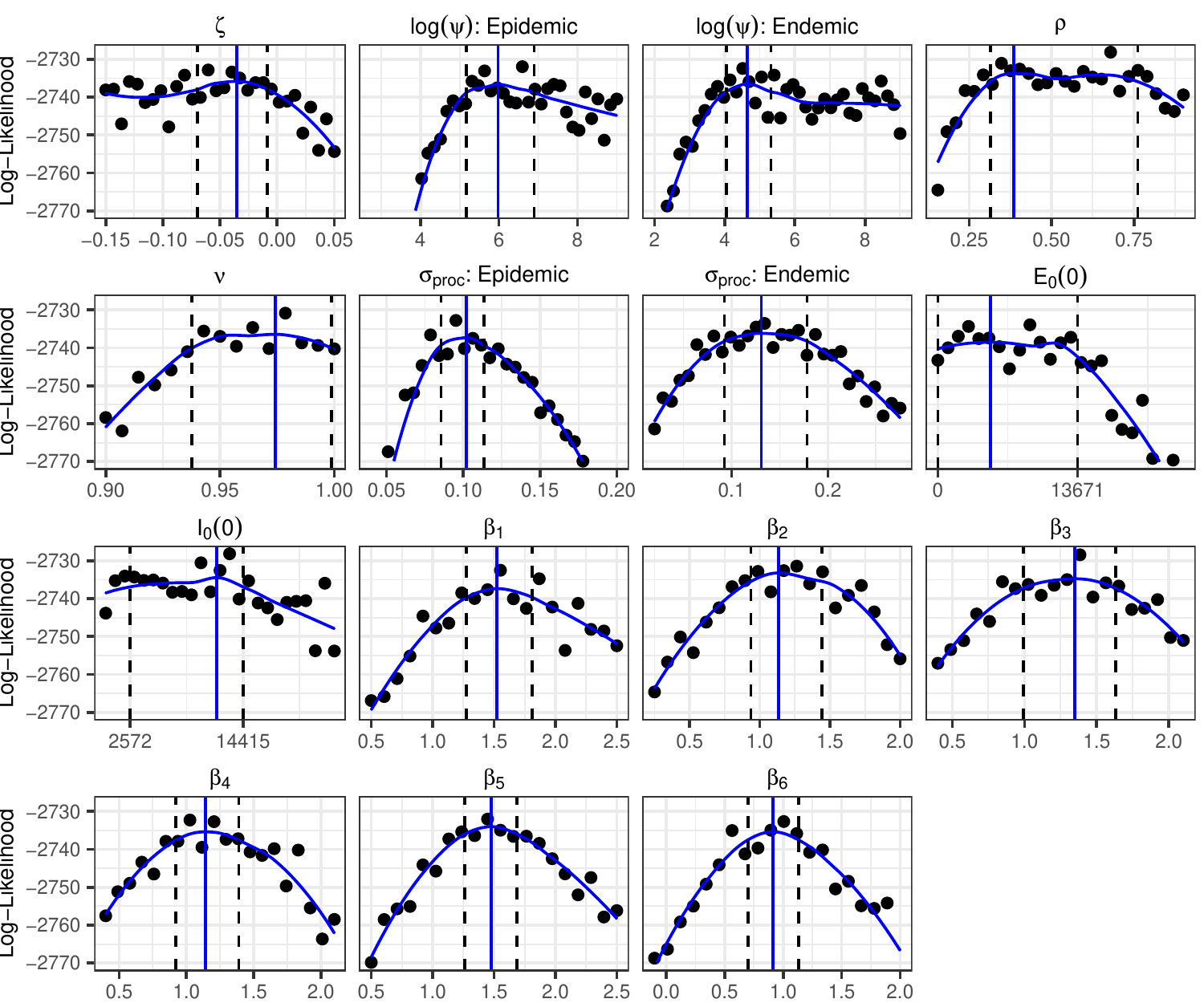} 

}

\end{knitrout}
\caption{\label{fig:m1Profs}MCAP confidence intervals for Model 1 parameters. The vertical blue line indicates the smoothed MLE.}
\end{figure}

Parameter estimates for Model~1, along with the MCAP confidence intervals for the estimate, are given in Table~\ref{tab:mod1CI}. 
Figure~\ref{fig:m1Profs} displays the Monte Carlo evaluations of the profile likelihood values, obtained using a particle filter.

\begin{table}[!h]
\centering
\caption{\label{tab:mod1CI}Model~1 parameter estimates and their corresponding confidence intervals, obtained via the MCAP algorithm.}
\vspace{2mm}
\begin{tabular}{|c|c|c|c|}
\hline
\textbf{Mechanism} & \textbf{Parameter} & \textbf{MLE} & $\bm{95\%}$ \textbf{Confidence Interval}
\\
\hline
\hline

 Seasonality & $\transmissionTrend$ & $-0.036$
   &
  $(-0.070, -0.008)$ 
\\
\hline

 Seasonality & $\transmission_1$ & $1.417$
   &
  $(1.277, 1.811)$ 
\\
\hline

 Seasonality & $\transmission_2$ & $1.169$
   &
  $(0.937, 1.445)$ 
\\
\hline

 Seasonality & $\transmission_3$ & $1.136$
   &
  $(0.990, 1.630)$ 
\\
\hline

 Seasonality & $\transmission_4$ & $1.140$
   &
  $(0.922, 1.389)$ 
\\
\hline

 Seasonality & $\transmission_5$ & $1.401$
   &
  $(1.261, 1.687)$ 
\\
\hline

 Seasonality & $\transmission_6$ & $0.988$
   &
  $(0.699, 1.132)$ 
\\
\hline

 Observation Variance & $\obsOverdispersion: \mathrm{Epi}$ & $279.147$
   &
  $(177.226, 990.191)$ 
\\
\hline

 Observation Variance & $\obsOverdispersion: \mathrm{End}$ & $78.326$
   &
  $(57.171, 204.654)$ 
\\
\hline

  Reporting Rate & $\reportRate$ & $0.679$
   &
  $(0.315, 0.761)$ 
\\
\hline

  Mixing Exponent & $\mixExponent$ & $0.978$
   &
  $(0.938, 0.999)$ 
\\
\hline

  Process noise {\small (wk\textsuperscript{1/2})} & $\sigmaProc: \mathrm{Epi}$ & $0.092$
   &
  $(0.085, 0.113)$ 
\\
\hline

  Process noise {\small (wk\textsuperscript{1/2})} & $\sigmaProc: \mathrm{End}$ & $0.118$
   &
  $(0.092, 0.179)$ 
\\
\hline

  Initial Values & $I_{0}(0)$ & $7298$
   &
  $(2572, \ensuremath{1.4415\times 10^{4}})$ 
\\
\hline

  Initial Values & $E_{0}(0)$ & $350$
   &
  $(1, \ensuremath{1.3671\times 10^{4}})$ 
\\
\hline

\end{tabular}
\end{table}

\subsection{Model~2 parameters}

\begin{figure}[ht]
\begin{knitrout}
\definecolor{shadecolor}{rgb}{0.969, 0.969, 0.969}\color{fgcolor}

{\centering \includegraphics[width=\maxwidth]{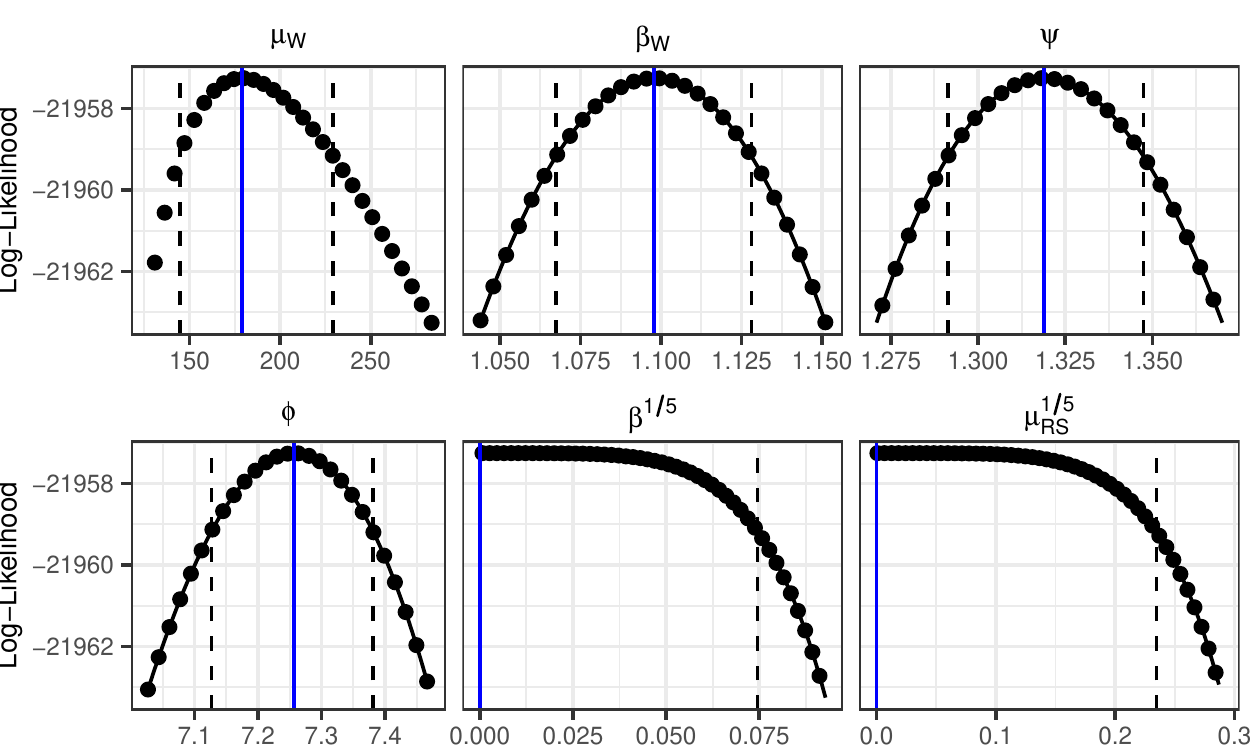} 

}

\end{knitrout}
\caption{\label{fig:m2Profs}MCAP confidence intervals for Model 2 parameters. The vertical blue line indicates the MLE.}
\end{figure}

Parameter estimates for Model~2, along with the profile likelihood confidence intervals for each estimate, are given in Table~\ref{tab:mod2CI}. 
Figure~\ref{fig:m2Profs} displays the profile log-likelihood curve near the MLE. 
In Table~\ref{tab:mod2CI}, the confidence interval for $\muRS^{-1}$, the duration of natural immunity due to cholera infection, is arbitrarily large (going to infinity). 
This is possible because the parameter that was estimated was $\muRS$, and the true MLE for this parameter is zero (see Figure~\ref{fig:m2Profs}). 
This suggests that the fitted model favors a regime where reinfection events are not possible. 
Similarly, the MLE for the parameter $\transmission$, which controls the amount of cholera transmission from human to human, is zero. 
Because Model~2 fails to describe the incidence data as well as a simple statistical benchmark, we must be careful to not interpret these results as evidence that reinfections and human-to-human infection events do not occur.
Instead, we may consider this as additional evidence of model mispecification.

\begin{table}[!h]
\centering
\caption{\label{tab:mod2CI}Model~2 parameter estimates and their corresponding confidence intervals, obtained via profile likelihood.}
\vspace{2mm}
\begin{tabular}{|c|c|c|c|}
\hline
\textbf{Mechanism} & \textbf{Parameter} & \textbf{MLE} & $\bm{95\%}$ \textbf{Confidence Interval}
\\
\hline
\hline

 Human to water shedding {\small (wk\textsuperscript{-1})} & $\Wshed$ & $179.2$
   &
  $(144.6, 229.4)$ 
\\
\hline

 Water to Human Infection {\small (yr\textsuperscript{-1})} & $\beta_W$ & $1.098$
   &
  $(1.067, 1.128)$ 
\\
\hline

 Observation Variance & $\obsOverdispersion$ & $1.319$
   &
  $(1.291, 1.347)$ 
\\
\hline

 Seasonality & $\phaseParm$ & $0.974$
   &
  $(7.127, 7.381)$ 
\\
\hline

Human to Human Infection {\small (yr\textsuperscript{-1})} & $\transmission$ & $\ensuremath{5.97\times 10^{-15}}$\textsuperscript{*}
   &
  $[0, \ensuremath{2.3\times 10^{-6}})$ 
\\
\hline

Immunity {\small (yr)} & $\muRS^{-1}$ & $\ensuremath{1.4\times 10^{11}}$\textsuperscript{*}
   &
  $(1410, \inf)$ 
\\
\hline

\end{tabular}
\begin{flushleft} 
\textsuperscript{*}As evident in Figure~\ref{fig:m2Profs}, the true MLE for these parameters is $0$ and $\infty$, respectively; this value could not be obtained numerically due to the parameter transformation applied to the parameter for the model fitting processes.
\end{flushleft}
\end{table}

\subsection{Model~3 parameters}

\begin{figure}[ht]
\begin{knitrout}
\definecolor{shadecolor}{rgb}{0.969, 0.969, 0.969}\color{fgcolor}

{\centering \includegraphics[width=\maxwidth]{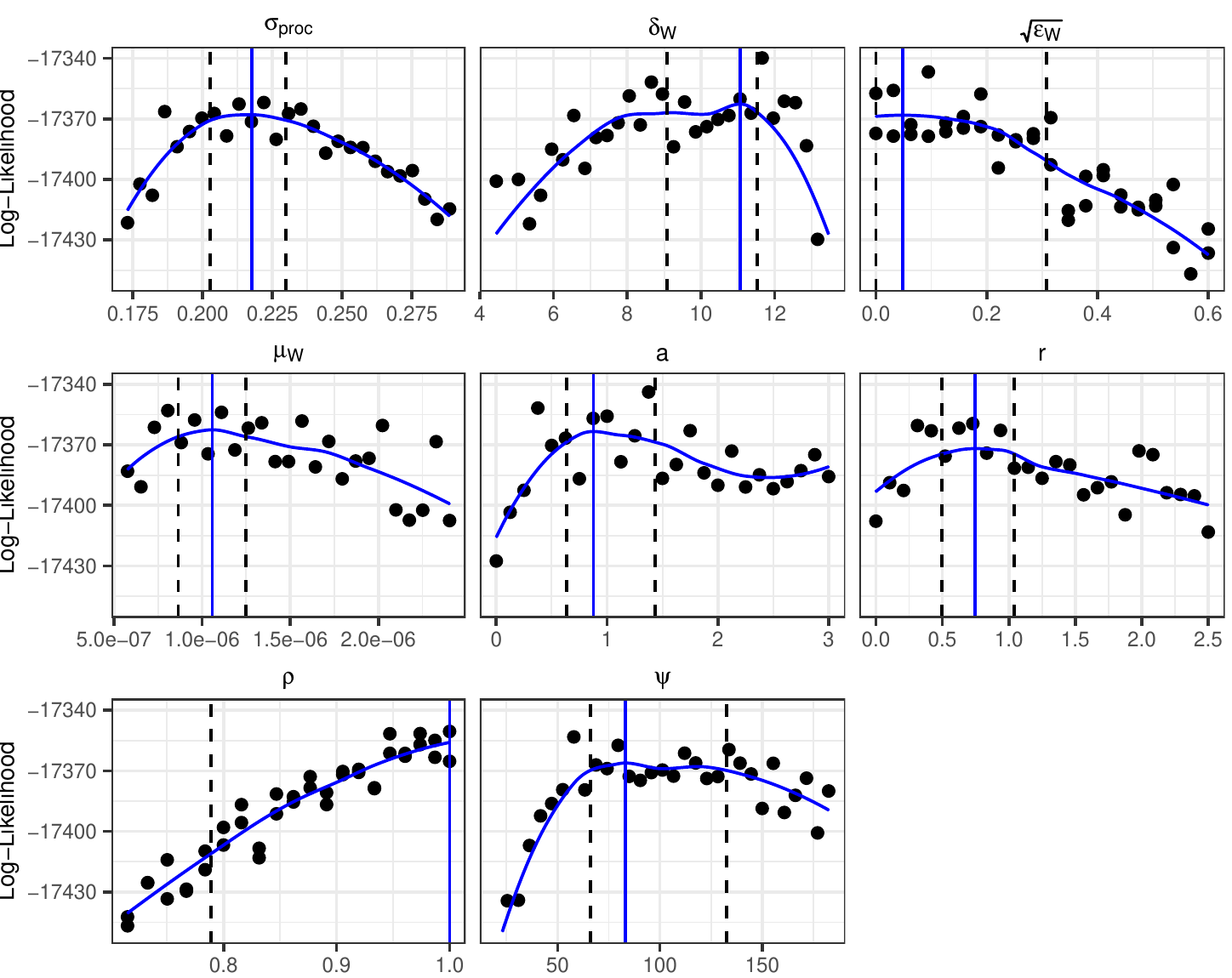} 

}

\end{knitrout}
\caption{\label{fig:m3Profs}MCAP confidence intervals for Model 3 parameters. The vertical blue line indicates the smoothed MLE.}
\end{figure}

Parameter estimates for Model~3, along with the MCAP confidence intervals for the estimate, are given in Table~\ref{tab:mod3CI}. 
Figure~\ref{fig:m3Profs} displays the Monte Carlo evaluations of the profile likelihood values, obtained using a particle filter.

\begin{table}[!h]
\centering
\caption{\label{tab:mod3CI}Model~3 parameter estimates and their corresponding confidence intervals, obtained via the MCAP algorithm.}
\vspace{2mm}
\begin{tabular}{|c|c|c|c|}
\hline
\textbf{Mechanism} & \textbf{Parameter} & \textbf{MLE} & $\bm{95\%}$ \textbf{Confidence Interval}
\\
\hline
\hline

 Process Noise {\small (wk\textsuperscript{1/2})} & $\sigmaProc$ & $0.218$
   &
  $(0.203, 0.230)$ 
\\
\hline

 Water Survival {\small (wk)} & $\Wremoval^{-1}$ & $0.108$
   &
  $(0.087, 0.110)$ 
\\
\hline

 Human to Water Shedding {\small $\frac{\mathrm{km^2}}{\mathrm{wk}}$} & $\Wshed$ & $\ensuremath{9.77\times 10^{-7}}$
   &
  $(\ensuremath{8.64\times 10^{-7}}, \ensuremath{1.25\times 10^{-6}})$ 
\\
\hline

 Asymptomatic Shedding & $\asymptomRelativeShed$ & $0.008$
   &
  $(0.0, 0.095)$ 
\\
\hline

 Seasonality & $\seasAmplitude$ & $1.000$
   &
  $(0.637, 1.432)$ 
\\
\hline

 Seasonality & $\rainfallExponent$ & $0.780$
   &
  $(0.498, 1.041)$ 
\\
\hline

 Reporting Rate & $\reportRate$ & $0.983$
   &
  $(0.789, 1.000)$ 
\\
\hline

 Observation Variance & $\obsOverdispersion$ & $88.578$
   &
  $(66.034, 132.563)$ 
\\
\hline

\end{tabular}
\end{table}

\section{Forecasting with parameter uncertainty}

Let $f_{Y_{1:N}}(y_{1:N} | \theta)$ denote the pdf of the model under consideration, were $\theta$ is a parameter vector that indexes the model.
Furthermore, denote the observed data as $y_{1:N}^*$.
Because the uncertainty in just a single parameter can lead to drastically different forecasts \citep{saltelli20},
parameter uncertainty should be considered when obtaining model forecasts when the goal is to influencing policy.
In a Bayesian modelling paradigm, the most natural way to account for parameter uncertainty in model forecasts is to suppose that $\theta$ comes from a distribution $f_{\Theta}$, and then to obtain $J$ forecasts from the model where each forecast is obtained using parameters drawn from the posterior distribution $\theta_{1:J} \mid Y_{1:N} = y_{1:N}^* \sim f_{\Theta}\big(\theta | Y_{1:N} = y_{1:N}^*\big)$.

When frequentist methods are used, however, there does not exist a posterior distribution from which one could sample.
A common approach could be to obtain a weighted average of the simulations from various models \citep{hoeting99}, but this can be problematic when forecasts from each model are very different from each other \citep{grueber11}.
A similar approach that has been taken \citep{king15} is to obtain model forecasts using multiple sets of parameter values and then sample from the resulting forecasts using weights proportional to the corresponding likelihoods of the parameter values.
This approach could be considered as empirical Bayes, as it is equivalent to using a discrete uniform prior where the set of values in the prior distribution is determined by a stochastic routine applied to the observed data, as discussed below.

For each $k \in 1:K$, let $\theta_k$ be a unique set of model parameters.
Letting $\Theta$ denote a random vector of model parameters, we endow $\Theta$ with a discrete uniform distribution on the set $\{\theta_1, \theta_2, \ldots, \theta_K\}$, such that $P\big(\Theta = \theta_k\big) = \frac{1}{K}$ for all values $k \in \seq{1}{K}$.
Using this as a prior distribution, the posterior distribution of $\Theta | Y_{1:N} = y_{1:N}^*$ can be expressed as:
$P\big(\Theta = \theta_k | Y_{1:N} = y_{1:N}^*\big) = \frac{f_{Y_{1:N}}(y_{1:N}^*| \theta_k)}{\sum_{l = 1}^K f_{Y_{1:N}}(y_{1:N}^*| \theta_l)}$.
In a standard empirical Bayes analysis, the values $\theta_1, \ldots, \theta_k$ of the prior distribution would be chosen using the observed data, resulting in a posterior distribution that weighs the prior parameter vectors proportional to their corresponding likelihoods.
We choose $\theta_k$ to be the output of a stochastic routine applied to the observed data by setting $\theta_k$ to be the output of an iterated filtering algorithm.
In practice, because the likelihood maximization routines of iterated filtering methods are stochastic, it is common to run the iterated filtering method multiple times $(K)$ for each model in order to obtain a maximum likelihood estimate for model parameters.
This results in a natural set of parameters near the MLE that could be used as the discrete prior distribution.
Alternatively, the set $\{\theta_1, \theta_2, \ldots, \theta_K\}$ could be determined by first obtaining marginal confidence intervals for each element of the parameter vector $\Theta$, and then creating a hypercube using the combination of marginal confidence intervals. 
The set $\{\theta_1, \theta_2, \ldots, \theta_K\}$ is then obtained by sampling uniformly $K$ values from the resulting hypercube, as was done by \citep{king15}.

\section{Translating to \citet{lee20} notation}

Since the models of \citet{lee20} were developed independently, the choice of notation varies inconsistently between models.
For our reanalysis, we rename parameters to provide a unified notation facilitating comparison between models.
Table~\ref{tab:translate} maps this notation back to the original notations, for reference.

\begin{table}
  \begin{center}
  \begin{tabular}{|c|c|c|c|c|}\hline
    \multirow{2}{*}{Parameter} & Our & \multicolumn{3}{c|}{Lee et al. (2020a)} \\\cline{3-5}
     & Notation & 1 & 2 & 3 \\
    \hline
    \hline
    Reporting Rate & $\reportRate$ & $\rho$ & $\rho$ & $\epsilon_1, \epsilon_2$ \\\hline
    Mixing Coefficient & $\mixExponent$ & $\nu$ & --- & --- \\\hline
    Measurement Over-Dispersion & $\obsOverdispersion$ & $\tau$ & --- & $p$ \\\hline
    Birth Rate & $\muBirth$ & $\mu$ & --- & --- \\\hline
    Natural Mortality Rate & $\muDeath$ & $\delta$ & --- & $\mu$ \\\hline
    Cholera Mortality Rate & $\choleraDeath$ & --- & --- & $\alpha$ \\\hline
    Latent Period & $1/\muEI$ & $1/\sigma$ & $1/\gamma_E$ & --- \\\hline
    Recovery Rate & $\muIR$ & $\gamma$ & $\gamma$ & $\gamma$ \\\hline
    Loss of Immunity & $\muRS$ & $\alpha$ & $\sigma$ & $\rho$ \\\hline
    Symptomatic Ratio & $\symptomFrac$ & $1 - \theta_0$ & $k$ & $\sigma$ \\\hline
    Asymptomatic Relative Infectiousness & $\asymptomRelativeInfect$ & $1-\kappa$ & $red_\beta$ & --- \\\hline
    Human-to-Water Shedding & $\Wshed$ & --- & $\mu$ & $\theta_I$ \\\hline
    Asymptomatic Relative Shedding & $\asymptomRelativeShed$ & --- & $red_\mu$ & $\theta_A/\theta_I$ \\\hline
    Seasonal Amplitude & $\seasAmplitude$ & --- & $\alpha_s$ & $\lambda$ \\\hline
    Transmission & $\transmission$ & $\beta$ & $\beta$ & $c$ \\\hline
    Water-to-Human & $\Wbeta{}$ & --- & $\beta_W$ & $\beta$ \\\hline
    Bacteria Mortality & $\Wremoval$ & --- & $\delta$ & $\mu_\beta$ \\\hline
    Vaccination Efficacy & $\vaccineEfficacy$ & $\theta_{vk}$ & $\theta_1, \theta_2, \theta_{1_5}, \theta_{2_5}$ & $\eta_{1d}, \eta_{2d}$ \\\hline
    Process Over-dispersion & $\sigmaProc$ & --- & --- & $\sigma_w$\\\hline
  \end{tabular}
  \end{center}
  \caption{\label{tab:translate}Translations between our common notation and notation used by Lee et al (2020).}
\end{table}